\documentclass[10pt,journal,compsoc]{IEEEtran}

\usepackage{graphicx}
\usepackage{amsmath}
\usepackage{amssymb,subfigure,cases}
\usepackage{setspace}
\usepackage{color,hyperref}
\usepackage{algorithm}
\usepackage{algorithmicx}
\usepackage{algpseudocode}
\usepackage{multirow}
\usepackage{booktabs} 
\usepackage{makecell}
\usepackage{url}
\usepackage{bm}
\usepackage{makecell}
\usepackage{amsthm}
\usepackage{booktabs} 
\usepackage{multirow}

\newtheorem{Definition}{Definition} 
\newtheorem{Lemma}{Lemma}
\newtheorem{theorem}{Theorem}

\ifCLASSOPTIONcompsoc
  \usepackage[nocompress]{cite}
\else
  \usepackage{cite}
\fi

\ifCLASSINFOpdf
\else
 \fi

\begin{document}
\title{Entropy Minimizing Matrix Factorization}

\author{Mulin~Chen and Xuelong~Li,~\IEEEmembership{Fellow,~IEEE}
\IEEEcompsocitemizethanks{

\IEEEcompsocthanksitem  All the authors are with  the Center for Optical Imagery Analysis and Learning (OPTIMAL), Northwestern Polytechnical University, Xi'an 710072, Shaanxi, China. E-mails: chenmulin@mail.nwpu.edu.cn, li@nwpu.edu.cn. X. Li is the corresponding author.
}}

\markboth{XXX,~Vol.~XXX, No.~XXX, XXX~XXX}%
{Shell \MakeLowercase{\textit{et al.}}: Bare Advanced Demo of IEEEtran.cls for IEEE Computer Society Journals}

\IEEEtitleabstractindextext{%
\begin{abstract}
Nonnegative Matrix Factorization (NMF) is a widely-used data analysis technique, and has yielded impressive results in many real-world tasks. Generally, existing NMF methods represent each sample with several centroids, and find the optimal centroids by minimizing the sum of the approximation errors. However, the outliers deviating from the normal data distribution may have large residues, and then dominate the objective value seriously. In this study, an Entropy Minimizing Matrix Factorization framework (EMMF) is developed to tackle the above problem. Considering that the outliers are usually much less than the normal samples, a new entropy loss function is established for matrix factorization, which minimizes the entropy of the residue distribution and allows a few samples to have large approximation errors. In this way, the outliers do not affect the approximation of the normal samples. The multiplicative updating rules for EMMF are also designed, and the convergence is proved both theoretically and experimentally. In addition, a Graph regularized version of EMMF (G-EMMF) is also presented to deal with the complex data structure. Clustering results on various synthetic and real-world datasets demonstrate the reasonableness of the proposed models, and the effectiveness is also verified through the comparison with the state-of-the-arts.

\end{abstract}

\begin{IEEEkeywords}
Nonnegative Matrix factorization, entropy loss, robustness, clustering
\end{IEEEkeywords}}

\maketitle

\IEEEpeerreviewmaketitle

\ifCLASSOPTIONcompsoc
\IEEEraisesectionheading{\section{Introduction}\label{sec:introduction}}
\else
\section{Introduction}
\label{sec:introduction}
\fi

\IEEEPARstart{N}{onnegative Matrix Factorization} (NMF) is a popular unsupervised machine learning technique for handling matrix data. Based on the matrix factorization theory~\cite{paatero1994positive}, Lee and Seung~\cite{nmf} imposed the nonnegative constraint to learn the parts-of-whole interpretations. After that, NMF has atrracted sufficent attention due to its simplicity and interpretability, and shown encouraging performance in many real-world tasks, such as face recognition~\cite{xiaofeigraph}, document analysis~\cite{docu}, hyperspectral imagery~\cite{zhangzihan} and recommendation systems~\cite{recom}. 

Specifically, NMF approximates the nonnegative data matrix with the product of two nonnegative factor matrices. One is consist of the basis vectors, and another one is regarded as the coefficient matrix. By minimizing the approximation error, each sample is represented by the linear combination of the basis vectors, and the sample is associated with the basis which contributes the most to the representation. Therefore, the basis vectors act as the cluster centroids, and the coefficient matrix can be regarded as the cluster indicator. In addition, benefited from the nonnegative constraint, NMF allows only additive operation. Consequently, a parts-based representation is achieved, which is able to provide an interpretable understanding about the input data. 

Over the past decades, NMF have been studied from a wide variety of perspectives. For example, researchers have proved the connection between NMF and some popular machine learning techniques~\cite{dingequivalence,rmnmf,ldanmf}, such as $k$-means, spectral clustering and linear discriminant analysis.  A number of techniques~\cite{lcf,srmcf,lccf,cf,tnnconcept} have been proposed to perform NMF in the subspace.  Some works improved NMF by exploiting the data geometry~\cite{gnmf,lefeigraph,chenneru,xiaofeigraph,tnnrse}, while some others deal with the missing data~\cite{missing1,missing2,missing3,missing4}. Recently, deep NMF~\cite{zhangzihan, recom,deep1,deep2} has became an attractive research area. Despite the its salient properties and wide usage, NMF has some major drawbacks. In this paper, we devote to tackle the robustness problem. 

NMF is sensitive to the outliers because it employs the least square error function as the objective. For the outliers, which deviate from the normal distribution, their approximation errors are squared and may dominate the objective function. As a result, they affect the final results seriously. To improve the robustness, some variants of NMF have been presented. In stead of using Frobenius-norm, Ke and Kanade~\cite{l1nmf} proposed the $\ell _{1}$-norm NMF. The influence of the outliers is alleviated, since the approximation errors are summed up directly without taking the square. To maintain the feature rotation invariance, Kong et al.~\cite{l21nmf} utilized the $\ell _{2,1}$-norm formulation. Recently, Qi et al.~\cite{hxnmf} calculated the residue with the logarithmic loss function. Therefore, the objective value increases more slowly with the approximation error. The above methods weaken the effect of the outliers by utilizing different loss functions. However, the effects of outliers still exist. If the error is extremely large, the outliers will affect the results as well. Gao et al.~\cite{gaocap} designed the capped norm matrix factorization model. They found the outliers directly by thresholding the approximation error, and set their residues as a constant. This strategy remove the outliers thoroughly, but it is unrealistic to find a suitable threshold for various real-world applications.

In this paper, an Entropy Minimizing Matrix Factorization framework (EMMF) is presented to improve the robustness. Different from the previous works, we do not approximate all the samples. A new entropy loss function is designed, which models the whole distribution of the residues and avoids the effect of outliers naturally. The proposed loss function could also be applied in other tasks involving matrix computation. In order to preserve the intrinsic geometry, the Graph regularized EMMF (G-EMMF) is also developed. The main contributions made in this study are summarized as follows.

\begin{itemize}
\item[1.] We design a general entropy loss function for matrix factorization. By minimizing the entropy of the residue distribution, the proposed EMMF allows a few samples to be with relatively large errors, and focuses on approximating the most of the rest. Therefore, the outliers do not affect the updating of centroids. 

\item[2.] We provide the efficient optimization algorithms for the proposed framework. The optimal solution can be obtained by the multiplicative updating rules with proved convergence. The computation costs of the optimization algorithm is almost the same as NMF, which guarantees the practicability for real-world tasks.

\item[3.] We conduct extensive experiments to validate the reasonableness and effectiveness of the proposed framework. As demonstrated by the results, the objective function is insensitive to the outliers with extremely large errors, and it works well for the data without outliers. The proposed G-EMMF also outperforms the existing graph-regularized NMF methods.
\end{itemize}

The paper is organized as follows. Section~\ref{pre} reviews some existing NMF algorithms. Section~\ref{sec:emmf} introduces the EMMF formulation, and provides the corresponding optimization algorithm. Section~\ref{sec:gemmf} presents the G-EMMF. Section~\ref{sec:experemmf} gives the experimental results of EMMF, and discusses its advantages. Section~\ref{sec:expergemmf} shows the clustering performance of G-EMMF. Section~\ref{sec:conclusion} concludes this article

\textbf{Notations:} in this paper, we write the matrices as uppercase and write the vectors as lowercase. For a matrix $\mathbf{A}$, its $(i,k)$-th element is defined as $\mathbf{A}_{ik}$. Its $i$-th row, column are denoted as $\mathbf{a}_{i,:}$ and $\mathbf{a}_{i}$ respectively. The trace of $\mathbf{A}$ is defined as ${\rm Tr} (\mathbf{A})$. The transpose of $\mathbf{A}$ and $\mathbf{a}_i$ are indicated by $\mathbf{A}^T$ and $\mathbf{a}_i^T$. $\mathbf{I}$ is the identity matrix. The $\ell _ \rho$ norm of $\mathbf{a}_i$ is calculated as $||\mathbf{a}_i||_\rho = \sqrt {\sum\limits_{k = 1}^d {|\mathbf{A}_{ki}|^\rho} }$ ($\rho>0$).

\section{Preliminary}
\label{pre}
In this section, we revisited the formulation of NMF and some existing methods. Numerous algorithms have been proposed to improve NMF from different aspects, and we mainly focus on the robust variants.

\subsection{Nonnegative Matrix Factorization}
Supposing the data matrix is $\mathbf{X} = [\mathbf{x}_1,\mathbf{x}_2,\cdots,\mathbf{x}_n ] \in \mathbb{R}^{d\times n}$ and the desired centroid number is $c$, NMF aims to find the nonnegative matrices $\mathbf{U}\in \mathbb{R}^{d\times c}$ and $\mathbf{V}\in \mathbb{R}^{n\times c}$ which satisfy  $\mathbf{X} \approx \mathbf{U}{\mathbf{V}^T}$. The least square error objective function is formulated as
\begin{equation}
\mathop {\min }\limits_{ \mathbf{U}\ge 0,\mathbf{V} \ge 0} ||\mathbf{X - U}{\mathbf{V}^T}||_F^2 = \sum\limits_{i = 1}^n {||{\mathbf{x}_i} - \mathbf{Uv}_{i,:}^T||_2^2} ,
\label{e-nmf}
\end{equation}
where $||\cdot||_F$ is the Frobenius norm. Lee and Seung derived the multiplicative updating rules for the above problem:
\begin{equation}
\begin{split}
{\mathbf{U}_{ik}} \leftarrow {\mathbf{U}_{ik}}\frac{{{{(\mathbf{XV})}_{ik}}}}{{{{(\mathbf{U}{\mathbf{V}^T}\mathbf{V})}_{ik}}}},\\
{\mathbf{V}_{jk}} \leftarrow {\mathbf{V}_{jk}}\frac{{{{({\mathbf{X}^T}\mathbf{U})}_{jk}}}}{{{{(\mathbf{V}{\mathbf{U}^T}\mathbf{U})}_{jk}}}}.
\end{split}
\nonumber
\end{equation}

Instead of using Euclidean distance, another commonly used NMF uses the "divergence" as the loss function:
\begin{equation}
\mathop {\min }\limits_{\mathbf{U} \ge 0,\mathbf{V} \ge 0} {\rm DIV} (\mathbf{X||UV}^T) ,
\label{e-nmfkl}
\end{equation}
where ${\rm DIV}(\mathbf{A||B})=\sum \limits _{i=1}^d {\sum \limits _{k=1}^n (\mathbf{A}_{ij}\log \frac{\mathbf{A}_{ij}}{\mathbf{B}_{ij}}-\mathbf{A}_{ij}+\mathbf{B}_{ij})}$ is called as the divergence measurement. Similarly, the updating rules are given as
\begin{equation}
\begin{split}
{\mathbf{U}_{ik}} \leftarrow {\mathbf{U}_{ik}}\frac{{\sum\limits_{j = 1}^n {{\mathbf{X}_{ij}}{\mathbf{V}_{jk}}/} {{(\mathbf{U}{\mathbf{V}^T})}_{ij}}}}{{\sum\limits_{j = 1}^n {{\mathbf{V}_{jk}}} }},\\
{\mathbf{V}_{jk}} \leftarrow {\mathbf{V}_{jk}}\frac{{\sum\limits_{i = 1}^d {{\mathbf{X}_{ij}}{\mathbf{U}_{ik}}/{{(\mathbf{U}{\mathbf{V}^T})}_{ij}}} }}{{\sum\limits_{i = 1}^d {{\mathbf{U}_{ik}}} }},
\end{split}
\nonumber
\end{equation}
By solving $\mathbf{U}$ and $\mathbf{V}$ iteratively, the local optimal solution of both problem~(\ref{e-nmf}) and (\ref{e-nmfkl}) will be found. Besides, the problems can be decompose into the approximation of each sample, i.e. $\mathbf{x}_i\approx \mathbf{Uv}_{i,:}^T$. The largest element in $\mathbf{v}_{i,:}$ indicates the closet centroid to $\mathbf{x}_i$. Therefore, $\mathbf{V}$ indicates the clustering results directly. 

Since NMF minimizes the residues of all the samples, outliers with large approximation errors will affect the optimization inevitably. Furthermore, the square of the errors compounds the problem severely. 

\subsection{Robust NMF}
To improve the robustness to outliers, some variants of NMF have been put forward. Ke and Kanade~\cite{l1nmf} replaced the Frobenius norm with the $\ell _1$ norm, which yields the following problem
\begin{equation}
\mathop {\min }\limits_{ \mathbf{U}\ge 0,\mathbf{V} \ge 0} ||\mathbf{X - U}{\mathbf{V}^T}||_{1} = \sum\limits_{i = 1}^n {|{\mathbf{x}_i} - \mathbf{Uv}_{i,:}^T|}, 
\label{e-nmfl1}
\end{equation}
and the model is solved by convex programming. Instead of squaring the approximation errors, problem~(\ref{e-nmfl1}) takes the $\ell _1$ norm as the objective function. Therefore, the large residues are depressed.

Considering that $\ell _1$ norm is sensitive to feature rotation, Kong et al.~\cite{l21nmf} proposed the $\ell _{2,1}$ norm NMF with the following formulation
\begin{equation}
\mathop {\min }\limits_{ \mathbf{U}\ge 0,\mathbf{V} \ge 0} ||\mathbf{X - U}{\mathbf{V}^T}||_{2,1} = \sum\limits_{i = 1}^n {||{\mathbf{x}_i} - \mathbf{Uv}_{i,:}^T||_2}.
\label{e-nmfl21}
\end{equation}
Similar with the $\ell _1$ norm NMF, problem~(\ref{e-nmfl21}) takes off the square operation. Besides, because $||\mathbf{A}||_{2,1}$ equals to $||\mathbf{AR}||_{2,1}$ for any rotation matrix $\mathbf{R}$, $\ell _{2,1}$ norm NMF is invariant to the feature rotation. Furthermore, it also achieves the structural sparsity. Considering the above advantages, $\ell _{2,1}$ norm NMF has been extensively studied in the literature~\cite{chenneru,lefeigraph,lefeigraph2,robust1}, and became one of the most popular robust NMF.
Based on $\ell _{2,1}$ norm NMF, Huang et al.~\cite{rmnmf} removed the nonnegative constraint on $\mathbf{U}$ to handle the negative data, leading to
\begin{equation}
\mathop {\min }\limits_{ \mathbf{V}\ge 0,\mathbf{VV}^T = I} ||\mathbf{\mathbf{X - U}}{\mathbf{\mathbf{V}}^T}||_{2,1} +\lambda {\rm Tr}(\mathbf{V}^T\mathbf{LV}).
\nonumber
\end{equation}
The Laplacian graph $\mathbf{L}$ is utilized to preserve the local manifold structure. Cluster indicator $\mathbf{V}$ is constrained to be orthogonal to keep the uniqueness of solution. Ding et al.~\cite{dingorth} further pointed out that the orthogonal constraint facilitates the interpretation of the clustering results.

Du et al.~\cite{correntropy} employed the correntropy induced metric to calculate the error, and proposed the following problem
\begin{equation}
\mathop {\max }\limits_{\beta,U \ge 0,V \ge 0 } \sum\limits_{i = 1}^d {\sum\limits_{k = 1}^n {\exp [ - {{({\mathbf{X}_{ik}} - \sum\limits_{j = 1}^c {{\mathbf{U}_{ij}}{\mathbf{\mathbf{V}}_{jk}}} )^2}}/2{\beta ^2}]} } ,
\nonumber
\end{equation}
where $\beta$ is the optimal Gaussian variance to be learned. The above model can handle the non-Gaussian outliers.
To further weaken the large residues, Qi et al.~\cite{hxnmf} designed the logarithmic loss function:
\begin{equation}
\mathop {\min }\limits_{\mathbf{U} \ge 0,{\bf{\mathbf{V}}} \ge 0} \sum\limits_{i = 1}^n {\log (1 + ||{\mathbf{x}_i} - \mathbf{Uv}_{i,:}^T|{|_2})}.
\nonumber
\end{equation}

There are also many other robust cost functions for NMF, such as the $\ell _{1,2}$ norm loss~\cite{l12,l12nmf2} and hypersurface loss~\cite{hypersurface}. However, all of them share the same drawback with the classical NMF that they minimize the errors of all the samples. Consequently, the outliers with extremely large errors may still affect the results. Gao et al.~\cite{gaocap} proposed to remove the sample if its error exceeds a certain threshold, which may be inappropriate for various kinds of tasks.

\section{Entropy Minimizing Matrix Factorization}
\label{sec:emmf}
In this section, the Entropy Minimizing Matrix Factorization framework (EMMF) is introduced. The optimization strategy and the convergence analysis are also given.

\subsection{Methodology}
Before describing the formulation of EMMF, we first introduce the concept of entropy. Defining $\{p_i\}$ as the probability distribution of a random variable, the Shannon entropy is given by
\begin{equation}
H =  - \sum\limits_i {{p_i}\log } {p_i}.
\nonumber
\end{equation}
According to the information theory~\cite{shannon}, the entropy is maximized when the distribution is uniform, i.e. all the probabilities are with the same value. Conversely, the less value of the entropy indicates the imbalance distribution.

Defining $\mathbf{M=X-UV^T}$, we define $p_i$ as
\begin{equation}
p_i=\frac{||\mathbf{m}_i||_2}{||\mathbf{M}||_{2,1}},i\in [1,n].
\nonumber
\end{equation}
According to the aforementioned definitions, it is manifest that $\sum\nolimits_{i = 1}^n {{p_i}}  = 1$. Therefore, $\{p_i\}$ is exact the samples' residue distribution, and the entropy is computed as
\begin{equation}
H(\mathbf{M}) =  - \sum\limits_{i = 1}^n {\frac{{||\mathbf{m}_i|{|_2}}}{{||\mathbf{M}|{|_{2,1}}}}\log \frac{{||\mathbf{m}_i|{|_2}}}{{||\mathbf{M}|{|_{2,1}}}}}.
\label{entropy-1}
\end{equation}
The value of $H(\mathbf{M})$ is minimized when the residue distribution is extremely imbalance. However, the distribution can not reflect the exact value of the residues, i.e. $H(\mathbf{M})$ equals to $H(\rho \mathbf{M})$ for any $\rho >0$.  To keep the uniqueness, the matrix residue $||\mathbf{M}||_{2,1}$ should be also minimized. Since both the entropy and matrix residue are with positive values, we propose to minimize their product
\begin{equation}
\begin{split}
&\mathop {\min }\limits_{\mathbf{U,V}} H(\mathbf{M}) \times ||\mathbf{M}|{|_{2,1}},\\
s.t.\mathbf{M} &= \mathbf{X - U}{\mathbf{V}^T},\mathbf{U} \ge 0,\mathbf{V} \ge 0.
\end{split}
\nonumber
\end{equation}
which yields the objective function of EMMF
\begin{equation}
\begin{split}
&\mathop {\min }\limits_{\mathbf{\mathbf{U,V}}}  - \sum\limits_{i = 1}^n {||{\mathbf{m}_i}|{|_2}\log \frac{{||{\mathbf{m}_i}|{|_2}}}{{||\mathbf{M}|{|_{2,1}}}}}, \\
s.&t.\mathbf{M = X - U}{\mathbf{V}^T},\mathbf{U} \ge 0,\mathbf{V} \ge 0.
\end{split}
\label{entropy-obj}
\end{equation}

If several samples are with large approximation errors, EMMF just let them be and moves the centroids towards the remaining ones. By searching an imbalance residue distribution, the effects of the outliers are avoided. One may doubt the correctness of the model for the data without outliers. In fact, since all the samples are with uniform distribution, the centroids will not change too much if a few samples are considered as outliers. Therefore, the mistaken outliers  can still be connected with the correct centroid. This statement will be verified in Section~\ref{sec:experemmftoy}.

\subsection{Optimization}
Due to the dependency between $\mathbf{M}$, $\mathbf{U}$ and $\mathbf{V}$, it is difficult to solve problem~(\ref{entropy-obj}) directly. In this part, we transform the objective into the trace from. 

Taking $\mathbf{m}_i$ as the variable to be optimized, the first order derivative of the objective is $- \frac{{{\mathbf{m}_i}}}{{||{\mathbf{m}_i}|{|_2}}}\log \frac{{||{\mathbf{m}_i}|{|_2}}}{{||\mathbf{M}|{|_{2,1}}}}\ge 0$, and the second order derivative is $\frac{1}{{||\mathbf{M}|{|_{2,1}}}} - \frac{1}{{||{\mathbf{m}_i}|{|_2}}} \le 0$. Thus, the proposed function is monotonic increasing w.r.t. $\mathbf{m}_i$. The Lagrange function is
\begin{equation}
{\cal L}_1(\mathbf{m}_i) =  - \sum\limits_{i = 1}^n {||{{\bf{m}}_i}|{|_2}\log \frac{{||{{\bf{m}}_i}|{|_2}}}{{||{\bf{M}}|{|_{2,1}}}} + } {\cal G}(\mathbf{m}_i,\Lambda _1 ),
\nonumber
\end{equation}
where ${\cal G}(\mathbf{m}_i,\Lambda _1)$ represents the constraint on $\mathbf{m}_i$, $\Lambda _1$ is the Lagrange multiplier. Computing the derivative of ${\cal L}_1(\mathbf{m}_i) $ and setting it as zero, we get the optimal solution as
\begin{equation}
{\mathbf{Q}_{ii}}{\mathbf{m}_i} + \frac{{\partial {\cal G}(\mathbf{m}_i,\Lambda _1 )}}{{\partial {\mathbf{m}_i}}}=0,
\label{lzero}
\end{equation}
where $\mathbf{Q}\in \mathbb{R}^{n\times n}$ is the diagonal matrix with
\begin{equation}
\mathbf{Q}_{ii} = - \frac{1}{{||{\mathbf{m}_i}|{|_2}}}\log \frac{{||{\mathbf{m}_i}|{|_2}}}{{||\mathbf{M}|{|_{2,1}}}}.
\label{updateq}
\end{equation}
In implementation, we add a small enough factor $\varepsilon$ on $||\mathbf{m}_i||_2$ to prevent it from being zero.
When $\mathbf{Q}$ is set as stationary, Eq.~(\ref{lzero}) is also the optimal solution to the following problem
\begin{equation}
\begin{split}
\min \limits_{\mathbf{\mathbf{\mathbf{M}}}} {\rm Tr}(\mathbf{MQM}^T) =\sum\limits_{i = 1}^n {{\mathbf{Q}_{ii}}||{\mathbf{m}_i}||_2^2}, \\
s.t.\mathbf{M = X - U}{\mathbf{V}^T},\mathbf{U} \ge 0,\mathbf{V} \ge 0.
\end{split}
\label{tra}
\end{equation}
Then the optimal $\mathbf{M}$ of objective~(\ref{entropy-obj}) can be obtained by solving problem~(\ref{tra}). Accordingly, we search the optimal $\mathbf{U}$ and $\mathbf{V}$ by solving
\begin{equation}
\begin{split}
\min \limits_{\mathbf{U} \ge 0,\mathbf{V} \ge 0} {\rm Tr}[(\mathbf{X-UV}^T)\mathbf{Q}(\mathbf{X-UV}^T)^T].
\end{split}
\label{trauv}
\end{equation}
In each iteration, $\mathbf{Q}$ is updated with the current $\mathbf{U,V}$ according to Eq.~(\ref{updateq}). The updating rules of $\mathbf{U}$ and $\mathbf{V}$ as follows.

\textbf{Updating} $\mathbf{U}$, problem~(\ref{trauv}) becomes
\begin{equation}
\min \limits_{\mathbf{U} \ge 0} {\rm Tr}(\mathbf{UV}^T\mathbf{QVU}^T) - 2{\rm Tr}(\mathbf{XQV}{\mathbf{U}^T}).
\label{fu}
\end{equation}
The above sub-problem is convex, and the Lagrangian function is
\begin{equation}
\begin{split}
{\cal L}_2(\mathbf{U})={\rm Tr}(\mathbf{UV}^T\mathbf{QVU}^T) - 2{\rm Tr}(\mathbf{XQV}{\mathbf{U}^T})+{\rm Tr}(\Lambda _2\mathbf{U}^T),
\end{split}
\label{lu}
\end{equation}
where $\Lambda _2\in \mathbb{R}^{d\times c}$ is the Lagrangian multiplier. Let $\frac{\partial {\cal L}_2({\bf{U}})}{\partial {\bf{U}}}$ to be zero, we have
\begin{equation}
2\mathbf{U}{\mathbf{V}^T}\mathbf{QV} - 2\mathbf{XQV} + {\Lambda _2}=0.
\label{holdsu}
\end{equation}
According to the KKT conditions $(\Lambda _2)_{ik} \mathbf{U}_{ik}=0$, we get
\begin{equation}
(\mathbf{U}{\mathbf{V}^T}\mathbf{QV)}_{ik} \mathbf{U}_{ik}^2 - (\mathbf{XQV})_{ik} \mathbf{U}_{ik}^2=0.
\label{updateu1}
\end{equation}
Then the updating rule of $\mathbf{U}$ is
\begin{equation}
{\mathbf{U}_{ik}} \leftarrow {\mathbf{U}_{ik}}\sqrt {\frac{{{{(\mathbf{XQV})}_{ik}}}}{{{{(\mathbf{U}{\mathbf{V}^T}\mathbf{QV})}_{ik}}}}}.
\label{updateu}
\end{equation}
At convergence, the equality holds for Eq.~(\ref{updateu}), so the condition in Eq.~(\ref{holdsu}) is satisfied.

\textbf{Updating} $\mathbf{V}$, the sub-problem is
\begin{equation}
\min \limits_{\mathbf{V} \ge 0} {\rm Tr}(\mathbf{V}^T\mathbf{QVU}^T\mathbf{U}) - 2{\rm Tr}(\mathbf{V}^T\mathbf{QX}^T\mathbf{U}).
\label{fv}
\end{equation}
The Lagrangian function is
\begin{equation}
\begin{split}
{\cal L}_3(\mathbf{V})={\rm Tr}(\mathbf{V}^T\mathbf{QVU}^T\mathbf{U}) - 2{\rm Tr}(\mathbf{V}^T\mathbf{QX}^T\mathbf{U})+{\rm Tr}(\Lambda _3\mathbf{V}^T),
\end{split}
\nonumber
\end{equation}
where $\Lambda _3\in\mathbb{R}^{n\times c}$ is the Lagrangian multiplier. Similar with Eq.~(\ref{updateu1}), we have
\begin{equation}
{(\mathbf{QV}{\mathbf{U}^T}\mathbf{U})_{ik}}\mathbf{V}_{ik}^2 - {(\mathbf{Q}{\mathbf{X}^T}\mathbf{U})_{ik}}\mathbf{V}_{ik}^2 = 0,
\nonumber
\end{equation}
and the updating rule of $\mathbf{V}$ is
\begin{equation}
{\mathbf{V}_{ik}} \leftarrow {\mathbf{V}_{ik}}\sqrt {\frac{{{{(\mathbf{Q}{\mathbf{X}^T}\mathbf{U})}_{ik}}}}{{{{(\mathbf{QV}{\mathbf{U}^T}\mathbf{U})}_{ik}}}}}.
\label{updatev}
\end{equation}
At convergence,  $\mathbf{V}$ satisfies the condition $\frac{\partial {\cal L}({\bf{V}},{\Lambda _3})}{\partial {\bf{V}}}=0$.

The details of the optimization for problem~(\ref{trauv}) is described in Algorithm~\ref{alg1}. In each iteration, the computation of the diagonal matrix $\mathbf{Q}$ requires ${\cal O}(ndc)$ operations. The costs  for updating $\mathbf{U}$ and $\mathbf{V}$ are also ${\cal O}(ndc)$. After $t$ iterations, the overall cost of EMMF is ${\cal O}(tndc)$.

\begin{algorithm}
        \caption{Optimization algorithm of EMMF}
        \begin{algorithmic}[1]
        \Require Data matrix $\mathbf{X}$, centroid number $c$.
           \State Initialize $\mathbf{U}$ and $\mathbf{V}$.
           
        	\Repeat
			\State Compute $\mathbf{Q}$ with Eq.~(\ref{updateq}).
			\State Update $\mathbf{U}$ with Eq.~(\ref{updateu}).
			\State Update $\mathbf{V}$ with Eq.~(\ref{updatev}).
			\Until{Converge}
		\Ensure Optimal $\mathbf{U}$, $\mathbf{V}$.
        \end{algorithmic}
        \label{alg1}
    \end{algorithm}

\subsection{Convergence}

The convergence analysis consists of two parts. First, we demonstrate that problem~(\ref{tra}) converges to the optimal solution to objective~(\ref{entropy-obj}). After that, since the updating rules of $\mathbf{U}$ and $\mathbf{V}$ are in similar form, we only prove the convergence of problem~(\ref{fu}). 

\textbf{Convergence of problem}~(\ref{tra}):
we introduce the following theorem.
\begin{theorem}
The optimization of problem~(\ref{trauv}) decreases the objective value of problem~(\ref{entropy-obj}) monotonically.
\end{theorem}

\begin{proof}
Denote the value of $\mathbf{M}$, $\mathbf{Q}$ at the $t$-th iteration are $\mathbf{M}^{(t)}$ and $\mathbf{Q}^{(t)}$, and suppose the value of problem~(\ref{tra}) decreases through the optimization, i.e.
\begin{equation}
\sum\limits_{i = 1}^n {{\bf{Q}}_{ii}^{(t)}||{\bf{m}}_i^{(t)}||_2^2} \ge \sum\limits_{i = 1}^n {{\bf{Q}}_{ii}^{(t+1)}||{\bf{m}}_i^{(t+1)}||_2^2}.
\nonumber
\end{equation}
With the definition of $\mathbf{Q}$ in Eq.~(\ref{updateq}), the above inequality is transformed into 
\begin{equation}
\begin{split}
&\sum\limits_{i = 1}^n { - \frac{{||{\mathbf{m}_i}^{(t)}||_2^2}}{{||{\mathbf{m}_i}^{(t)}|{|_2}}}\log \frac{{||{\mathbf{m}_i}^{(t)}|{|_2}}}{{||{\mathbf{M}^{(t)}}|{|_{2,1}}}}}  \ge \\
& \sum\limits_{i = 1}^n { - \frac{{||{\mathbf{m}_i}^{(t + 1)}||_2^2}}{{||{\mathbf{m}_i}^{(t)}|{|_2}}}\log \frac{{||{\mathbf{m}_i}^{(t)}|{|_2}}}{{||{\mathbf{M}^{(t)}}|{|_{2,1}}}}} ,
\end{split}
\nonumber
\end{equation}
which leads to
\begin{equation}
\sum\limits_{i = 1}^n {(||\mathbf{m}_i^{(t + 1)}|{|_2} - ||\mathbf{m}_i^{(t)}|{|_2})\log \frac{{||\mathbf{m}_i^{(t)}|{|_2}}}{{||\mathbf{M}^{(t)}|{|_{2,1}}}}}  \ge 0.
\label{e-conv1}
\end{equation}
According to the log sum inequality~\cite{logsum}, we know that
\begin{equation}
\sum\limits_{i = 1}^n {||\mathbf{m}_i^{(t + 1)}|{|_2}(\log \frac{{||\mathbf{m}_i^{(t + 1)}|{|_2}}}{{||\mathbf{m}_i^{(t)}|{|_2}}} - \log \frac{{||\mathbf{M}^{(t + 1)}|{|_{2,1}}}}{{||{\mathbf{M}^{(t)}}|{|_{2,1}}}})}  \ge 0.
\label{e-conv2}
\end{equation}
Summing up Eq.~(\ref{e-conv1}) and~(\ref{e-conv2}), the following inequality holds:
\begin{equation}
\begin{split}
&\sum\limits_{i = 1}^n { - ||\mathbf{m}_i^{(t)}|{|_2}\log \frac{{||\mathbf{m}_i^{(t)}|{|_2}}}{{||{\mathbf{M}^{(t)}}|{|_{2,1}}}}}  \ge \\
& \sum\limits_{i = 1}^n { - ||\mathbf{m}_i^{(t + 1)}|{|_2}\log \frac{{||\mathbf{m}_i^{(t + 1)}|{|_2}}}{{||{\mathbf{M}^{(t + 1)}}|{|_{2,1}}}}},
\end{split}
\nonumber
\end{equation}
which completes the proof.
\end{proof}

\textbf{Convergence of problem}~(\ref{fu}):
to demonstrate that the updating rule~(\ref{updateu}) decreases the value of problem~(\ref{fu}), the following definition~\cite{nmf} is introduced.
\begin{Definition}
$g(\mathbf{U},{ \mathbf{\tilde U}})$ is the auxiliary function for $f(\mathbf{U})$ if for any $\mathbf{U}$ and ${\mathbf{\tilde U}}$ it satisfies
\begin{equation}
g(\mathbf{U},{\mathbf{\tilde U}})\ge f(\mathbf{U}), g(\mathbf{U,U})=f(\mathbf{U}).
\label{auxilariyu}
\end{equation}
\end{Definition}

As proved by Lee and Seung~\cite{nmf}], we have the following lemma:
\begin{Lemma}
Given the auxiliary function $g(\mathbf{U},{\mathbf{\tilde U}})$, $f(\mathbf{U}^{(t+1)})\le f(\mathbf{U}^{(t)})$ holds if $\mathbf{ U}^{(t+1)}$ is the solution to
\begin{equation}
\min \limits _{\mathbf{ U}} g(\mathbf{U},{\mathbf{ U}^{(t)}}).
\label{guu}
\end{equation}
\label{lemmaguu}
\end{Lemma}

We propose the following theorem to demonstrate the convergence of problem~(\ref{fv}).

\begin{theorem}
Updating rule~(\ref{updateu}) decreases the Lagrangian function ${\cal L}_2(\mathbf{U})$ in Eq.~(\ref{lu}) monotonically.
\label{theorem2}
\end{theorem}
\begin{proof}
As Ding et al.~\cite{dingconvex} pointed out, for any nonnegative matrices $\mathbf{A}\in\mathbb{R}^{d\times d}$, $\mathbf{B}\in\mathbb{R}^{c\times c}$, $\mathbf{C}\in\mathbb{R}^{d\times c}$, $\mathbf{G} \in\mathbb{R}^{d\times c}$, if $\mathbf{A}$ and $\mathbf{B}$ are symmetric, the following inequality holds:
\begin{equation}
{\rm Tr}(\mathbf{C}^T\mathbf{ACB})\le  \sum\limits_{i = 1}^d {\sum\limits_{k = 1}^c {\frac{{{{(\mathbf{AGB})}_{ik}}\mathbf{C}_{ik}^2}}{{{\mathbf{G}_{ik}}}}} } .
\nonumber
\end{equation}
Based on the above equation, the upper bound of the first term of $f(\mathbf{U})$ is written as
\begin{equation}
\begin{split}
{\rm Tr}(\mathbf{U}{\mathbf{V}^T}\mathbf{QVU}^T) &= {\rm Tr}({\mathbf{U}^T}\mathbf{U}{\mathbf{V}^T}\mathbf{QV})\\
&\le  \sum\limits_{i = 1}^d {\sum\limits_{k = 1}^c} {\frac{{{{({\mathbf{U}^{(t)}}{\mathbf{V}^T}\mathbf{QV})}_{ik}}\mathbf{U}_{ik}^2}}{{\mathbf{U}_{ik}^{(t)}}}} .
\end{split}
\nonumber
\end{equation}
For any scalar $\rho \ge 0$, we have $\rho\ge 1+\log \rho$, which leads to the following lower bound 
\begin{equation}
\begin{split}
Tr(\mathbf{XQVU}^T) &= \sum\limits_{i = 1}^d {\sum\limits_{k = 1}^c} {{{(\mathbf{XQV})}_{ik}}{\mathbf{U}_{ik}}} \\
 &\ge \sum\limits_{i = 1}^d {\sum\limits_{k = 1}^c} {{{(\mathbf{XQV})}_{ik}}\mathbf{U}_{ik}^{(t)}(1 + \log \frac{{{\mathbf{U}_{ik}}}}{{\mathbf{U}_{ik}^{(t)}}})} .
\end{split}
\nonumber
\end{equation}
The last term of ${\cal L}_2(\mathbf{U})$ equals to zero, so we do not consider it. Based on the bounds of the first two terms, the auxiliary function of ${\cal L}_2(\mathbf{U})$ is
\begin{equation}
\begin{split}
g(\mathbf{U},{\mathbf{U}^{(t)}}) =& \sum\limits_{i = 1}^d {\sum\limits_{k = 1}^c} {\frac{{{{({\mathbf{U}^{(t)}}{\mathbf{V}^T}\mathbf{QV})}_{ik}}\mathbf{U}_{ik}^2}}{{\mathbf{U}_{ik}^{(t)}}}} \\
 &- 2\sum\limits_{i = 1}^d {\sum\limits_{k = 1}^c} {{{(\mathbf{XQV})}_{ik}}\mathbf{U}_{ik}^{(t)}(1 + \log \frac{{{\mathbf{U}_{ik}}}}{{\mathbf{U}_{ik}^{(t)}}})},
\end{split}
\nonumber
\end{equation}
which satisfies the conditions in Eq.~(\ref{auxilariyu}).

The first-order derivative of ${g}(\mathbf{U},{\mathbf{U}^{(t)}})$ is
\begin{equation}
\frac{{\partial {g}(\mathbf{U},{\mathbf{U}^{(t)}})}}{{\partial {\mathbf{U}_{ik}}}} = \frac{{2{{({\mathbf{U}^{(t)}}{\mathbf{V}^T}\mathbf{QV})}_{ik}}{\mathbf{U}_{ik}}}}{{\mathbf{U}_{ik}^{(t)}}} - \frac{{2{{(\mathbf{XQV})}_{ik}}\mathbf{U}_{ik}^{(t)}}}{{{\mathbf{U}_{ik}}}},
\nonumber
\end{equation}
The Hessian matrix is
\begin{equation}
\frac{{{\partial ^2}{g}(\mathbf{U},{\mathbf{U}^{(t)}})}}{{\partial {\mathbf{U}_{ik}}\partial {\mathbf{U}_{jl}}}} = 2{\delta _{ij}}{\delta _{kl}}(\frac{{{{(\mathbf{U}^{(t)}{\mathbf{V}^T}\mathbf{QV})}_{ik}}}}{{\mathbf{U}_{ik}^{(t)}}} + \frac{{{{(\mathbf{XQV})}_{ik}}{\mathbf{U}^{(t)}_{ik}}}}{{\mathbf{U}_{ik}^2}}),
\nonumber
\end{equation}
where $\delta _{ij}$ is the defined as
\begin{equation}
{\delta _{ij}} = \left\{ \begin{array}{l}
1,\quad {\rm{if }} \ i = j\\
0,\quad {\rm{otherwise}}
\end{array} \right..
\nonumber
\end{equation}
The Hessian matrix is a positive definite diagonal matrix, so ${g}(\mathbf{U},{\mathbf{U}^{(t)}}$ is convex on $\mathbf{U}$. The global optimal solution $\mathbf{U}^{(t+1)}$ to $\min \limits _{\mathbf{ U}} g(\mathbf{U},{\mathbf{ U}^{(t)}})$ is computed by setting the first-order derivative to zero:
\begin{equation}
{\mathbf{U}_{ik}^{(t+1)}} = \mathbf{U}_{ik}^{(t)}\sqrt {\frac{{{{(\mathbf{XQV})}_{ik}}}}{{{{({\mathbf{U}^{(t)}}{\mathbf{V}^T}\mathbf{QV})}_{ik}}}}} .
\nonumber
\end{equation}
According to Lemma~\ref{lemmaguu}, ${\cal L}_2(\mathbf{U})$ is non-increasing with the above updating rule.
\end{proof}

\section{Graph regularized EMMF}
\label{sec:gemmf}
EMMF uses the global centroids to represent the samples, so it cannot the data with complex manifold structures. To explore the local data relationship, the graph-regularized EMMF (G-EMMF) is introduced. Since this research mainly focus on the robustness, we simply incorporate a graph regularization term into EMMF to improve the performance.

\subsection{Methodology}
Supposing $\mathbf{S}\in \mathbb{R}^{n\times n}$ is the similarity graph of the data matrix $\mathbf{X}$, a large value of $\mathbf{S}_{ij}$ indicates the high similarity between $\mathbf{x} _i$ and $\mathbf{x} _j$. Intuitively, if $\mathbf{x} _i$ is similar to $\mathbf{x} _j$, their coefficient vectors should also be similar. Using the inner product to measure the distance between the vectors, the graph regularization term is given as
\begin{equation}
\mathop {\min }\limits_\mathbf{V\ge 0} ||\mathbf{S - V}{\mathbf{V}^T}||_2^2.
\label{svv}
\end{equation}
Ideally, we can obtain a block diagonal $\mathbf{VV}^T$
Kuang et al.~\cite{sym} proved that the above term is equivalent to spectral clustering if $\mathbf{S}$ is doubly stochastic and $\mathbf{V}$ is orthogonal. Normalizing the graph $\mathbf{S}=\mathbf{D}^{-\frac{1}{2}}\mathbf{S}\mathbf{D}^{-\frac{1}{2}}$, where $\mathbf{D}$ is the degree matrix of $\mathbf{S}$, problem~(\ref{svv}) becomes
\begin{equation}
\mathop {\min }\limits_\mathbf{\mathbf{V}\ge 0,\mathbf{V}^T\mathbf{V=I}} ||\mathbf{S - V}{\mathbf{V}^T}||_2^2.
\label{svv2}
\end{equation}
As mentioned in Section~\ref{pre}, the orthogonal constraint also facilitates the clustering interpretation. Combining Eq.~(\ref{svv2}) with the objective~(\ref{entropy-obj}), the model of G-EMMF is
\begin{equation}
\begin{split}
&\mathop {\min }\limits_{\mathbf{\mathbf{U,V}}}  - \sum\limits_{i = 1}^n {||{\mathbf{m}_i}|{|_2}\log \frac{{||{\mathbf{m}_i}|{|_2}}}{{||\mathbf{M}|{|_{2,1}}}}}+\lambda ||\mathbf{S - V}{\mathbf{V}^T}||_2^2, \\
&s.t.\mathbf{M = X - U}{\mathbf{V}^T},\mathbf{U} \ge 0,\mathbf{V} \ge 0, \mathbf{V}^T\mathbf{V=I},
\end{split}
\label{gentropy-obj}
\end{equation}
where $\lambda$ is the regularization parameter. With the above formulation, the coefficient matrix $\mathbf{V}$ preserves the local correlations between the samples.

\subsection{Optimization}
The objective of G-EMMF is equivalent to
\begin{equation}
\begin{split}
\min \limits_{\mathbf{U} ,\mathbf{V}}{\rm Tr}[(\mathbf{X}-&\mathbf{UV}^T)\mathbf{Q}(\mathbf{X-UV})]+\lambda ||\mathbf{S - V}{\mathbf{V}^T}||_2^2,\\
&s.t.\mathbf{U} \ge 0,\mathbf{V} \ge 0, \mathbf{V}^T\mathbf{V=I}.
\end{split}
\nonumber
\end{equation}
Removing the irrelevant terms, the above problem is simplified into
\begin{equation}
\begin{split}
\min \limits_{\mathbf{U} ,\mathbf{V}}{\rm Tr}[(\mathbf{X}-&\mathbf{UV}^T)\mathbf{Q}(\mathbf{X-UV})]-2\lambda {\rm Tr}(\mathbf{V}^T\mathbf{SV}),\\
&s.t.\mathbf{U} \ge 0,\mathbf{V} \ge 0, \mathbf{V}^T\mathbf{V=I}.
\end{split}
\label{trav}
\end{equation}
The updating rule of $\mathbf{U}$ is the same with EMMF, so we only give the updating rule of $\mathbf{V}$.

\textbf{Updating} $\mathbf{V}$, problem~(\ref{trav}) is transformed into
\begin{equation}
\begin{split}
\min \limits_{\mathbf{V}} {\rm Tr}(\mathbf{V}^T\mathbf{QVU}^T&\mathbf{U}) - 2{\rm Tr}(\mathbf{V}^T\mathbf{QX}^T\mathbf{U})-2\lambda {\rm Tr}(\mathbf{V}^T\mathbf{SV}),\\
& s.t.\mathbf{V} \ge 0,\mathbf{V}^T\mathbf{V}=\mathbf{I}.
\end{split}
\label{g-fv}
\end{equation}
The Lagrangian function is
\begin{equation}
\begin{split}
{\cal L}_4(\mathbf{V}) =& {\rm Tr}(\mathbf{V}^T\mathbf{QVU}^T\mathbf{U}) - 2{\rm Tr}(\mathbf{V}^T\mathbf{QX}^T\mathbf{U})\\
&-2\lambda {\rm Tr}(\mathbf{V}^T\mathbf{SV})+{\rm Tr}(\Lambda _4 \mathbf{V}^T)\\
&+{\rm Tr}[\Lambda _5 (\mathbf{V}^T\mathbf{V}-\mathbf{I})^T]
\end{split}
\label{lv}
\end{equation}
where $\Lambda _4 \in \mathbb{R}^{n\times c}$ and $\Lambda _5 \in \mathbb{R}^{c\times c}$ are Lagrangian multipliers. Given $(\Lambda _4)_{ik} \mathbf{V}_{ik} =0$, setting $\frac{\partial {\cal L}_4({\bf{V}})}{\partial {\bf{V}}}=0$ gives rise to
\begin{equation}
{(\mathbf{QV}{\mathbf{U}^T}\mathbf{U} - \mathbf{Q}{\mathbf{X}^T}\mathbf{U} - 2\lambda \mathbf{SV }+ \mathbf{V}{\Lambda _5})_{ik}}\mathbf{V}_{ik}^2 = 0,
\nonumber
\end{equation}
so the updating rule is
\begin{equation}
{\mathbf{V}_{ik}} \leftarrow {\mathbf{V}_{ik}}\sqrt {\frac{{{{(\mathbf{Q}{\mathbf{X}^T}\mathbf{U} + 2\lambda \mathbf{SV + V}\Lambda _5^ - )}_{ik}}}}{{{{(\mathbf{QV}{\mathbf{U}^T}\mathbf{U + V}\Lambda _5^ + )}_{ik}}}}},
\label{rulev1}
\end{equation}
where $\Lambda _5^ - $, $\Lambda _5^ +$ are the negative and positive parts of $\Lambda _5$, i.e. $(\Lambda _5^ - )_{ik}= \frac{|(\Lambda _5)_{ik}|-(\Lambda _5)_{ik}}{2}$ and $(\Lambda _5^ + )_{ik}= \frac{|(\Lambda _5)_{ik}|+(\Lambda _5)_{ik}}{2}$.
Since we also have
\begin{equation}
{(\mathbf{QV}{\mathbf{U}^T}\mathbf{U} - \mathbf{Q}{\mathbf{X}^T}\mathbf{U} - 2\lambda \mathbf{SV }+ \mathbf{V}{\Lambda _5})_{ik}}\mathbf{V}_{ik} = 0,
\nonumber
\end{equation}
$\Lambda _5$ is computed as
\begin{equation}
{\Lambda _5} = {\mathbf{V}^T}\mathbf{Q}{\mathbf{X}^T}\mathbf{U} + 2\lambda {\mathbf{V}^T}\mathbf{SV} - {\mathbf{V}^T}\mathbf{QV}{\mathbf{U}^T}\mathbf{U}.
\nonumber
\end{equation}
and ${\Lambda _5^-}$ and ${\Lambda _5^+}$ are computed as
\begin{equation}
\begin{split}
{\Lambda _5^-} &= {\mathbf{V}^T}\mathbf{QV}{\mathbf{U}^T}\mathbf{U},\\
{\Lambda _5^+} &= {\mathbf{V}^T}\mathbf{Q}{\mathbf{X}^T}\mathbf{U} + 2\lambda {\mathbf{V}^T}\mathbf{SV} .
\end{split}
\nonumber
\end{equation}
Therefore, the updating rule in Eq.~(\ref{rulev1}) is rewritten as
\begin{equation}
{\mathbf{V}_{ik}} \leftarrow {\mathbf{V}_{ik}}\sqrt {\frac{{{{(\mathbf{Q}{\mathbf{X}^T}\mathbf{U} + 2\lambda \mathbf{SV} + \mathbf{V}{\mathbf{V}^T}\mathbf{QV}{\mathbf{U}^T}\mathbf{U})}_{ik}}}}{{{{(\mathbf{QV}{\mathbf{U}^T}\mathbf{U} + \mathbf{V}{\mathbf{V}^T}\mathbf{Q}{\mathbf{X}^T}\mathbf{U} +2 \lambda \mathbf{V}{\mathbf{V}^T}\mathbf{SV})}_{ik}}}}} .
\label{g-updatev}
\end{equation}

The optimization strategy for G-EMMF is described in Algorithm~\ref{alg2}. G-EMMF takes ${\cal O}(n^2d)$ additional complexity to construct and normalize the similarity graph, and the remaining costs are the same with EMMF, i.e. ${\cal O}(tndc)$. After $t$ iterations, the overall computational cost of G-EMMF is ${\cal O}(n^2d+tndc)$.

\begin{algorithm}
        \caption{Optimization algorithm of EMMF}
        \begin{algorithmic}[1]
        \Require Data matrix $\mathbf{X}$, centroid number $c$, graph $\mathbf{S}$.
           \State Initialize $\mathbf{U}$ and $\mathbf{V}$.
           \State Normalize $\mathbf{S}$ as $\mathbf{D}^{-\frac{1}{2}}\mathbf{S}\mathbf{D}^{-\frac{1}{2}}$.
        	\Repeat
			\State Compute $\mathbf{Q}$ with Eq.~(\ref{updateq}).
			\State Update $\mathbf{U}$ with Eq.~(\ref{updateu}).
			\State Update $\mathbf{V}$ with Eq.~(\ref{g-updatev}).
			\Until{Converge}
		\Ensure Optimal $\mathbf{U}$, $\mathbf{V}$.
        \end{algorithmic}
        \label{alg2}
    \end{algorithm}

\subsection{Convergence}
Here we demonstrate the convergence of problem~(\ref{g-fv}). We fist introduce the following theorem.

\begin{theorem}
Updating rule ~(\ref{rulev1}) decreases Lagrangian function ${\cal L}_4(\mathbf{V})$ in Eq.~(\ref{lv}) monotonically. 
\end{theorem}
\begin{proof}

According to the proof of Theorem~\ref{theorem2}, denoting the value of $\mathbf{V}$ at the $t$-th iteration as $\mathbf{V}^{(t)}$, the bounds of the non-zero terms in ${\cal L}_4(\mathbf{V})$ are as follows.
\begin{equation}
\begin{split}
&{\rm Tr}({\mathbf{V}^T}\mathbf{QVU}^T\mathbf{U}) \le \sum\limits_{i = 1}^n {\sum\limits_{k = 1}^c {\frac{{{{(\mathbf{QV}^{(t)}{\mathbf{U}^T}\mathbf{U})}_{ik}}\mathbf{V}_{ik}^2}}{{\mathbf{V}_{ik}^{(t)}}}} },\\
&{\rm Tr}({\mathbf{V}^T}\mathbf{QX}^T\mathbf{U}) \ge \sum\limits_{i = 1}^n {\sum\limits_{k = 1}^c{{{(\mathbf{Q}{\mathbf{X}^T}\mathbf{U})}_{ik}}\mathbf{V}_{ik}^{(t)}(1 + \log \frac{{{\mathbf{V}_{ik}}}}{{{{\mathbf{V}^{t}_{ik}}}}})}}, \\
&{\rm Tr}({\mathbf{V}^T}\mathbf{SV}) \ge \sum\limits_{i = 1}^n {\sum\limits_{k = 1}^c {\sum\limits_{l = 1}^n {{\mathbf{S}_{il}}\mathbf{V}_{ik}^{(t)}\mathbf{V}_{lk}^{(t)}(1 + \log \frac{{{\mathbf{V}_{ik}}{\mathbf{V}_{lk}}}}{{\mathbf{V}_{ik}^{(t)}\mathbf{V}_{lk}^{(t)}}})} } }, \\
&{\rm Tr}(\Lambda _5^ -{\mathbf{V}^T} \mathbf{V}) \ge \sum\limits_{i = 1}^n {\sum\limits_{k = 1}^c {\sum\limits_{l = 1}^c {{{(\Lambda _5^ - )}_{kl}}\mathbf{V}_{ik}^{(t)}\mathbf{V}_{il}^{(t)}(1 + \log \frac{{{\mathbf{V}_{ik}}{\mathbf{V}_{il}}}}{{\mathbf{V}_{ik}^{(t)}\mathbf{V}_{il}^{(t)}}})} } },  \\
 &{\rm Tr}(\Lambda _5^ + {\mathbf{V}^T}\mathbf{V}) \le \sum\limits_{i = 1}^n {\sum\limits_{k = 1}^c {\frac{{{{({\mathbf{V}^{(t)}}\Lambda _5^ + )}_{ik}}\mathbf{V}_{ik}^2}}{{\mathbf{V}_{ik}^{(t)}}}} }. \\
\end{split}
\nonumber
\end{equation}
Combining the bounds, the auxiliary function for ${\cal L}_4(\mathbf{V})$ is
\begin{equation}
\begin{split}
&g(\mathbf{V,V}^{(t)})\\
=& \sum\limits_{i = 1}^n {\sum\limits_{k = 1}^c {\frac{{{{(\mathbf{QV}^{(t)}{\mathbf{U}^T}\mathbf{U})}_{ik}}\mathbf{V}_{ik}^2}}{{\mathbf{V}_{ik}^{(t)}}}} }\\
 &- 2\sum\limits_{i = 1}^n {\sum\limits_{k = 1}^c{{{(\mathbf{Q}{\mathbf{X}^T}\mathbf{U})}_{ik}}\mathbf{V}_{ik}^{(t)}(1 + \log \frac{{{\mathbf{V}_{ik}}}}{{{{\mathbf{V}^{t}_{ik}}}}})}} \\
 & - 2\lambda \sum\limits_{i = 1}^n {\sum\limits_{k = 1}^c {\sum\limits_{l = 1}^n {{\mathbf{S}_{il}}\mathbf{V}_{ik}^{(t)}\mathbf{V}_{lk}^{(t)}(1 + \log \frac{{{\mathbf{V}_{ik}}{\mathbf{V}_{lk}}}}{{\mathbf{V}_{ik}^{(t)}\mathbf{V}_{lk}^{(t)}}})} } } \\
   &- \sum\limits_{i = 1}^n {\sum\limits_{k = 1}^c {\sum\limits_{l = 1}^c {{{(\Lambda _5^ - )}_{kl}}\mathbf{V}_{ik}^{(t)}\mathbf{V}_{il}^{(t)}(1 + \log \frac{{{\mathbf{V}_{ik}}{\mathbf{V}_{il}}}}{{\mathbf{V}_{ik}^{(t)}\mathbf{V}_{il}^{(t)}}})} } }   \\
  & +\sum\limits_{i = 1}^n {\sum\limits_{k = 1}^c {\frac{{{{({\mathbf{V}^{(t)}}\Lambda _5^ + )}_{ik}}\mathbf{V}_{ik}^2}}{{\mathbf{V}_{ik}^{(t)}}}} }.
\end{split}
\nonumber
\end{equation}
The Hessian matrix is positive definite, so the optimal solution $\mathbf{V}^{(t+1)}$ that minimizes $g(\mathbf{V,V}^{(t)})$ is calculated by setting the first-order derivative to zero:
\begin{equation}
{\mathbf{V}_{ik}^{(t+1)}} = \mathbf{V}_{ik}^{(t)}\sqrt {\frac{{{{(\mathbf{Q}{\mathbf{X}^T}\mathbf{U} + \lambda \mathbf{S}{\mathbf{V}^{(t)}} + {\mathbf{V}^{(t)}}\Lambda _5^ - )}_{ik}}}}{{{{(\mathbf{Q}{\mathbf{V}^{(t)}}{\mathbf{U}^T}\mathbf{U} + {\mathbf{V}^{(t)}}\Lambda _5^ + )}_{ik}}}}} .
\nonumber
\end{equation}
According to Lemma~\ref{lemmaguu}, the above updating rule decreases ${\cal L}_4(\mathbf{V})$.
\end{proof}

\section{Evaluation of EMMF}
\label{sec:experemmf}
In this section, the proposed EMMF is evaluated on several synthetic and real-world datasets.

\subsection{Experiments on Synthetic Datasets}
\label{sec:experemmftoy}
Synthetic datasets are constructed to validate the robustness of EMMF. Some static properties of EMMF are also discussed.

\begin{figure}
\begin{center}
\subfigure[samples]{
\includegraphics[width=0.40\textwidth]{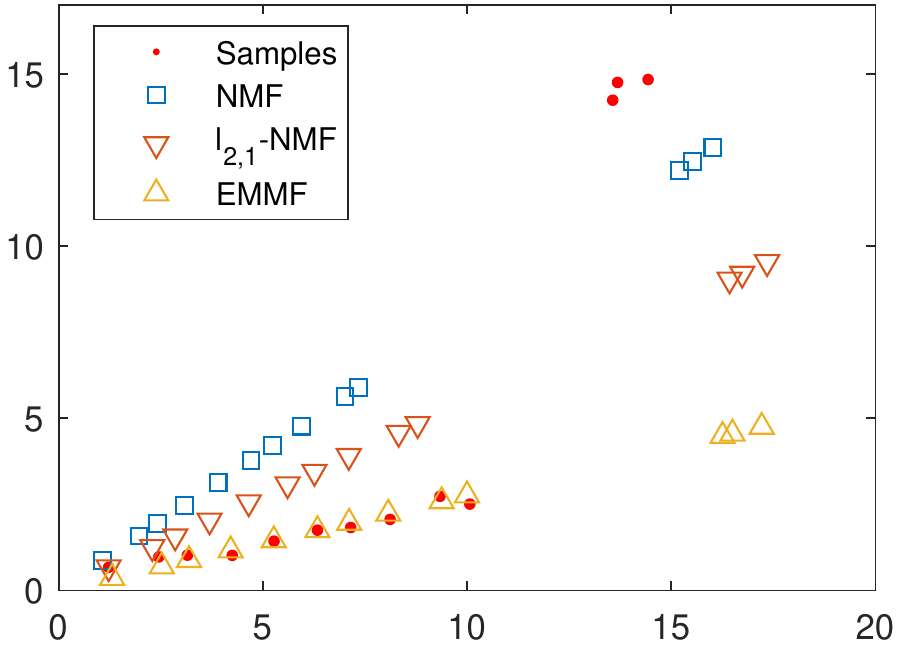}}

\subfigure[NMF]{
\includegraphics[width=0.15\textwidth]{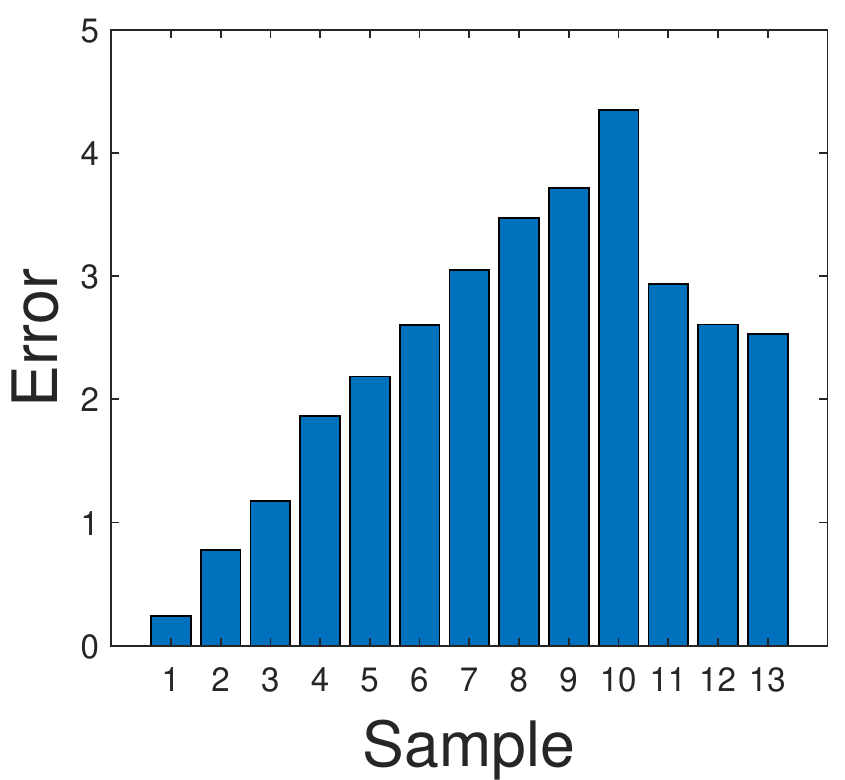}}
\subfigure[$\ell _{2,1}-$NMF]{
\includegraphics[width=0.15\textwidth]{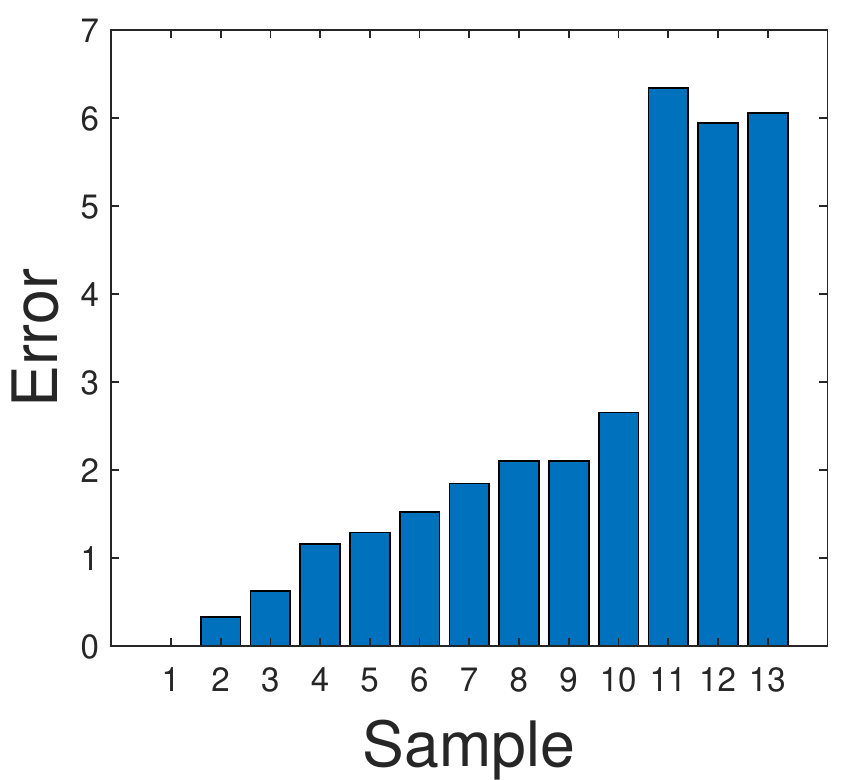}}
\subfigure[EMMF]{
\includegraphics[width=0.15\textwidth]{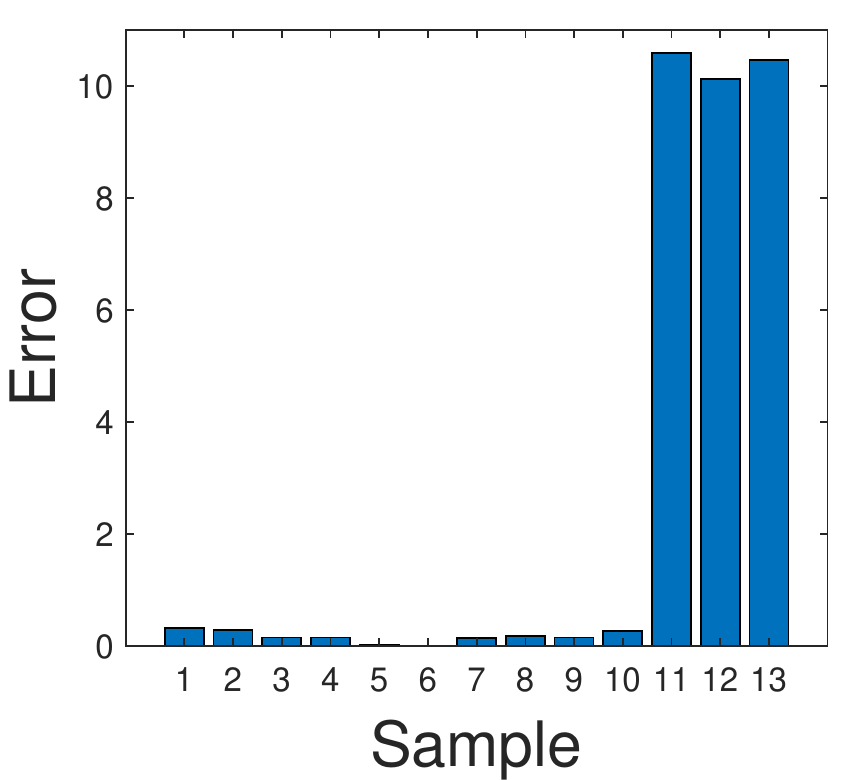}}
\end{center}

\caption{(a) Results on the synthetic dataset with outliers. (b-d) Approximation error $||\mathbf{x}_i-\mathbf{Uv}_{i,:}^T||_2$ calculated by different methods. }
\label{toy1}
\end{figure}

As shown in Fig.~\ref{toy1} (a), the first dataset consists of 13 two dimensional samples. The first ten samples are with normal distribution, and the last three samples are outliers. Fig.~\ref{toy1} (a) also visualizes the approximation results of NMF~\cite{nmf}, $\ell _{2,1}$-NMF~\cite{l21nmf} and EMMF. The approximated samples of NMF deviates the normal distribution largely, which indicates the approximation is dominated by the outliers. As shown in Fig.~\ref{toy1} (b), the errors of some normal samples are larger than the outliers. $\ell _{2,1}$-NMF shows better performance because the outliers are depressed. However, the outliers still affect the approximation. EMMF just let the outliers to be with large errors, as shown in Fig.~\ref{toy1} (d), such that the normal samples are approximated correctly. Therefore, it shows more robustness.

\begin{figure}
\begin{center}
\subfigure[samples]{
\includegraphics[width=0.40\textwidth]{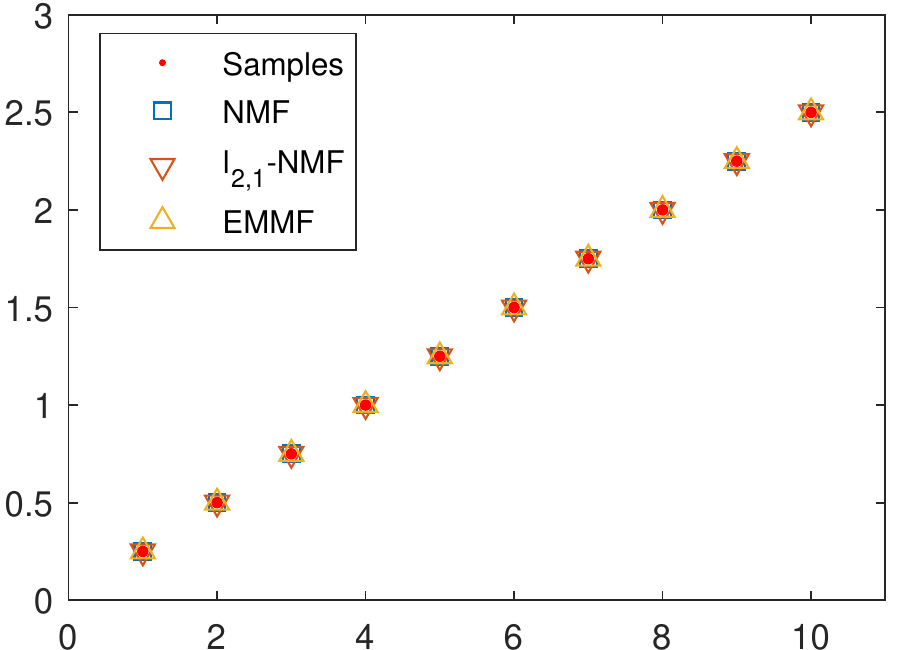}}

\subfigure[NMF]{
\includegraphics[width=0.145\textwidth]{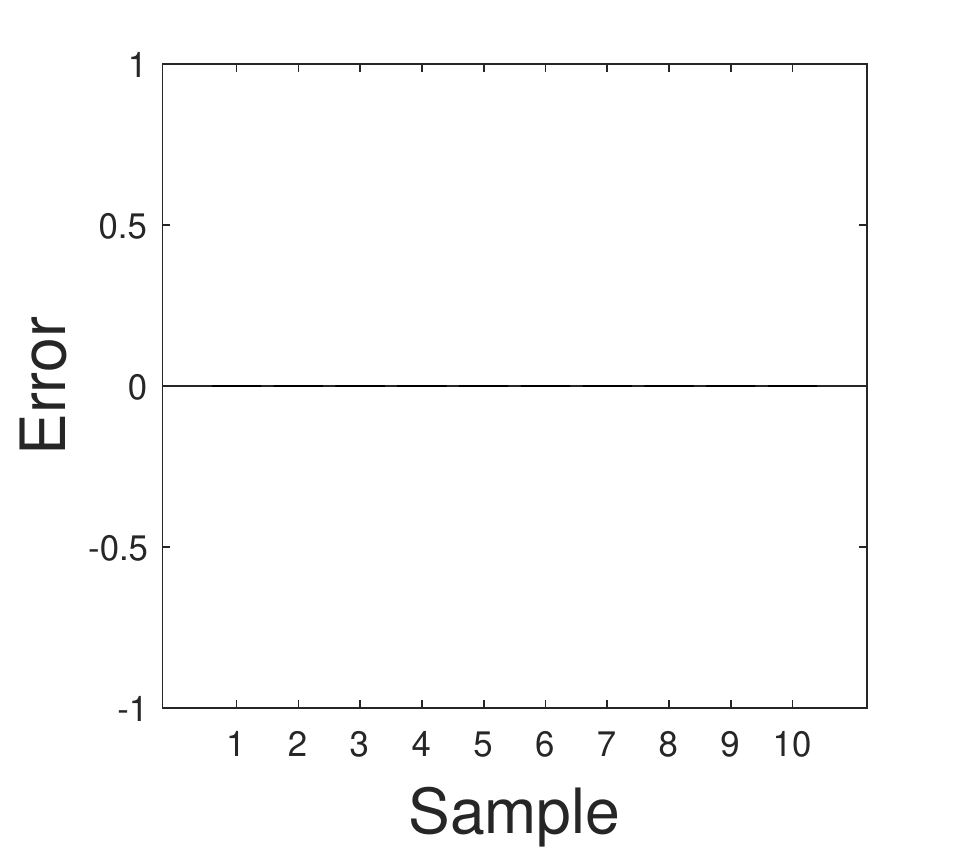}}
\subfigure[$\ell _{2,1}$-NMF]{
\includegraphics[width=0.145\textwidth]{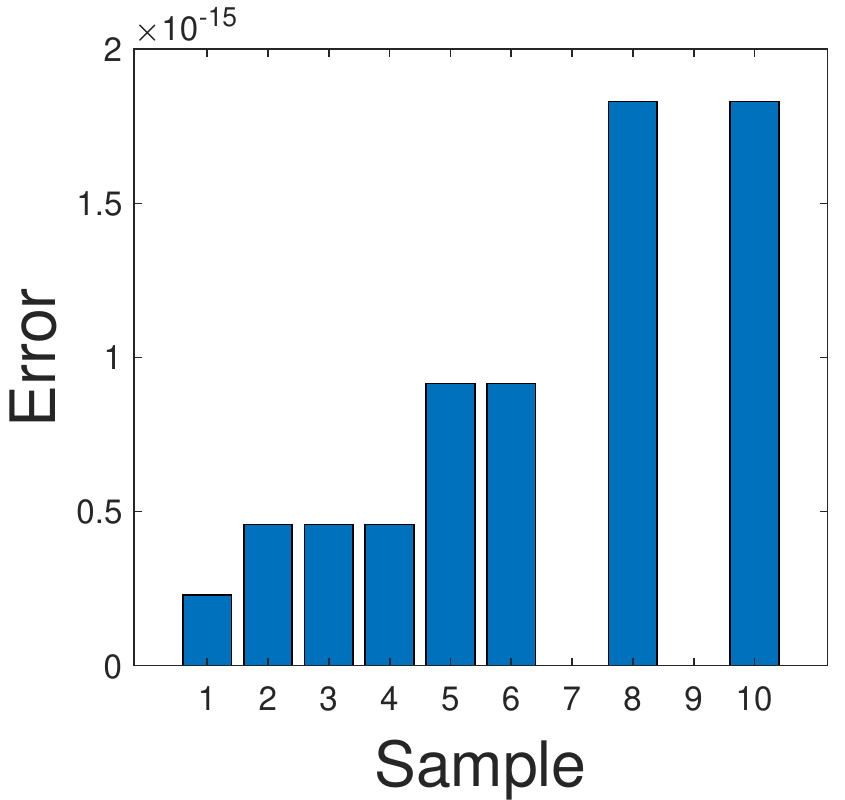}}
\subfigure[EMMF]{
\includegraphics[width=0.145\textwidth]{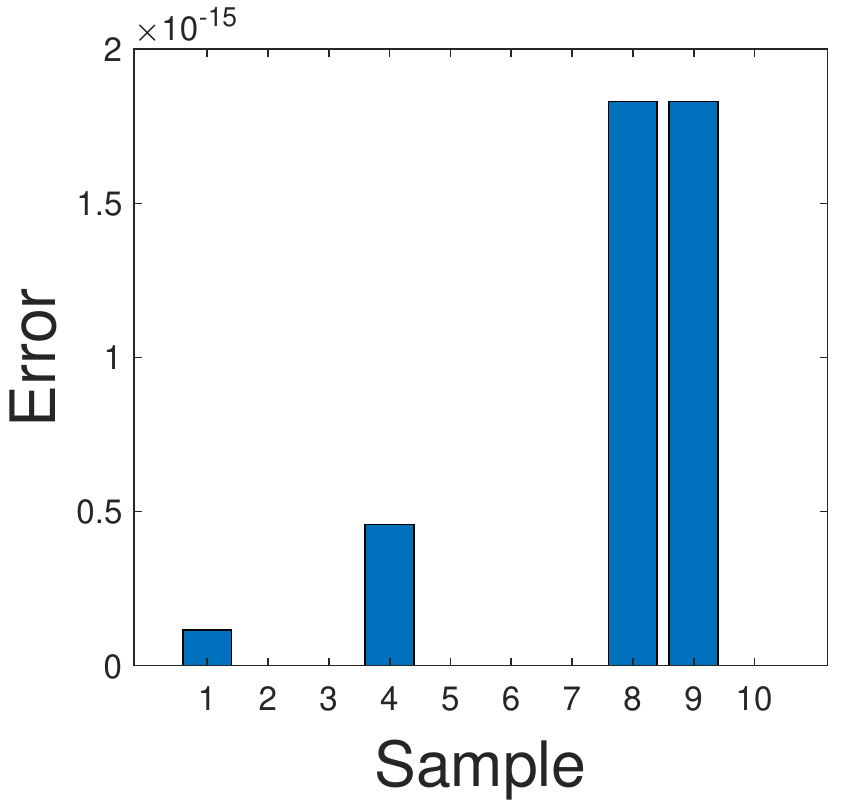}}
\end{center}

\caption{(a) Results on the synthetic dataset without outliers. (b-d) Approximation error $||\mathbf{x}_i-\mathbf{Uv}_{i,:}^T||_2$ calculated by different methods. }
\label{toy2}
\end{figure}

Fig.~\ref{toy2} shows the approximation results for the data without outliers. It can be seen that all the methods approximate the samples successfully. As visualized in Fig.~\ref{toy2} (c) and (d), both $\ell _{2,1}$-NMF and EMMF cannot achieve the zero error due to the factor $\varepsilon$ added on $\mathbf{Q}$. In Fig.~\ref{toy2} (d), some samples are with relative large errors because EMMF treat them as outliers. However, the errors of most samples are close to zero, so the overall error is still less than $\ell _{2,1}$-NMF. That is to say, a few mistaken outliers can not affect the approximation largely, and EMMF works well even when there is no outlier.

\begin{figure}
\begin{center}
\includegraphics[width=0.40\textwidth]{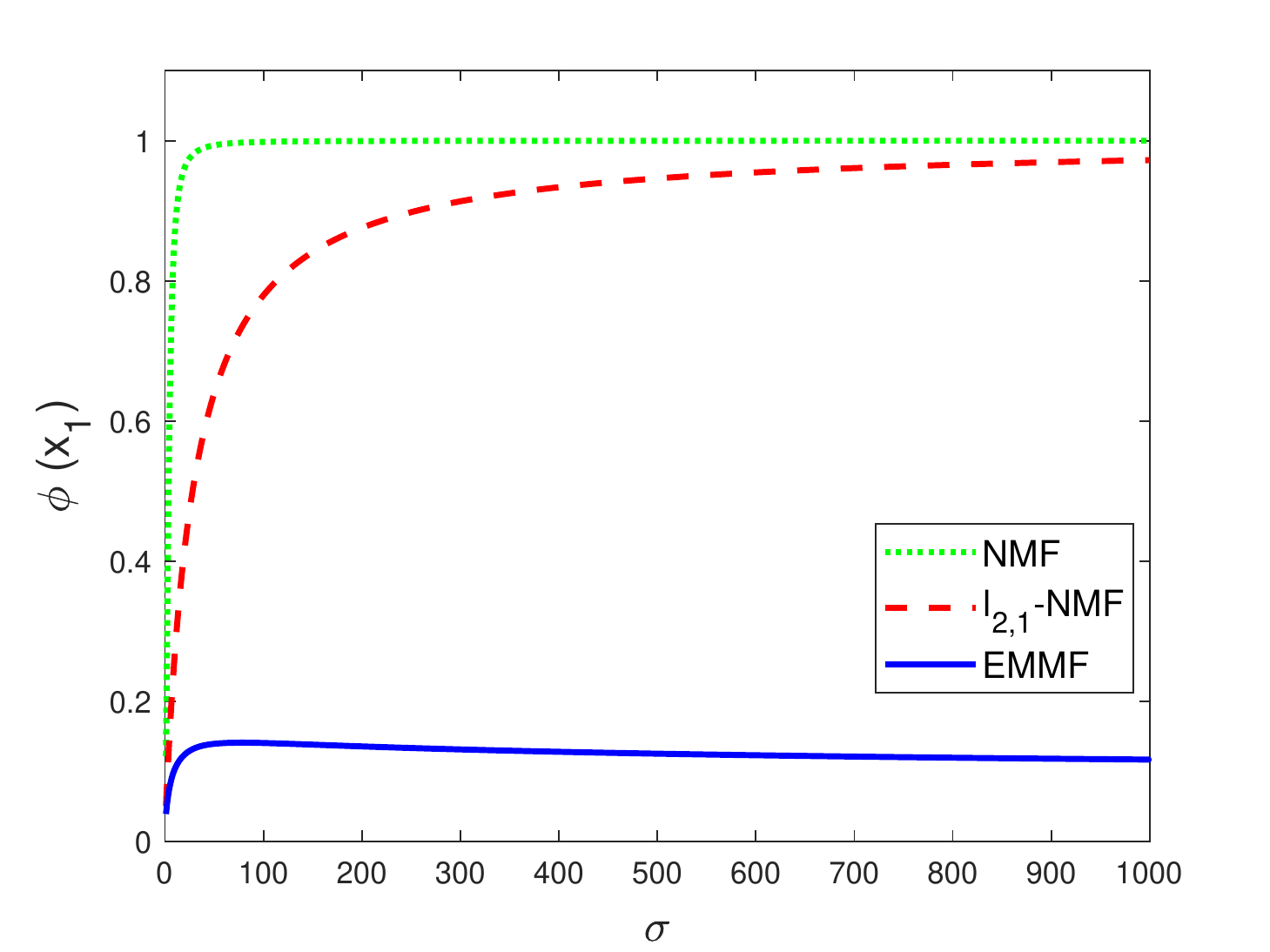}
\end{center}
\caption{Outlier effects of different methods with different value of $\sigma$ on the randomly generated data $\mathbf{X}\in \mathbb{R}^{3\times 3}$. The ratio of EMMF decreases when $\sigma$ exceeds a certain value.}
\label{ratio}
\end{figure}

We also demonstrate that the objective of EMMF is unlikely to be dominant by outliers. Supposing $x _i$ is the outlier, its effects on the objectives of NMF, $\ell _{2,1}$-NMF and EMMF are computed as
\begin{equation}
\begin{split}
&{\phi _{{\rm{NMF}}}}({\mathbf{x}_i}) = \frac{{||{{\bf{m}}_i}||_2^2}}{{\sum\limits_{i = 1}^n {||{{\bf{m}}_i}||_2^2} }},\\
&{\phi _{{\ell _{2,1}}{\rm{NMF}}}}({\mathbf{x}_i}) = \frac{{||{{\bf{m}}_i}|{|_2}}}{{\sum\limits_{i = 1}^n {||{{\bf{m}}_i}|{|_2}} }},\\
&{\phi _{{\rm{EMMF}}}}({\mathbf{x}_i}) = \frac{{||{{\bf{m}}_i}|{|_2}\log \frac{{||{{\bf{m}}_i}|{|_2}}}{{||{\bf{M}}|{|_{2,1}}}}}}{{\sum\limits_{i = 1}^n {||{{\bf{m}}_i}|{|_2}\log \frac{{||{{\bf{m}}_i}|{|_2}}}{{||{\bf{M}}|{|_{2,1}}}}} }}.
\end{split}
\nonumber
\end{equation}
We randomly generate a matrix $\mathbf{X}\in \mathbb{R}^{50\times 50}$, where $ 0\le \mathbf{X} _{ik}\le 1$. Adding a noise factor $\sigma$ on $\mathbf{X}_{11}$,  we investigate the effect of outlier $\mathbf{x} _1$ on the objective value with varying $\sigma$. As shown in Fig.~\ref{ratio}, the outlier effect on NMF increases dramatically. Compared with NMF, ${\phi _{{\ell _{2,1}}{\rm{NMF}}}}(\mathbf{x}_1)$ increases slower. But both of them reach to 1 when $\sigma$ is very large. Meanwhile, $\phi _{\rm{EMMF}}(\mathbf{x}_1)$ decreases when $\sigma$ exceeds a certain value. Therefore, EMMF is insensitive to the outliers with extremely large errors.

\begin{table*}
\caption{Description on the real-world datasets.}
\label{dataset}
\centering
\renewcommand\arraystretch{1.2}
\small
\begin{tabular}{|p{1.6cm}<{\centering}|p{1.4cm}<{\centering}|p{1.4cm}<{\centering}|p{1.4cm}<{\centering}|p{1.4cm}<{\centering}|p{1.4cm}<{\centering}|p{1.4cm}<{\centering}|p{1.4cm}<{\centering}|p{1.5cm}<{\centering}|}

\hline
 &COIL20 & COIL100 & YALE &JAFFE&UMIST&Mfeat&BA&Movement\\
\hline
Samples &1440 &7200& 165 &213& 575& 2000 &1404& 360\\
\hline
Dimension &1024 &1024 &256 &676& 644 &240 &320 &90\\
\hline
Class &20 &100 &15 &10 &20& 10& 36 &15\\
\hline
\end{tabular}
\end{table*}

\begin{table*}
\caption{Performance of EMMF on real-world datasets. Best results are in bold face.}
\label{emmftable}
\centering
\renewcommand\arraystretch{1.2}
\small
\begin{tabular}{|p{0.6cm}<{\centering}|p{1.5cm}<{\centering}|p{1.4cm}<{\centering}|p{1.4cm}<{\centering}|p{1.4cm}<{\centering}|p{1.4cm}<{\centering}|p{1.4cm}<{\centering}|p{1.4cm}<{\centering}|p{1.4cm}<{\centering}|p{1.5cm}<{\centering}|}

\hline
\multirow{7}*{ACC}&& COIL20 & COIL100 & YALE &JAFFE&UMIST&Mfeat&BA&Movement\\
\cline{2-10}
&NMF  &0.4486 &0.2532&0.3412 &0.6390 &0.3403 &  0.4233 
&0.1603 &0.3297\\
\cline{2-10}
&NMF-DIV &0.5097 &0.3954&0.4188  &0.9136 &0.3831 & 0.5945 
 &0.2369 &0.4372\\
\cline{2-10}
&PNMF&0.4882 &0.2297&0.2909  &0.3991 &0.3443 &0.5245  &0.1538 &0.3694 \\
\cline{2-10}
&GSNMF  &0.3431 &0.2507&0.3152 &0.2723 &0.2226 & 0.4005  &0.2486 &0.3806\\
\cline{2-10}
&$\ell _{2,1}$-NMF  &0.5826 &0.4256 &0.4182&0.9329 &0.4026 &  0.5821 &0.2203 &0.4453 \\
\cline{2-10}
&Hx-NMF   &0.5808 &0.4374&0.4097 &0.9357 &0.4090 &0.5790&0.2211 &0.4475\\
\cline{2-10}
&EMMF  &\textbf{0.5972} &\textbf{0.4542} &\textbf{0.4327}&\textbf{0.9423} &\textbf{0.4132} &\textbf{0.6195} &\textbf{0.2553} &\textbf{0.4542}\\
\hline
\hline
\multirow{7}*{NMI}& & COIL20 & COIL100 & YALE&JAFFE&UMIST&Mfeat&BA&Movement\\
\cline{2-10}
&NMF &0.5731 &0.5273 &0.4197 &0.6702 &0.4804 &0.3878 &0.2779 &0.3882\\
\cline{2-10}
&NMF-DIV &0.6413 &0.6534&0.4589  &0.8909 &0.5497 &0.5553  &0.3966 &0.5554\\
\cline{2-10}
&PNMF  &0.5682 &0.4760 &0.3276&0.4654 &0.4342 &  0.4903 &0.2589 &0.4347\\
\cline{2-10}
&GSNMF &0.4459 &0.5028 &0.3047 &0.2248 &0.2921 & 0.3133 &0.3549 &0.4661\\
\cline{2-10}
&$\ell _{2,1}$-NMF  &0.6937 &0.6653 &0.4541&0.9114 &0.5665 &0.5549 &0.3698 &0.5638\\
\cline{2-10}
&Hx-NMF  &0.6909 &0.6687 &0.4485&0.9134 &0.5658 &0.5591&0.3695 &0.5633\\
\cline{2-10}
&EMMF &\textbf{0.7059} &\textbf{0.6889} &\textbf{0.4686}&\textbf{0.9238} &\textbf{0.5827} &\textbf{0.5771} &\textbf{0.4075} &\textbf{0.5813}\\
\hline
\end{tabular}

\end{table*}

\begin{figure}
\begin{center}
\includegraphics[width=0.42\textwidth]{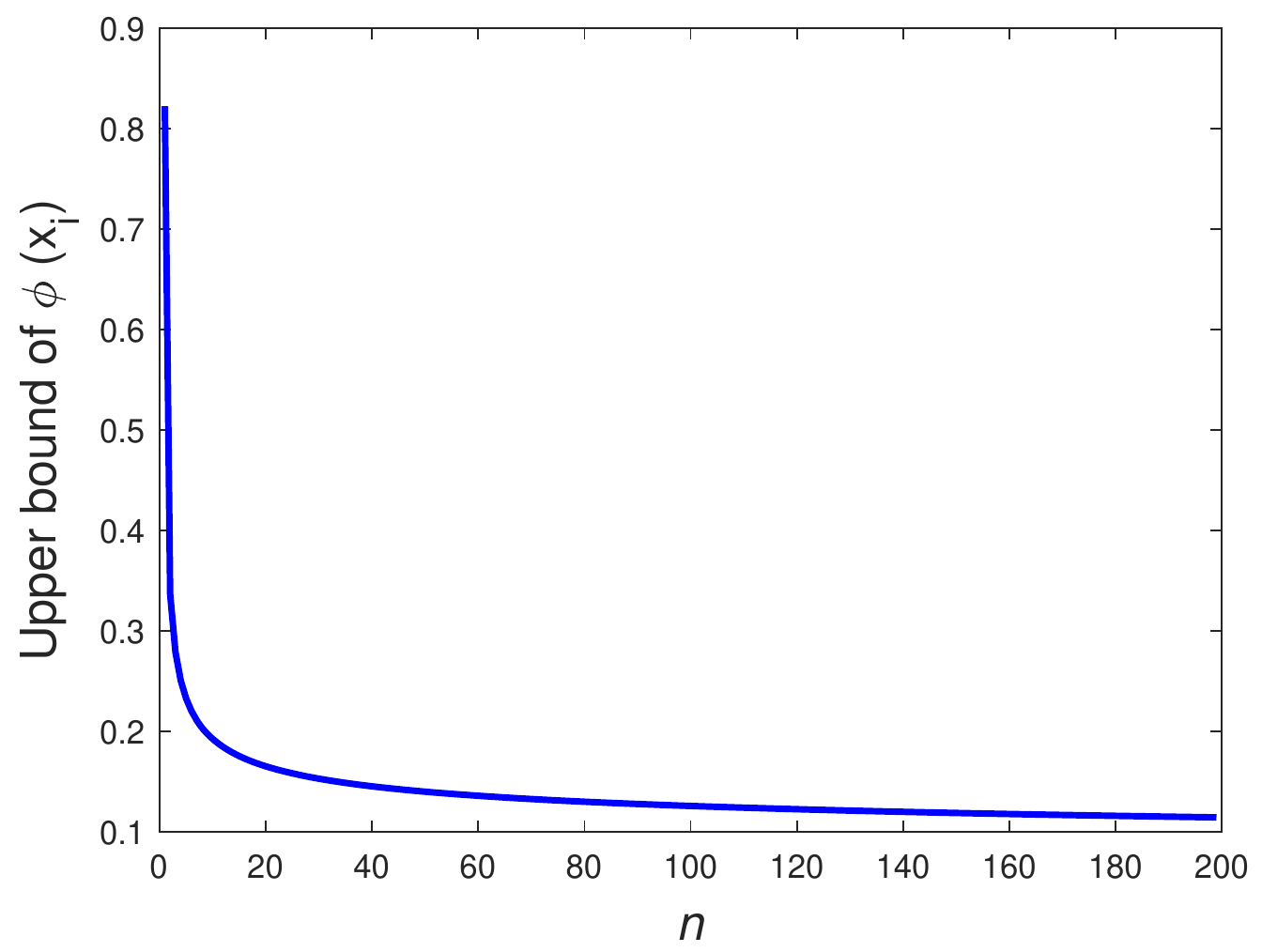}
\end{center}
\caption{Upper bound of $\phi _{\rm EMMF}(\mathbf{x}_i)$ with different value of $n$. The upper bound decreases with $n$.}
\label{bound}
\end{figure}

Suppose there is only one outlier $\mathbf{x} _i$, and its residue ratio is $p$, i.e. $\frac{||\mathbf{m}_i||_2}{||\mathbf{M}||_{2,1}}=p$. The entropy of the distribution is minimized when all the remaining samples are with the same residue ratio, i.e. $\frac{1-p}{n-1}$. In such situation, $\phi _{\rm{EMMF}}(\mathbf{x}_i)$ is
\begin{equation}
\phi _{\rm{EMMF}}(\mathbf{x}_i)=\frac{{p\log (p)}}{{p\log (p) + (1 - p)[\log(1 - p) - \log (n - 1)]}}.
\end{equation}
Given a certain $n$, the upper bound is calculated as the maximum value when varying $p$ within the range $(0, 1]$. Taking the step length of $p$ as 0.01 and increasing $n$, the upper bound curve is plotted in Fig.~\ref{bound}. We can see that the upper bound of $\phi _{\rm{EMMF}}(\mathbf{x}_i)$ decreases monotonically with the value of $n$, which complies with the human perception that the effect of an outlier should be small when there are many normal samples.

\subsection{Experiments on Real-world Datasets}
\label{sec:experemmfreal}

The clustering performance of EMMF is evaluated on real-world datasets. Clustering accuracy (ACC) and Normalized Mutual Information (NMI) are used as measurements.

\textbf{Datasets}: eight benchmarks for clustering are employed, including two object image datasets, i.e. COIL20 and COIL100~\cite{gnmf}, three face image datasets, i.e. YALE~\cite{yale}, JAFFE~\cite{jaffe} and UMIST~\cite{umist}, a multi-feature handwritten dataset, i.e. Multiple features (Mfeat)~\cite{uci}, a handwritten digit dataset, i.e. Binary Alphabet (BA)~\cite{ba}, and a hand movement dataset, i.e. Movement~\cite{uci}. All the samples are normalized as the unit vectors in the experiments. The details of the datasets are exhibited in Table~\ref{dataset}.

\begin{figure*}
\begin{center}

\subfigure[COIL20]{
\includegraphics[width=0.23\textwidth]{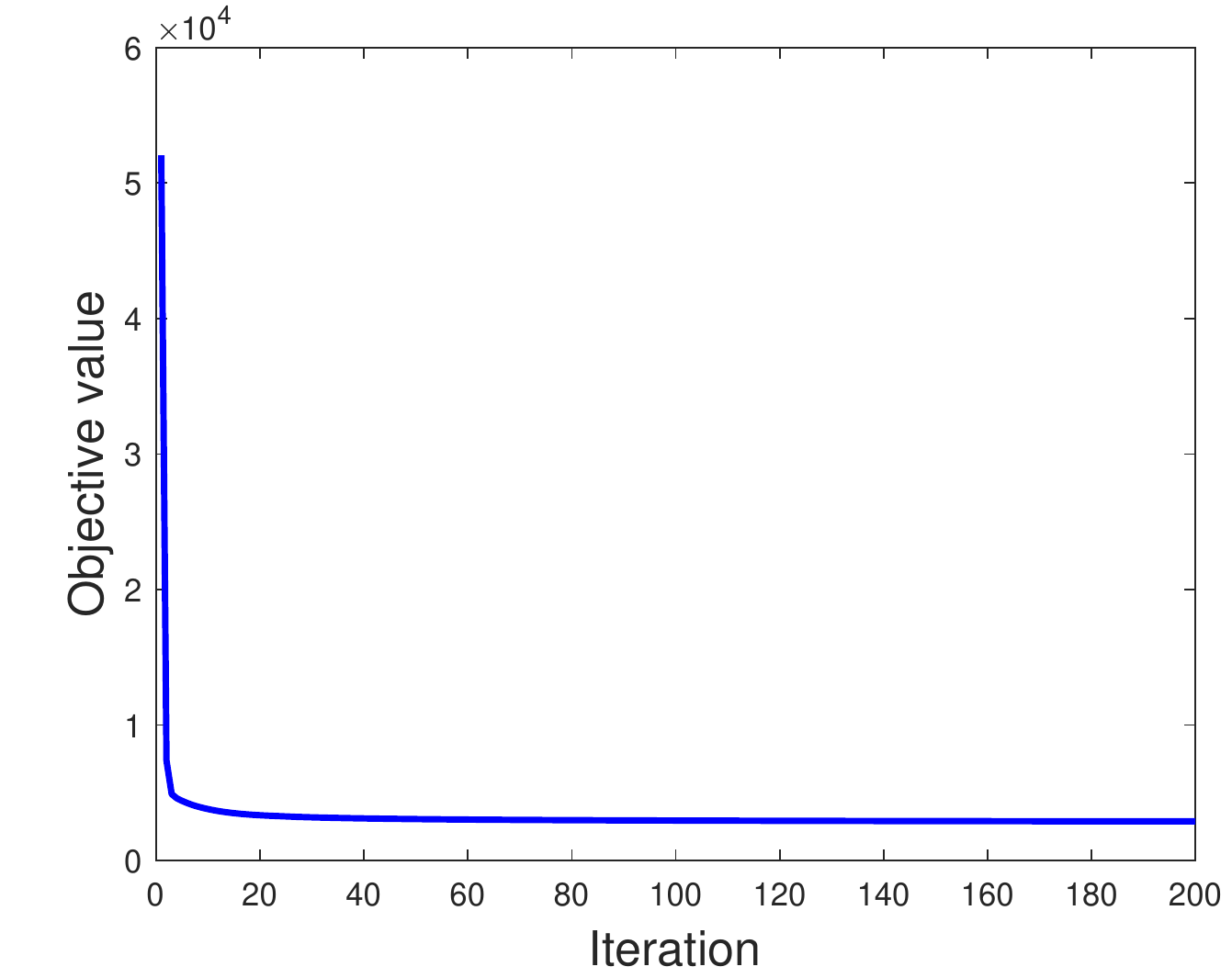}}
\hspace{1pt}
\subfigure[COIL100]{
\includegraphics[width=0.23\textwidth]{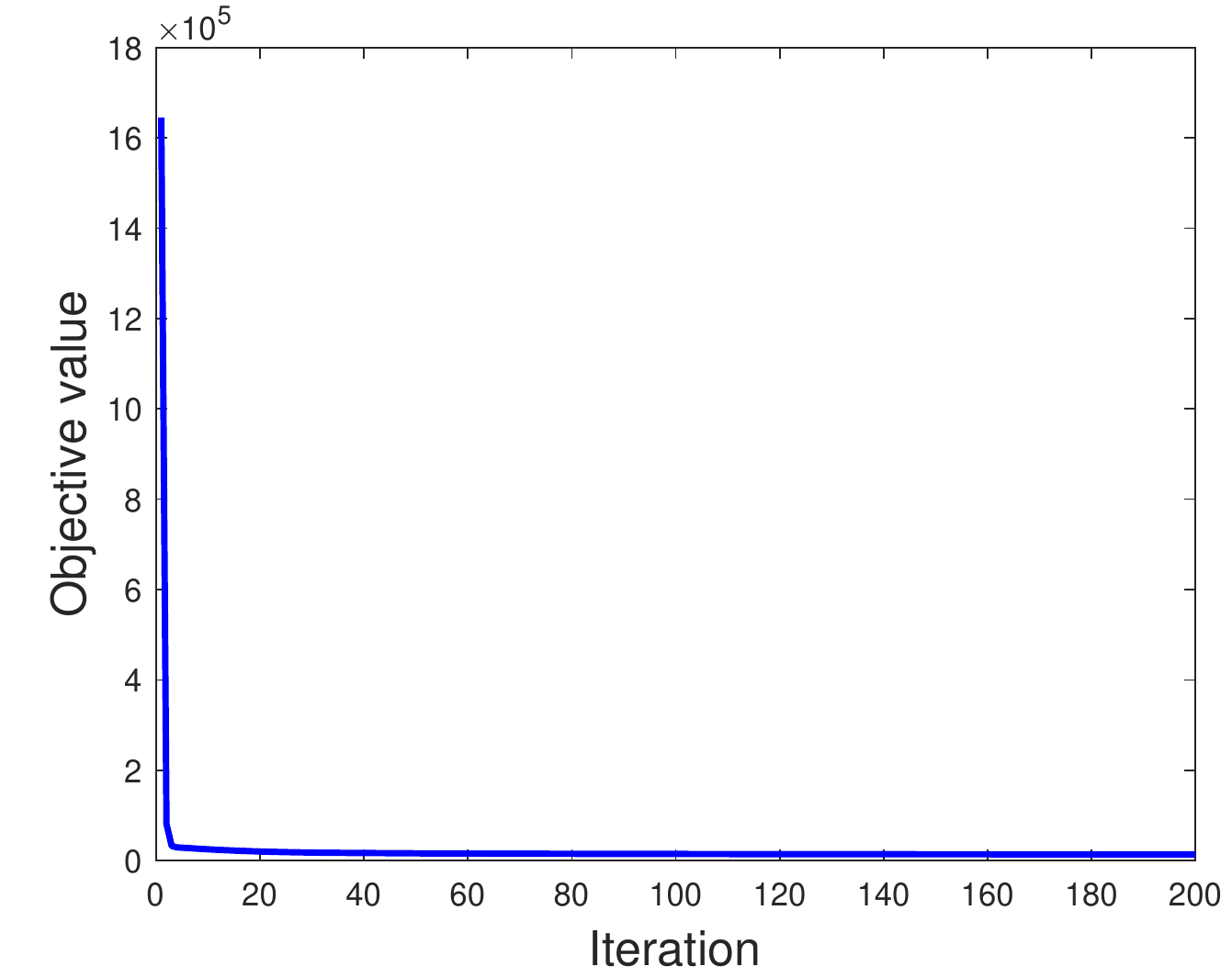}}
\hspace{1pt}
\subfigure[YALE]{
\includegraphics[width=0.23\textwidth]{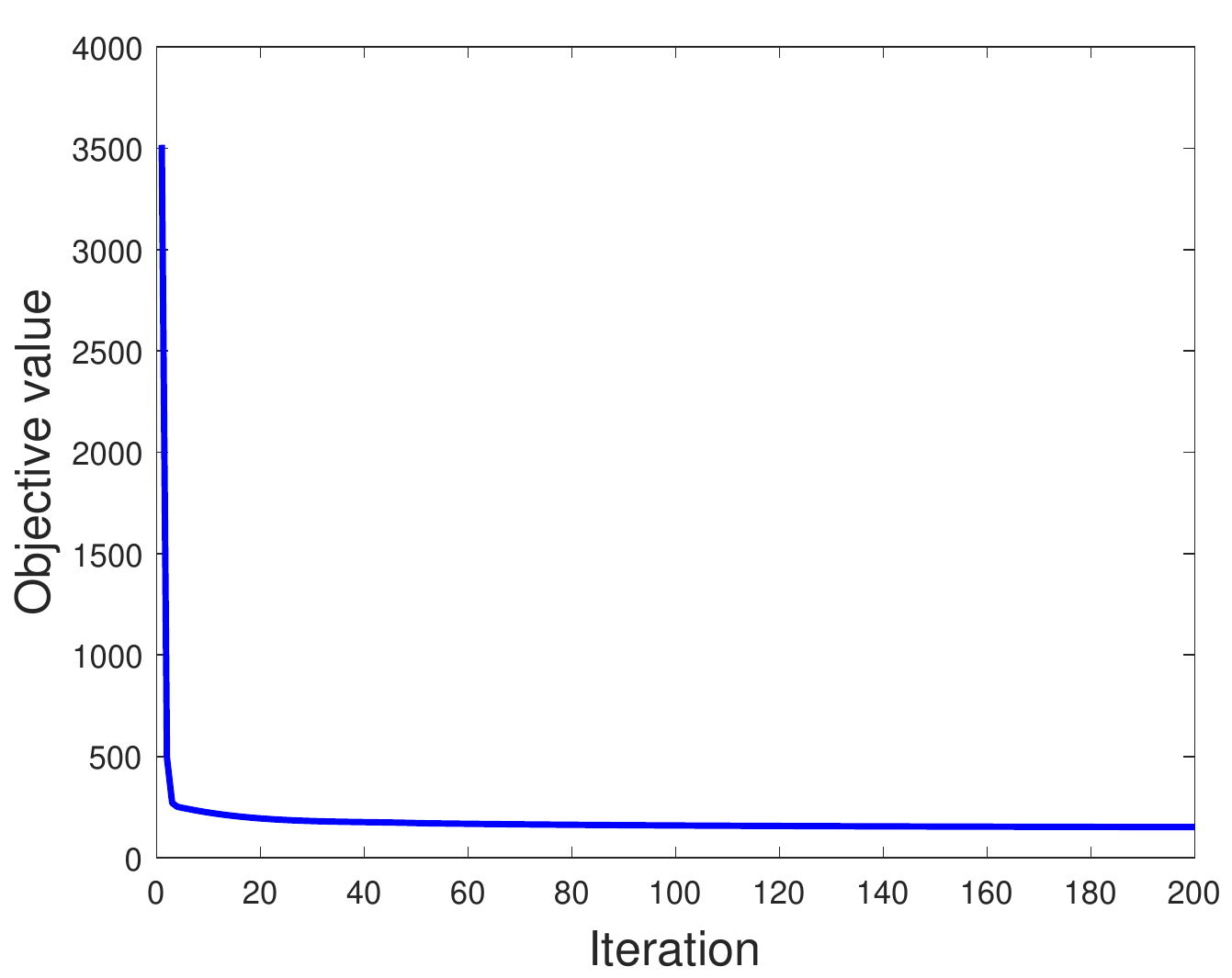}}
\hspace{1pt}
\subfigure[JAFFE]{
\includegraphics[width=0.23\textwidth]{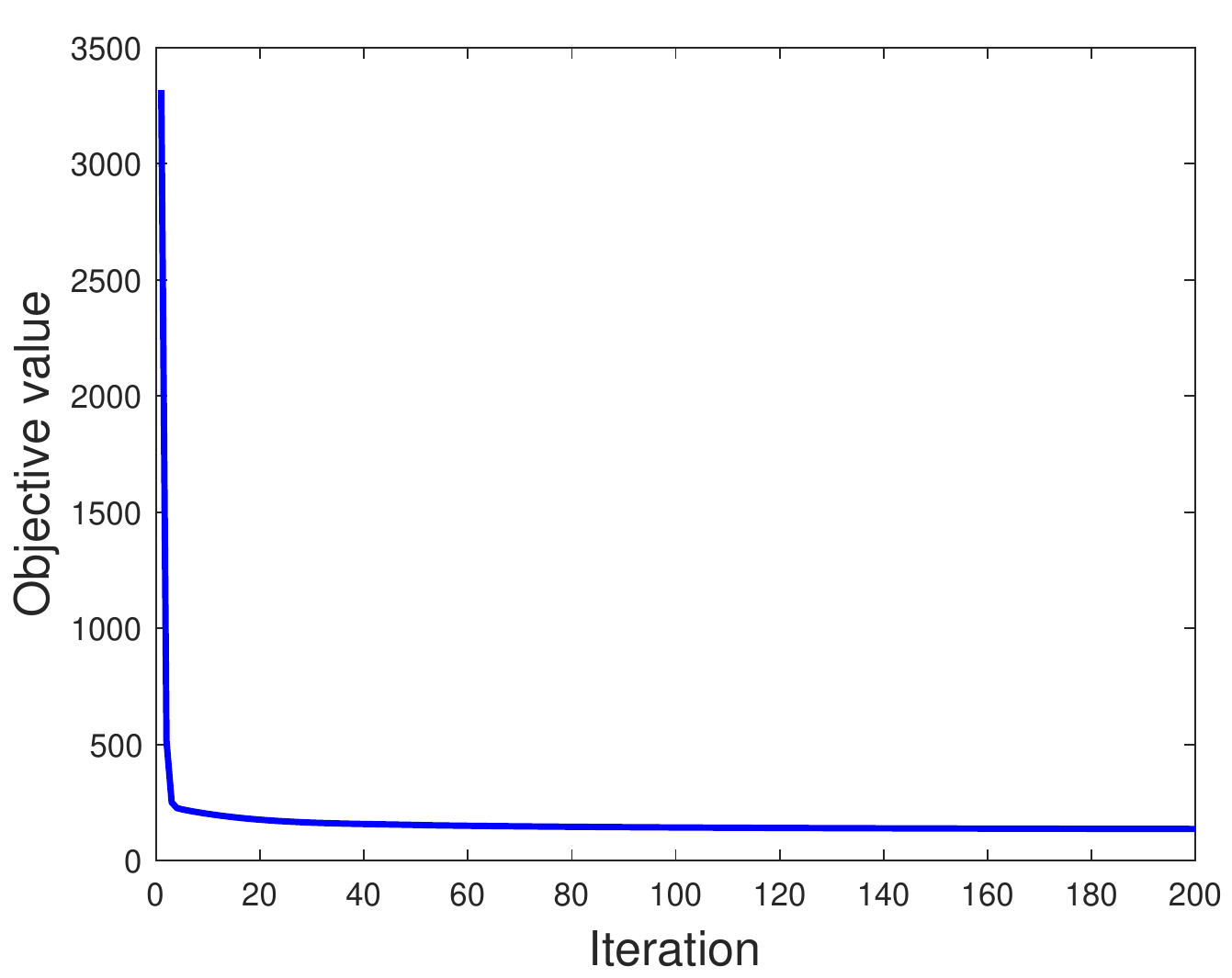}}
\end{center}

\begin{center}
\subfigure[UMIST]{
\includegraphics[width=0.23\textwidth]{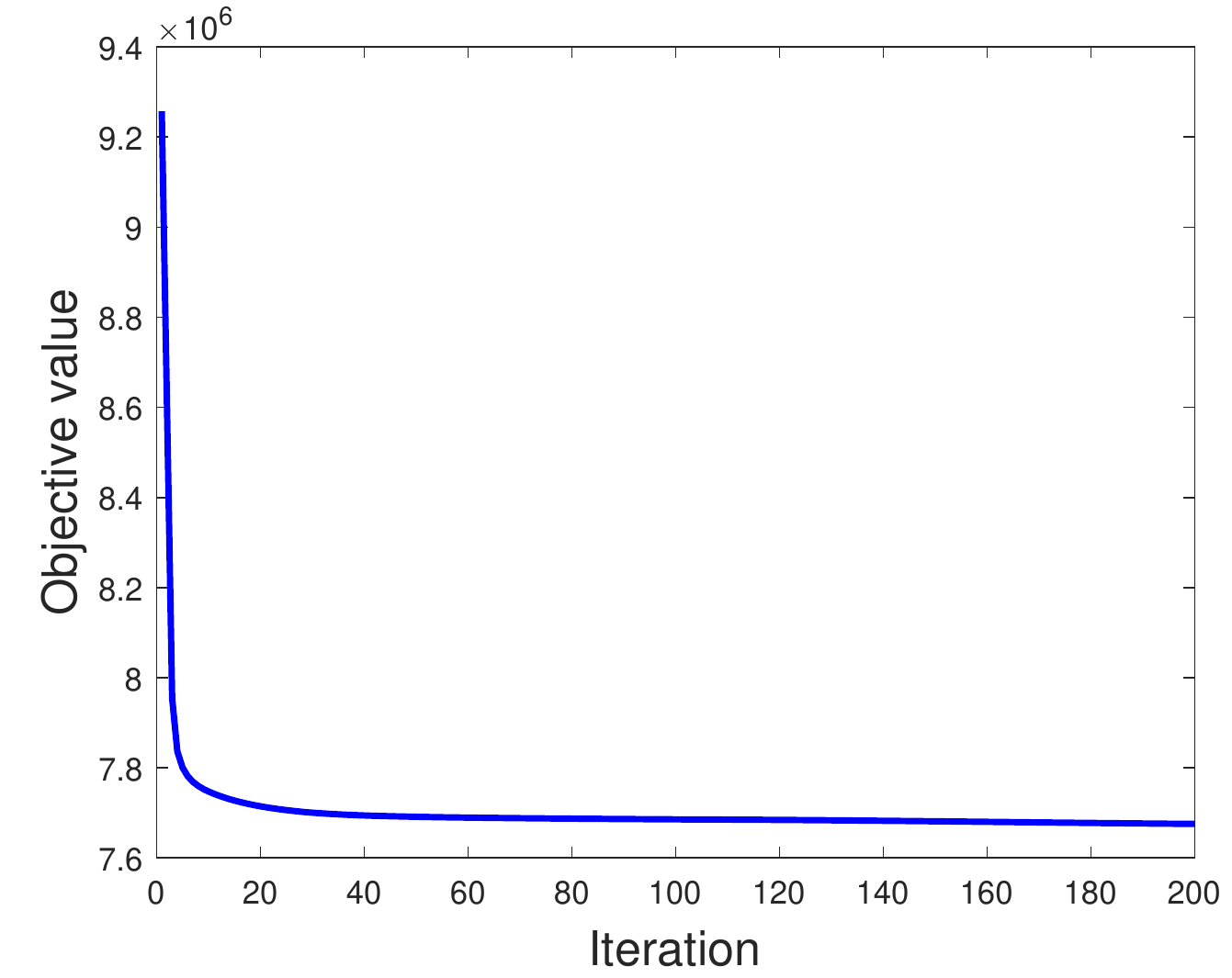}}
\hspace{1pt}
\subfigure[Mfeat]{
\includegraphics[width=0.23\textwidth]{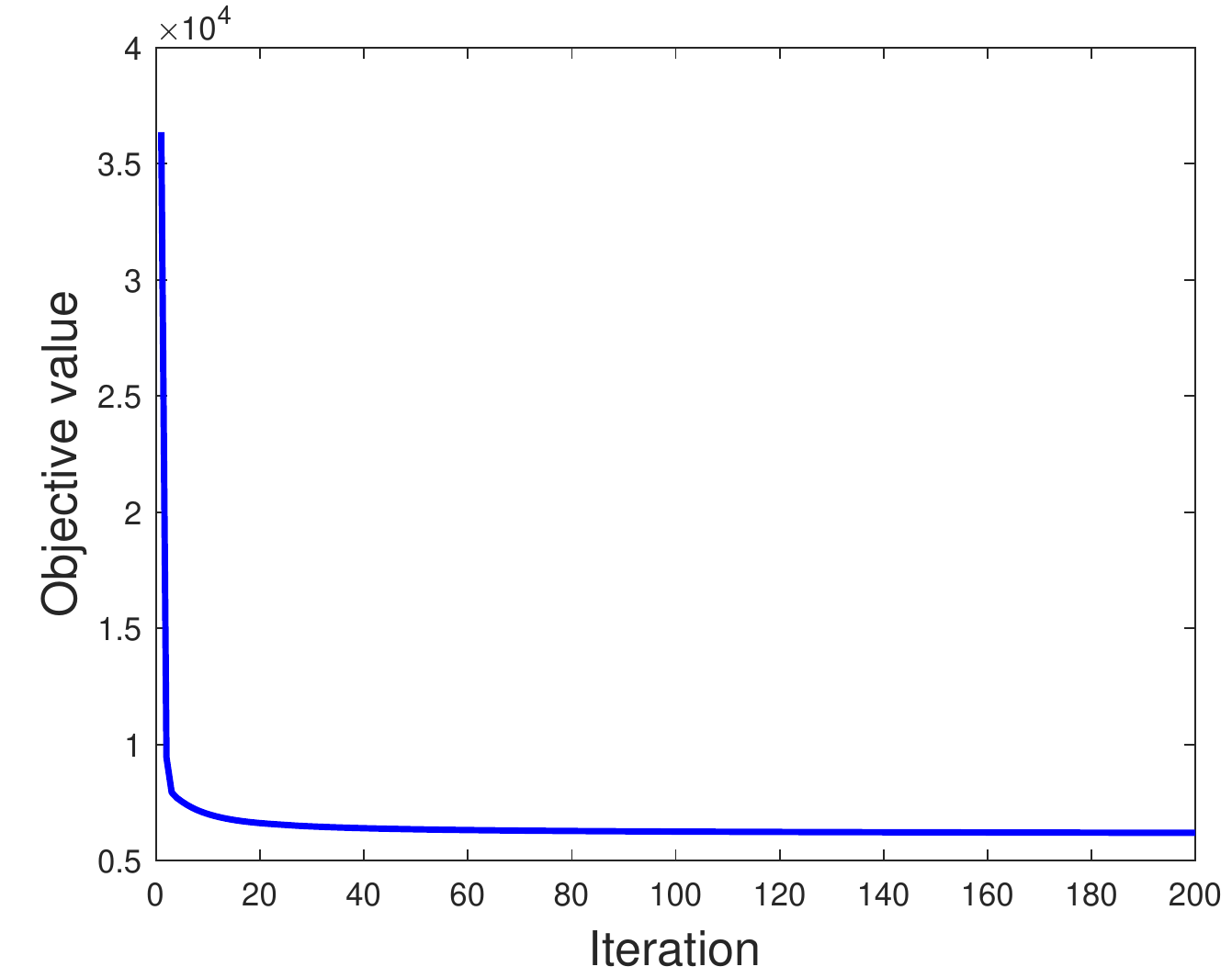}}
\hspace{1pt}
\subfigure[BA]{
\includegraphics[width=0.23\textwidth]{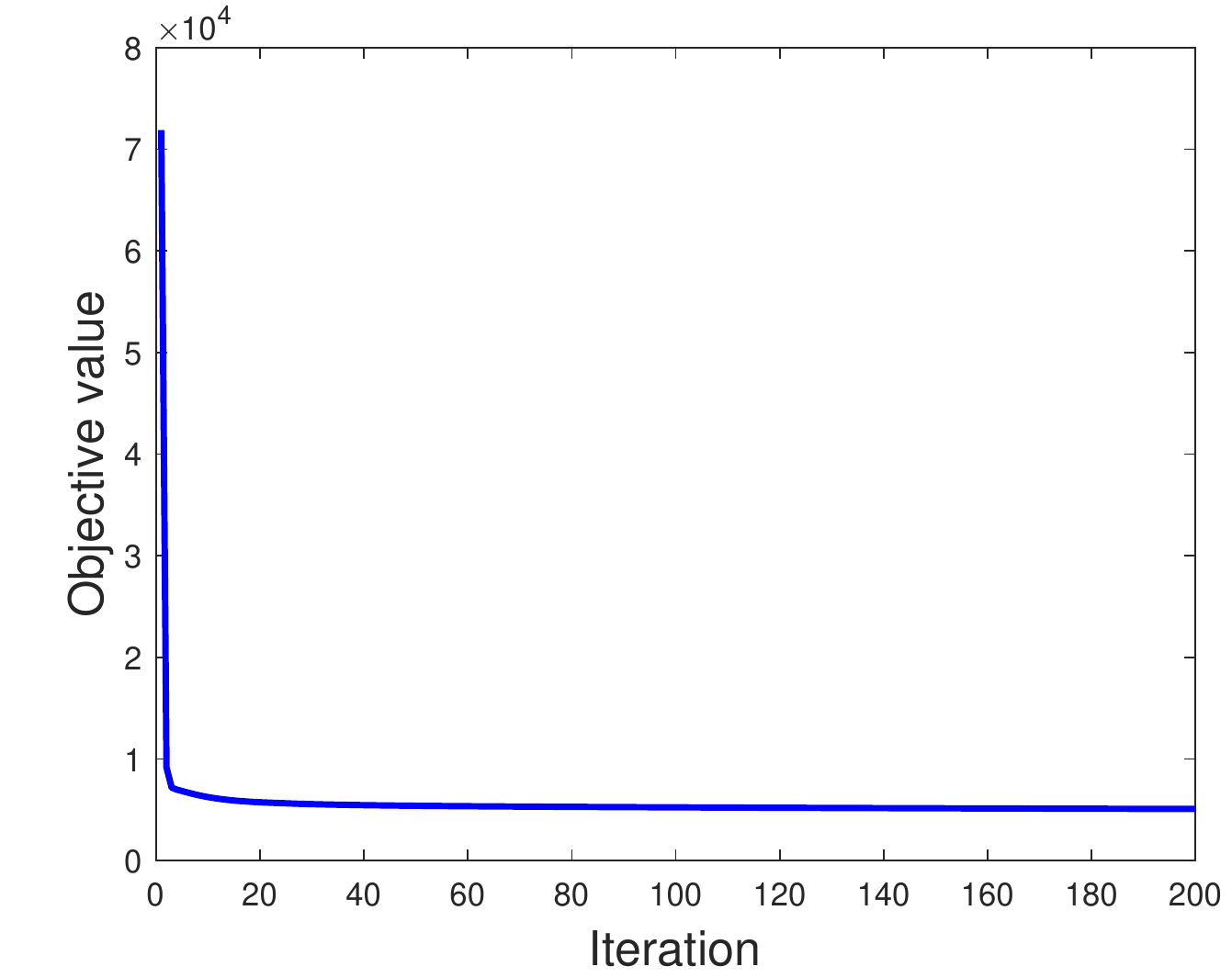}}
\hspace{1pt}
\subfigure[Movement]{
\includegraphics[width=0.23\textwidth]{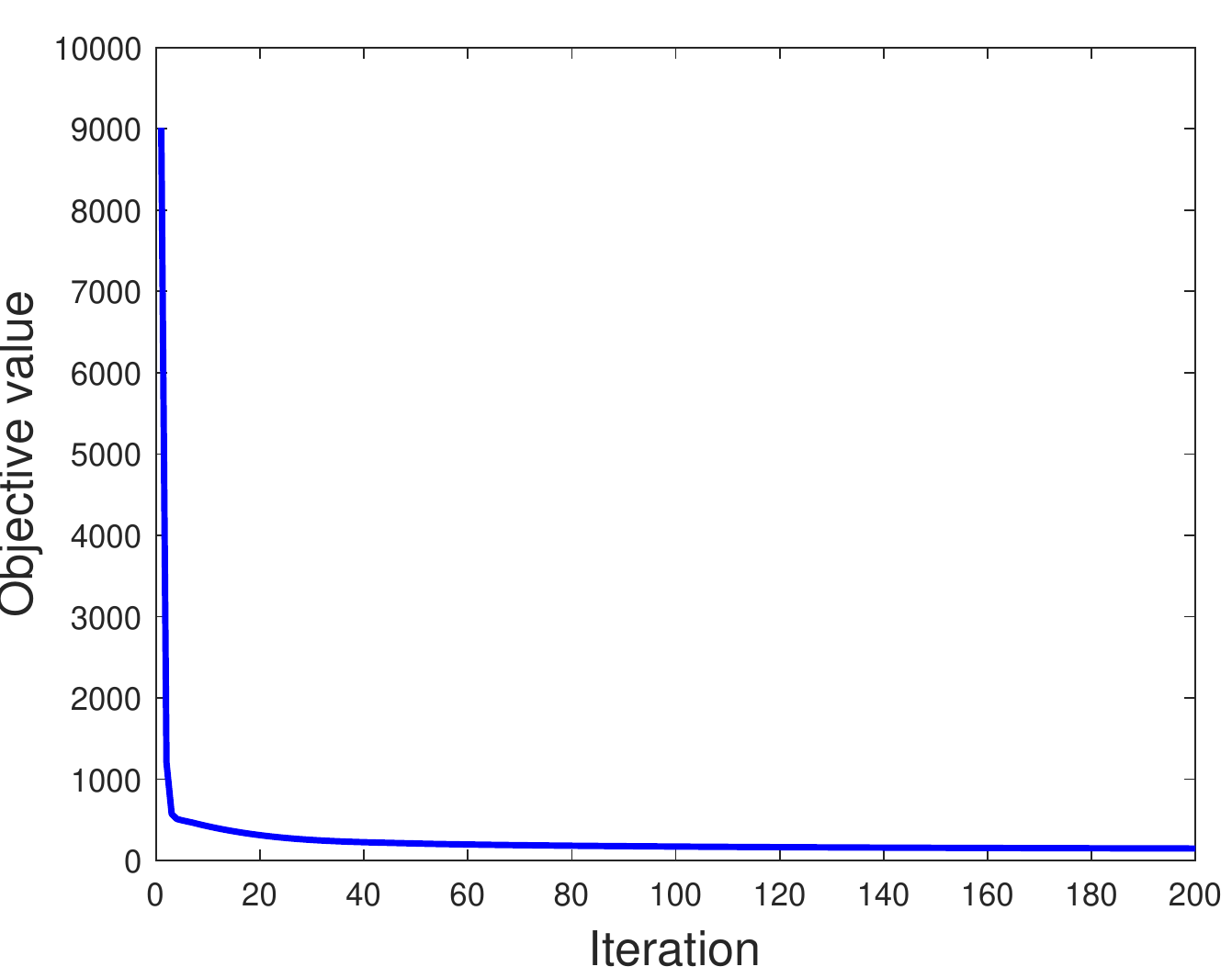}}
\end{center}
\caption{Convergence curves of EMMF.}
\label{convcurve}
\end{figure*}

\begin{figure*}
\begin{center}

\subfigure[COIL20]{
\includegraphics[width=0.23\textwidth]{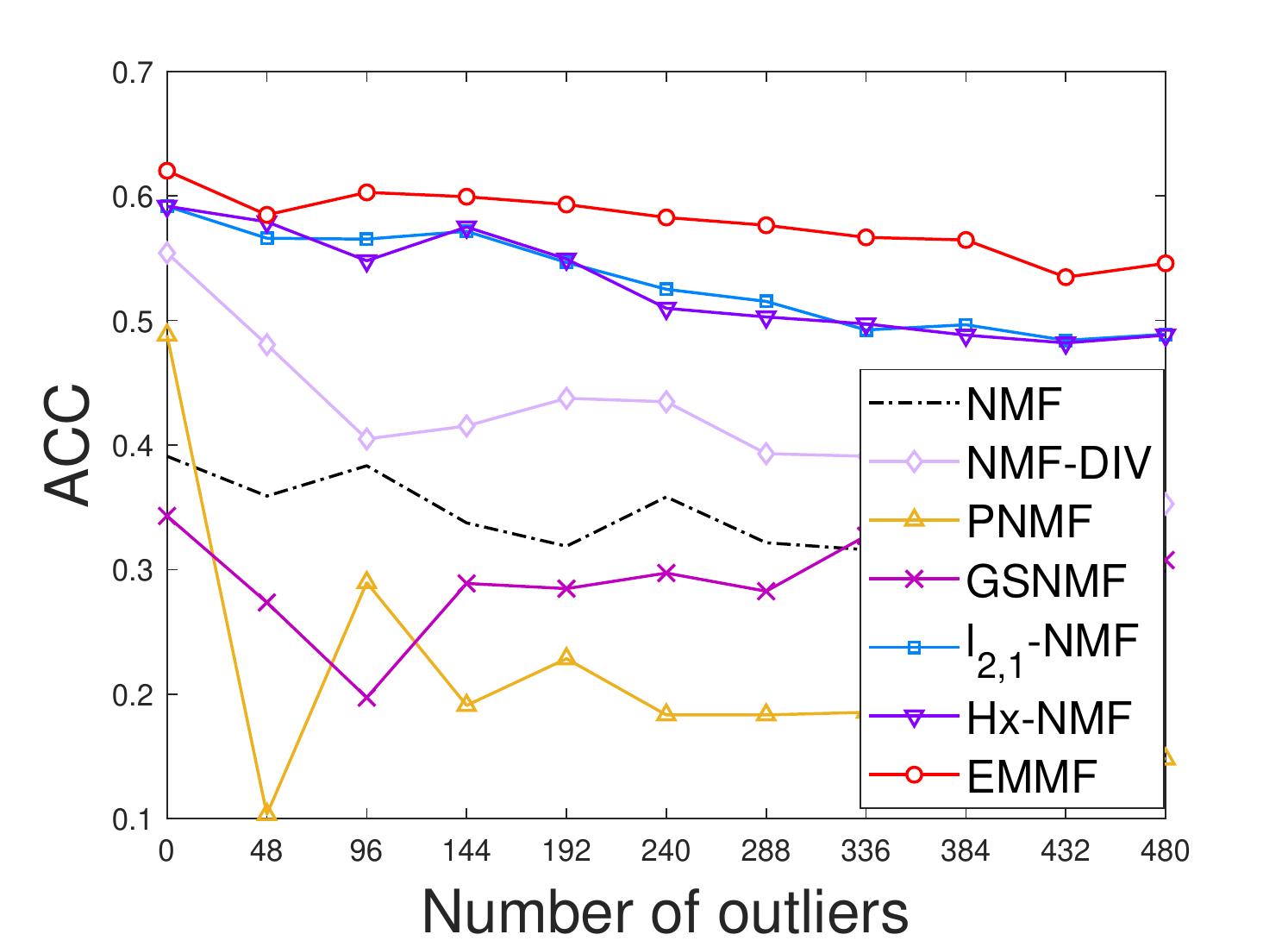}}
\hspace{1pt}
\subfigure[COIL100]{
\includegraphics[width=0.23\textwidth]{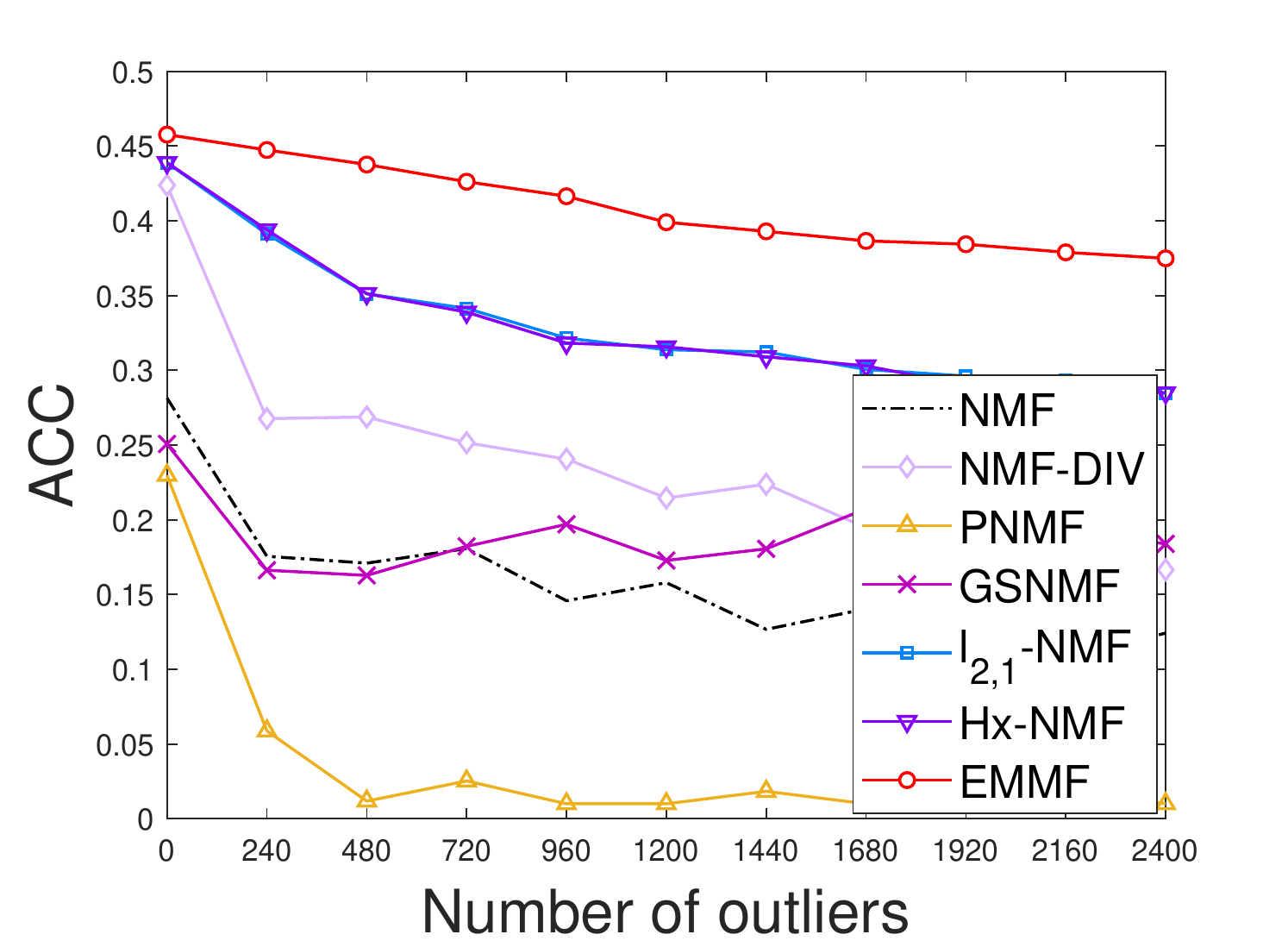}}
\hspace{1pt}
\subfigure[YALE]{
\includegraphics[width=0.23\textwidth]{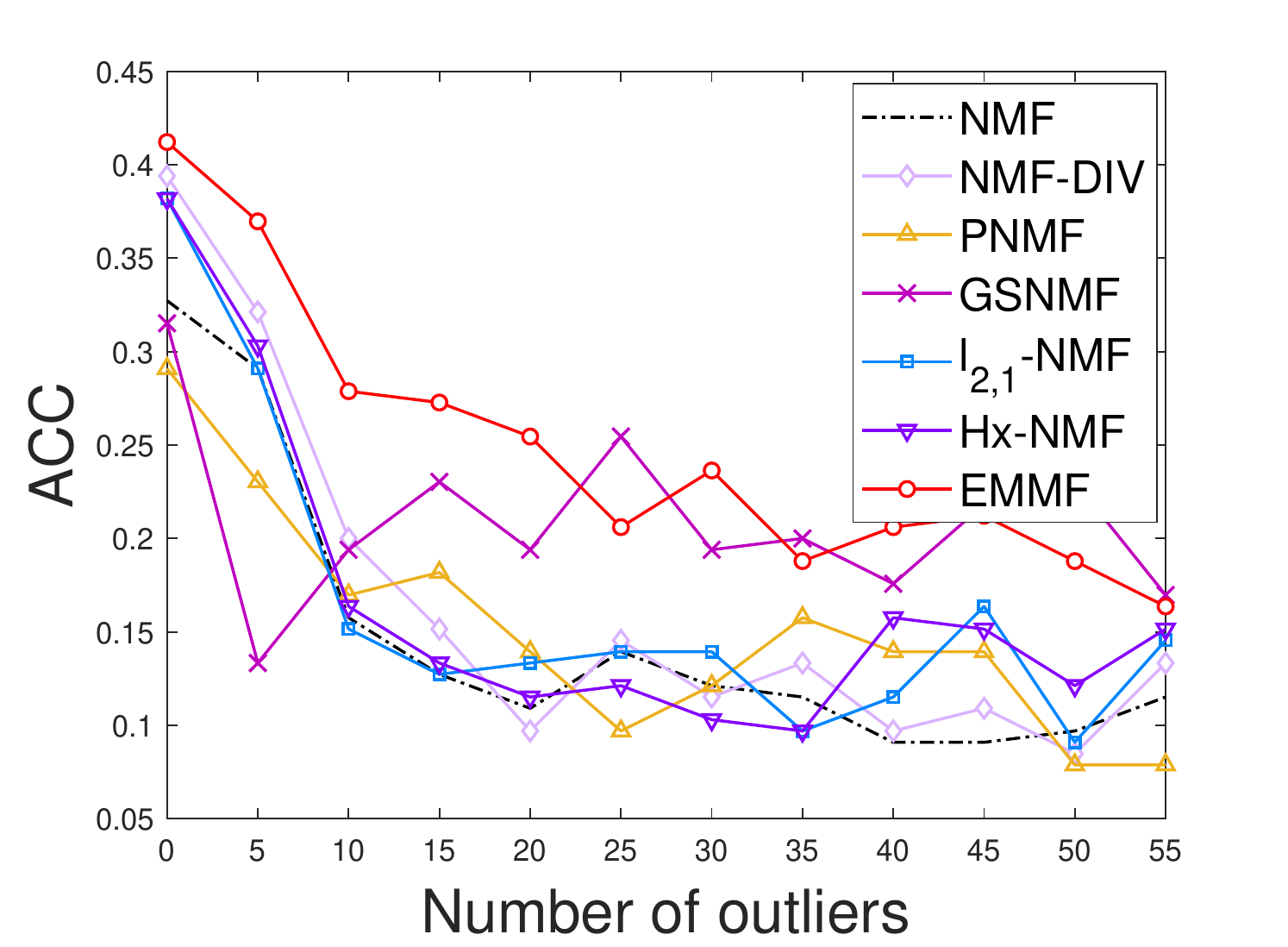}}
\hspace{1pt}
\subfigure[JAFFE]{
\includegraphics[width=0.23\textwidth]{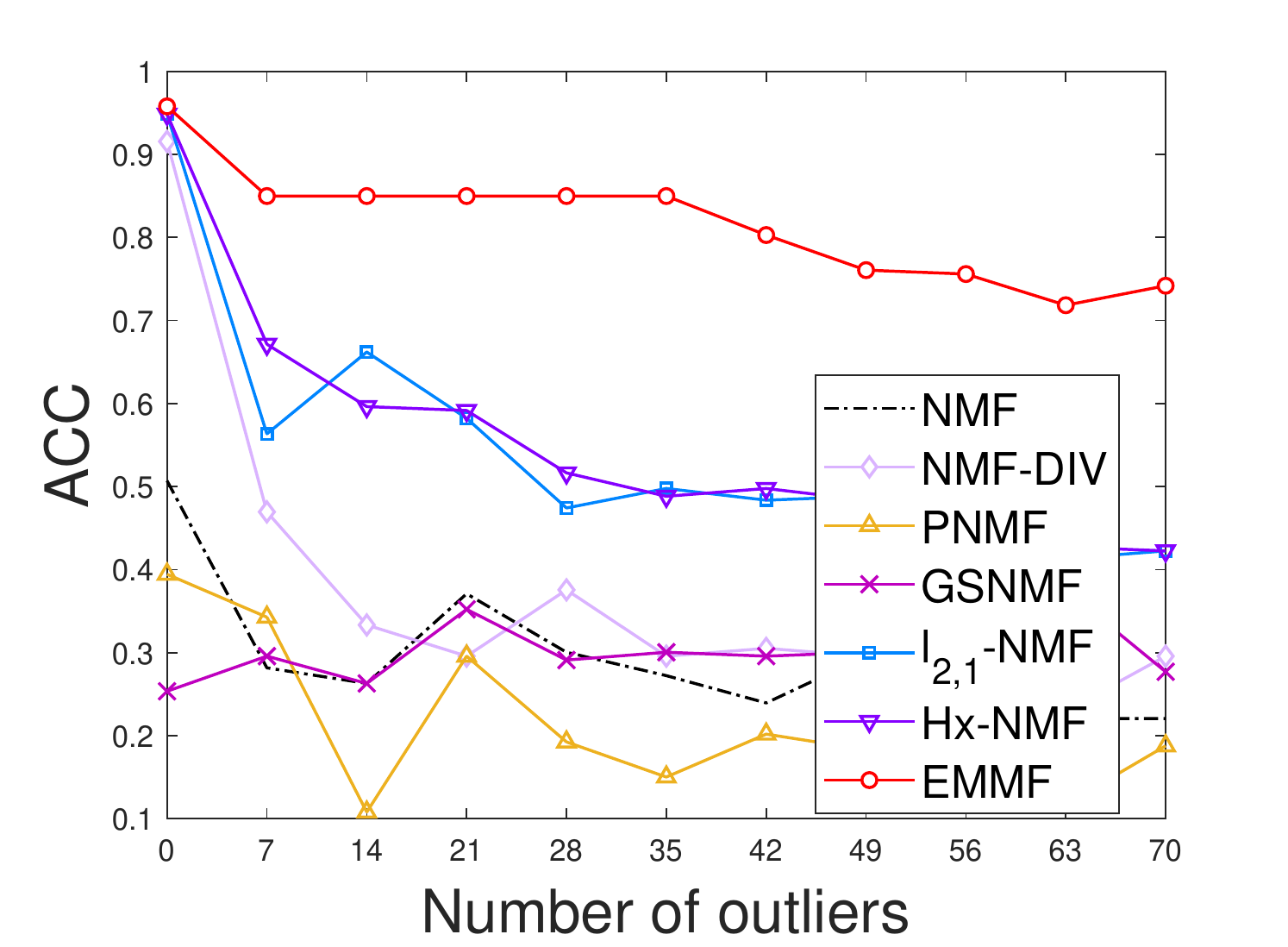}}
\end{center}

\begin{center}
\subfigure[UMIST]{
\includegraphics[width=0.23\textwidth]{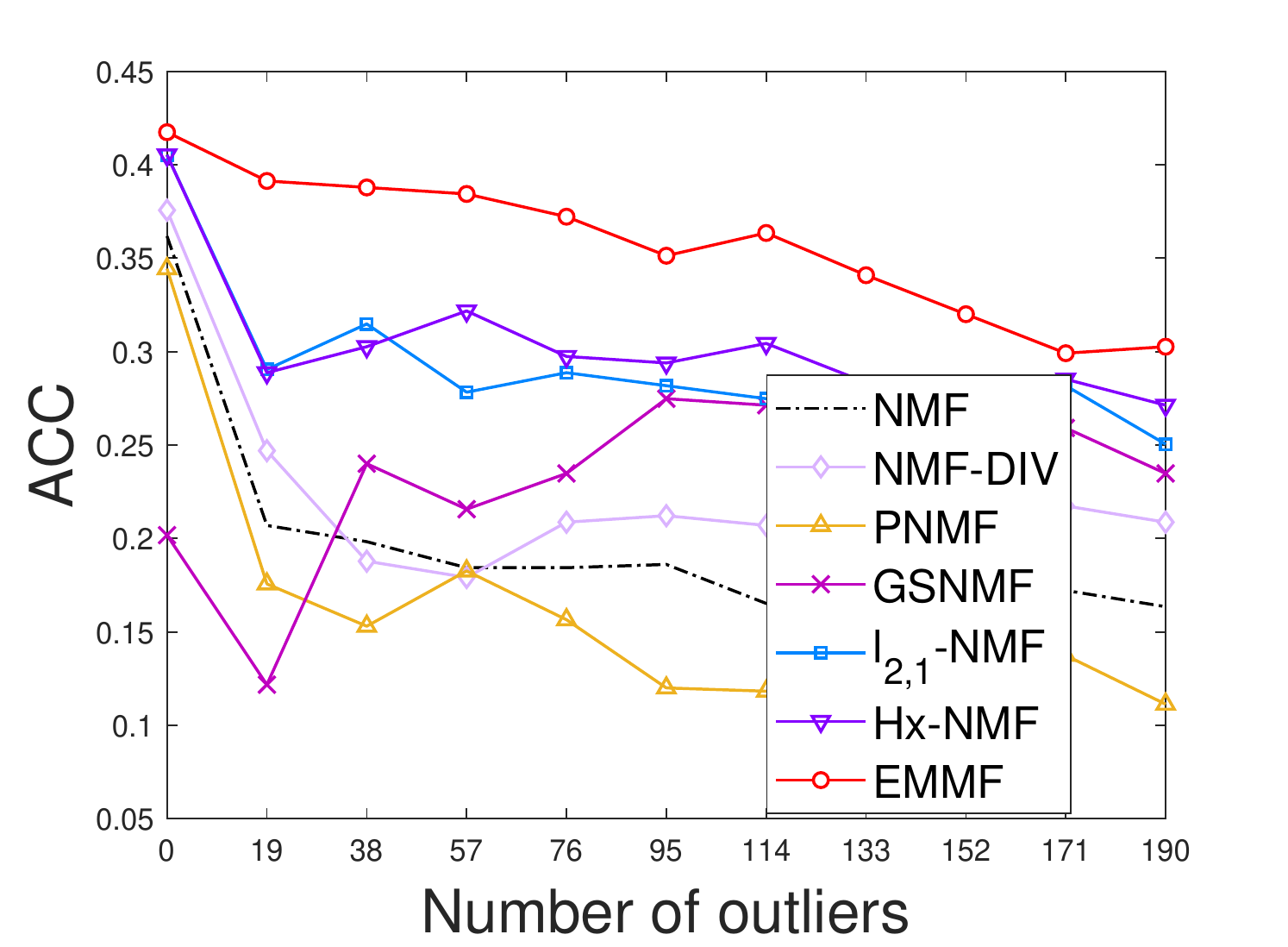}}
\hspace{1pt}
\subfigure[Mfeat]{
\includegraphics[width=0.23\textwidth]{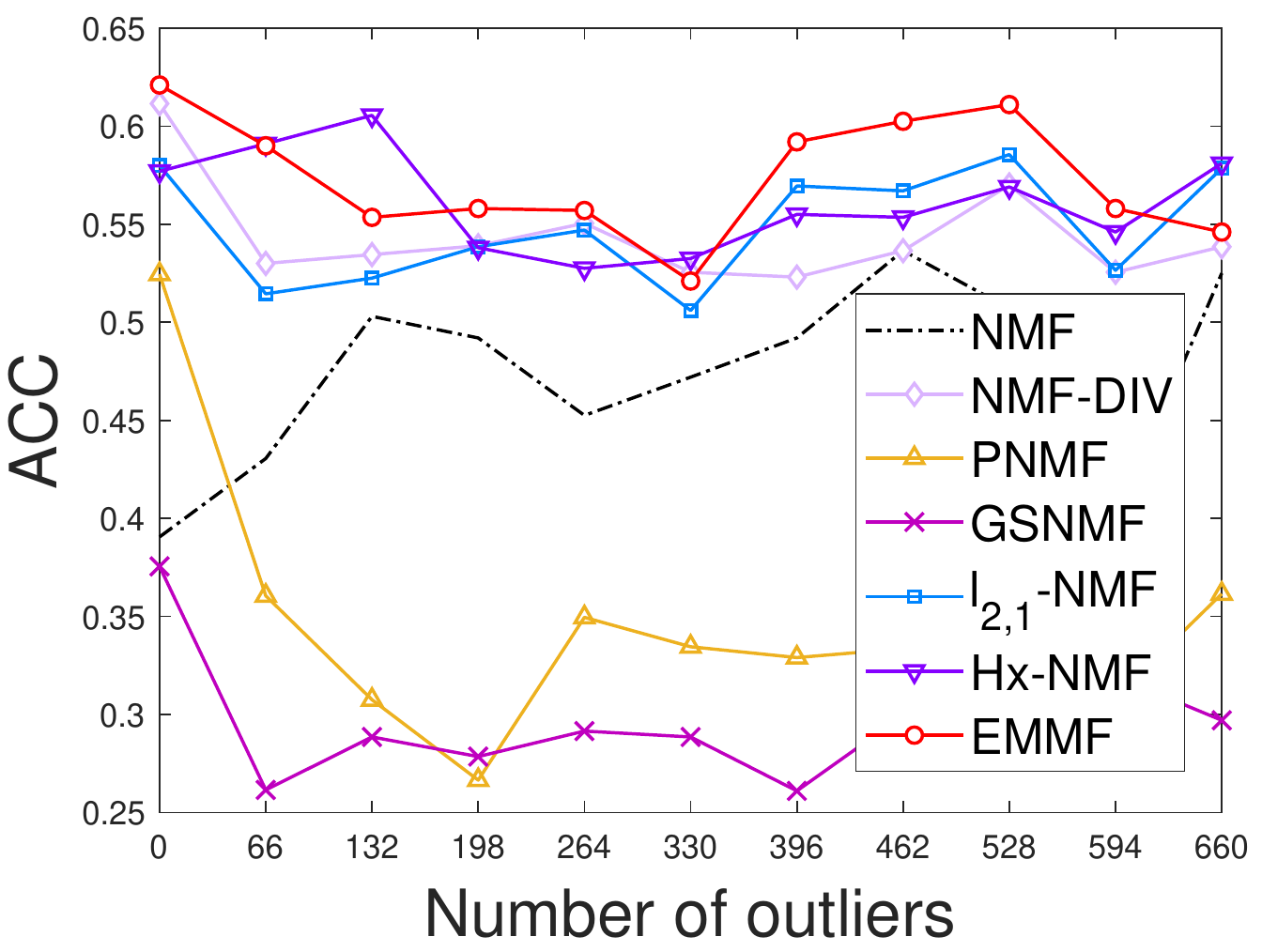}}
\hspace{1pt}
\subfigure[BA]{
\includegraphics[width=0.23\textwidth]{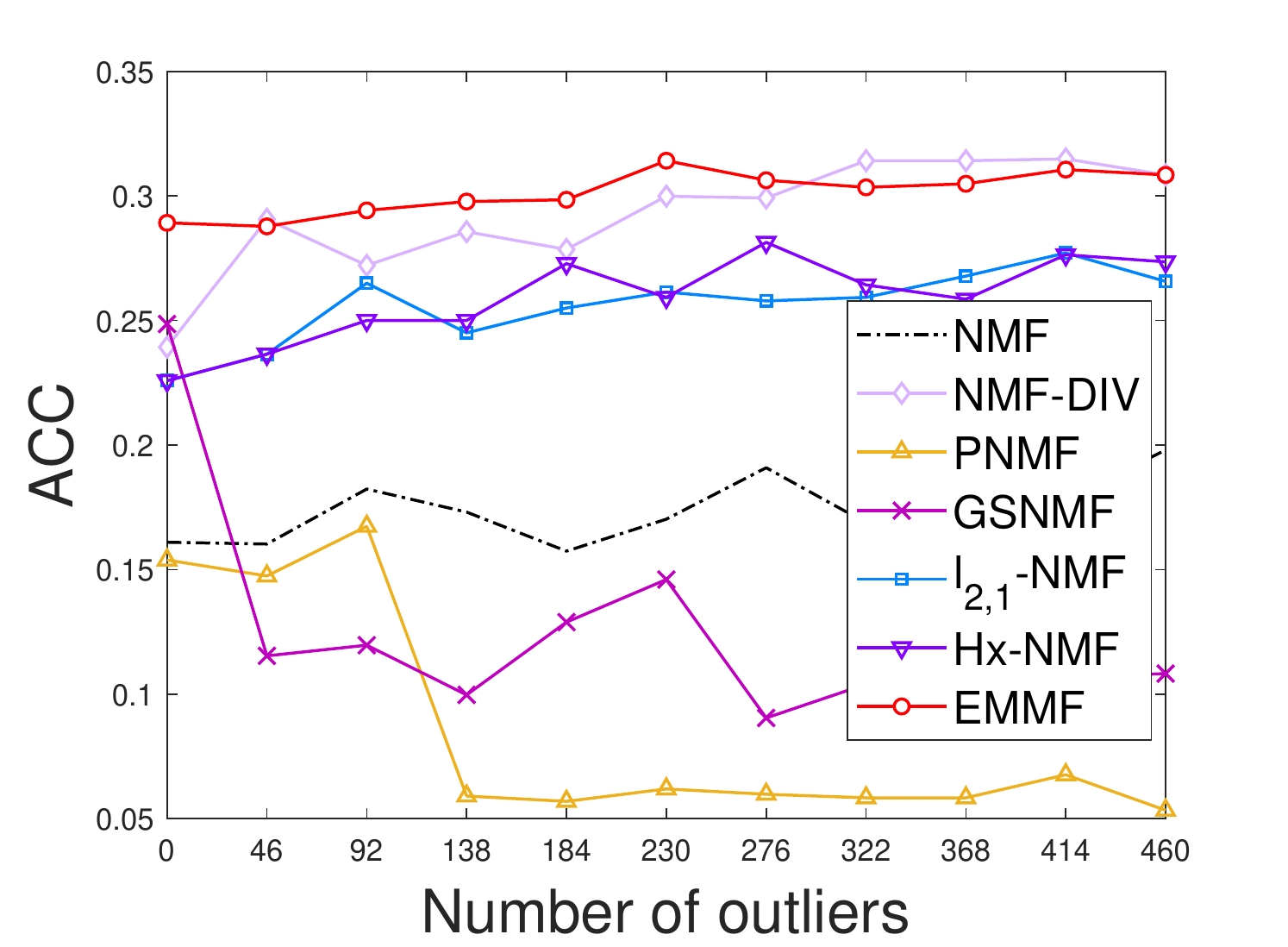}}
\hspace{1pt}
\subfigure[Movement]{
\includegraphics[width=0.23\textwidth]{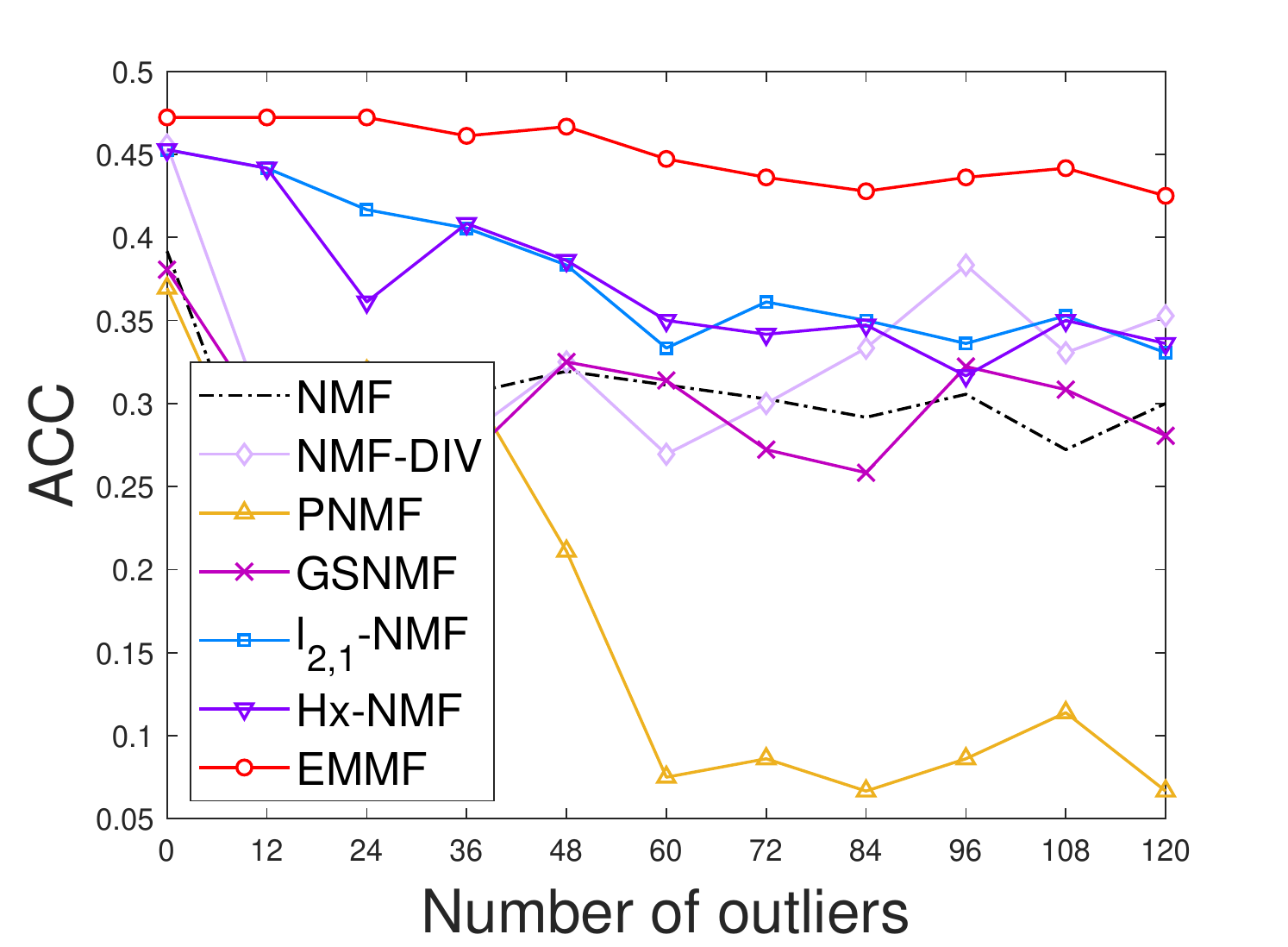}}
\end{center}
\caption{Performance of different methods on datasets with increasing number of outliers.}
\label{outliercurve}
\end{figure*}

\textbf{Competitors}: six state-of-the-arts are taken for comparison, including 
\begin{itemize}
\item NMF~\cite{nmf}: NMF with the Frobenius norm formulation.
\item NMF-DIV~\cite{nmf}: NMF with divergence formulation.
\item PNMF~\cite{pnmf}: probabilistic NMF solved by variational Bayesian.
\item GSNMF~\cite{gsnmf}: generalized separable NMF, which approximates both the rows and columns of the data matrix.
\item $\ell _{2,1}$-NMF~\cite{l21nmf}: NMF with the $\ell _{2,1}$ norm loss function.
\item Hx-NMF~\cite{hxnmf}: NMF with the logarithmic loss function.
\end{itemize}
For PNMF and GSNMF, the best parameters  are obtained by searching the grid $\{10^{-3},10^{-2},\cdots,10^{3}\}$, and they are initialized with the approaches suggested by the authors. For the other methods, including the proposed EMMF, we initialize $\mathbf{U}$ and $\mathbf{V}$ with $k$-means~\cite{kmeans}. For PNMF, variational Bayesian gives the best results after 100 iterations. The maximum iteration number for all the other methods is set as 500. To alleviate the influence of initiation condition, all the methods are performed for twenty repetitions, and the averaged results are reported.

\begin{table*}
\caption{Performance of G-EMMF on real-world datasets. Best results are in bold face.}
\label{gemmftable}
\centering
\renewcommand\arraystretch{1.2}
\small
\begin{tabular}{|p{0.6cm}<{\centering}|p{1.5cm}<{\centering}|p{1.4cm}<{\centering}|p{1.4cm}<{\centering}|p{1.4cm}<{\centering}|p{1.4cm}<{\centering}|p{1.4cm}<{\centering}|p{1.4cm}<{\centering}|p{1.4cm}<{\centering}|p{1.5cm}<{\centering}|}

\hline
\multirow{7}*{ACC}& & COIL20 & COIL100 & YALE&JAFFE&UMIST&Mfeat&BA&Movement\\
\cline{2-10}
&RMNMF  &0.5792 &0.4149 &0.4448&0.9108 &0.4191 & 0.5386 &0.3154 &0.4167\\
\cline{2-10}
&NLCF &0.6431 &0.4153 &0.4352 &0.9315 &0.3910 & 0.7472 &0.3652 &0.4444\\
\cline{2-10}
&LCF &0.6583 &0.4596&0.4158  &0.9268 &0.3920 &  0.7538 &0.3929 &0.4667\\
\cline{2-10}
&LSNMF  &0.6528 &0.4651 &0.2485&0.7502 &0.3986 &0.6166   &0.0879 &0.3278 \\
\cline{2-10}
&SRMCF &0.6660 &0.3990  &0.4121&0.9615 &0.4254 & 0.7036 &0.4103 &0.4583\\
\cline{2-10}
&NMFAN &0.6500 &0.4558&0.4230  &0.9305 &0.3969 & 0.7064   &0.3111 &0.4417\\
\cline{2-10}
& G-EMMF  &\textbf{0.6882} &\textbf{0.4803} &\textbf{0.4558}&\textbf{0.9812} &\textbf{0.4588} &\textbf{0.7833} &\textbf{0.4795} &\textbf{0.4778}\\
\hline
\hline
\multirow{7}*{NMI}& & COIL20 & COIL100 & YALE&JAFFE&UMIST&MNIST&BA&Movement\\
\cline{2-10}
&RMNMF  &0.6944 &0.6605&0.4909 &0.8894 &0.5938 & 0.5206  &0.4780 &0.5217\\
\cline{2-10}
&NLCF &0.7364 &0.6616&0.4867  &0.9197 &0.5883 &0.7112  & 0.4810  &0.6158\\
\cline{2-10}
&LCF   &0.7483 &0.7213 &0.4809&0.9037 &0.5901 &  0.7203 &0.5288 &0.6237\\
\cline{2-10}
&LSNMF  &0.7699 &0.6633&0.2510 &0.8404 &0.5310 & 0.6551  &0.0827 &0.4393\\
\cline{2-10}
&SRMCF  &0.7699 &0.6686&0.4745 &0.9556 &0.6292 &0.6464 &0.5453 &0.5900\\
\cline{2-10}
&NMFAN  &0.7492 &0.7330 &0.4682&0.9208 &0.5963 &  0.6690 &0.4433 &0.6167\\
\cline{2-10}
& G-EMMF &\textbf{0.8055} &\textbf{0.7547} &\textbf{0.5203} &\textbf{0.9731} &\textbf{0.6696} &\textbf{0.7888} &\textbf{0.6191} &\textbf{0.6362}\\
\hline
\end{tabular}

\end{table*}

\begin{figure*}
\begin{center}

\subfigure[COIL20]{
\includegraphics[width=0.23\textwidth]{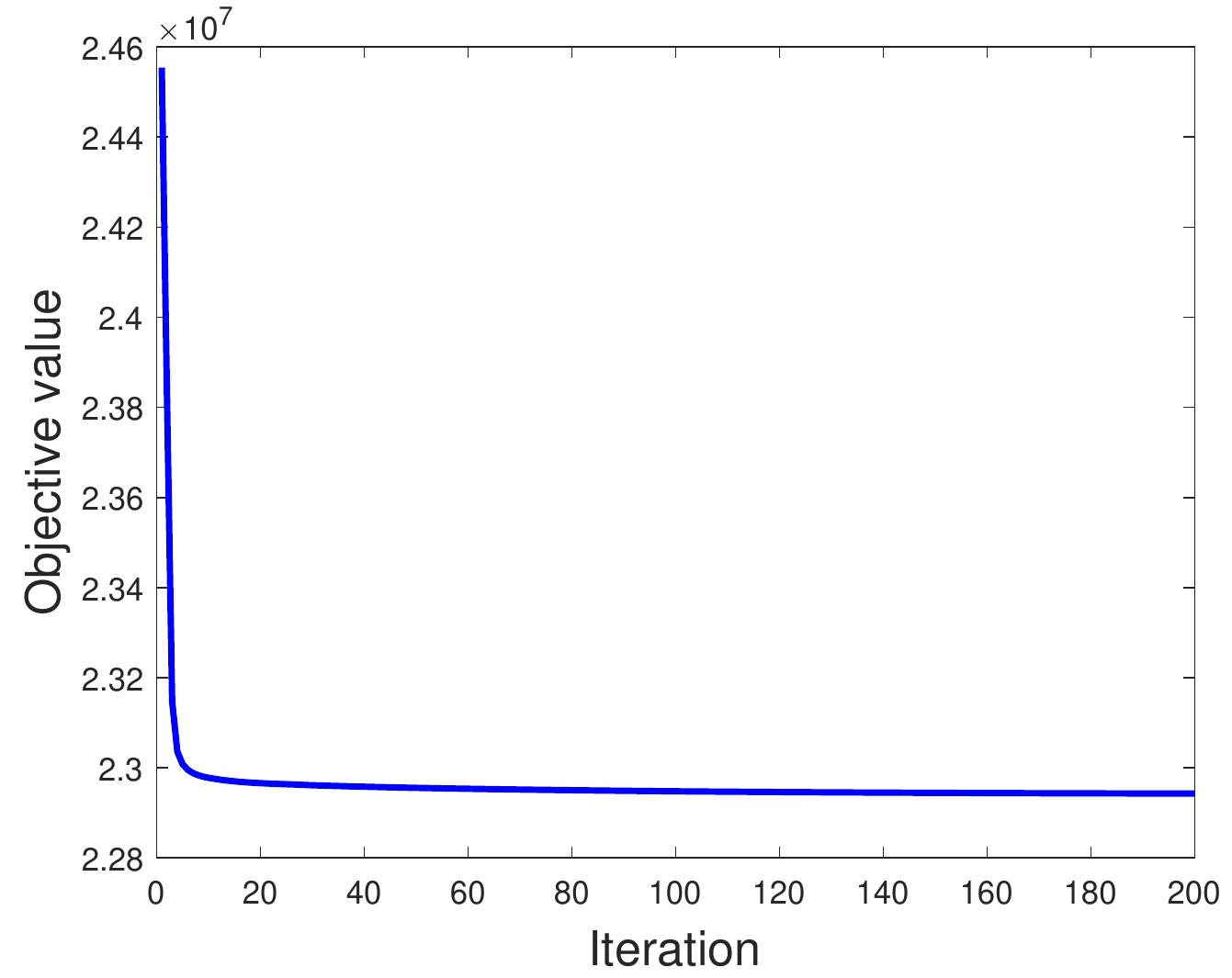}}
\hspace{1pt}
\subfigure[COIL100]{
\includegraphics[width=0.23\textwidth]{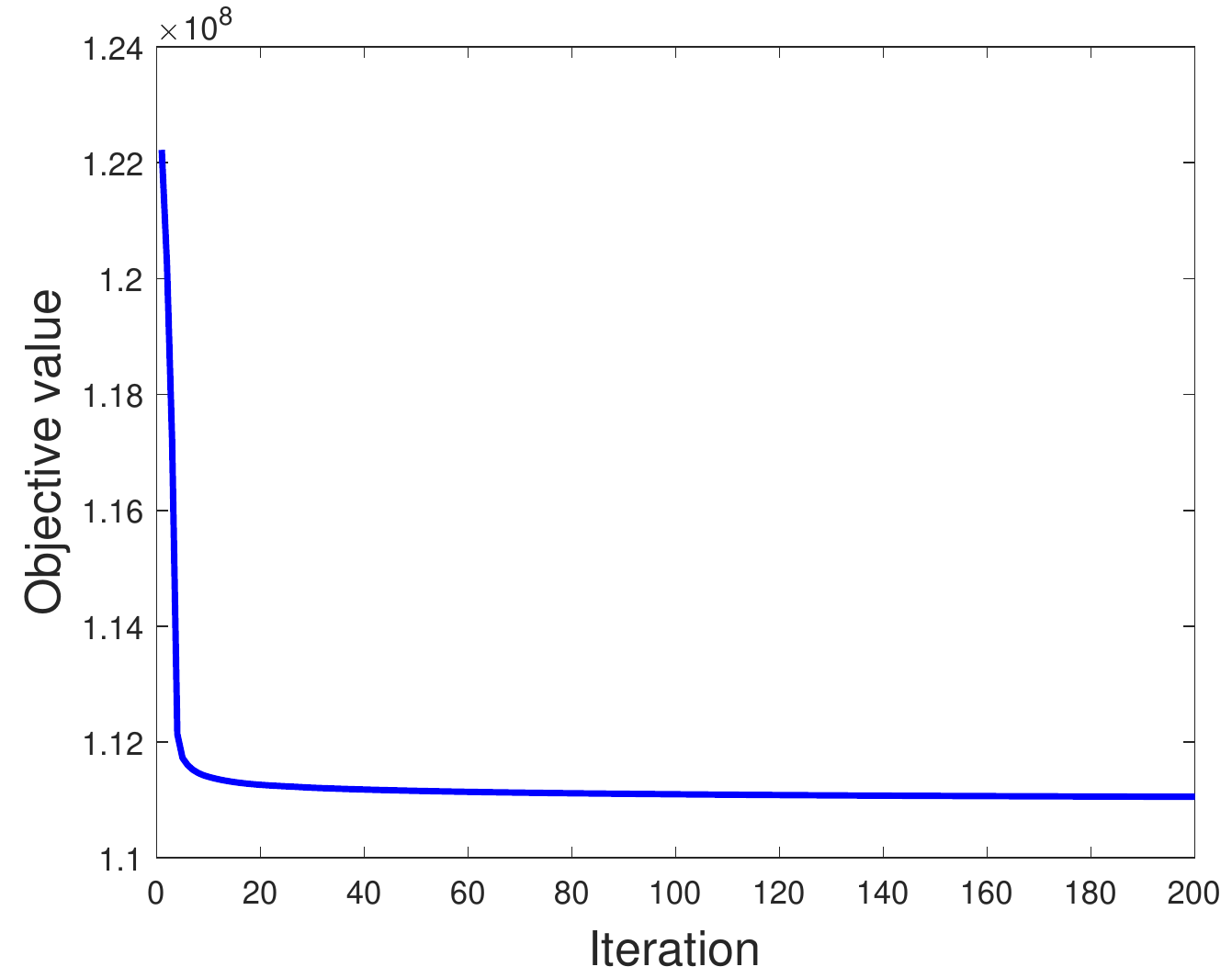}}
\hspace{1pt}
\subfigure[YALE]{
\includegraphics[width=0.23\textwidth]{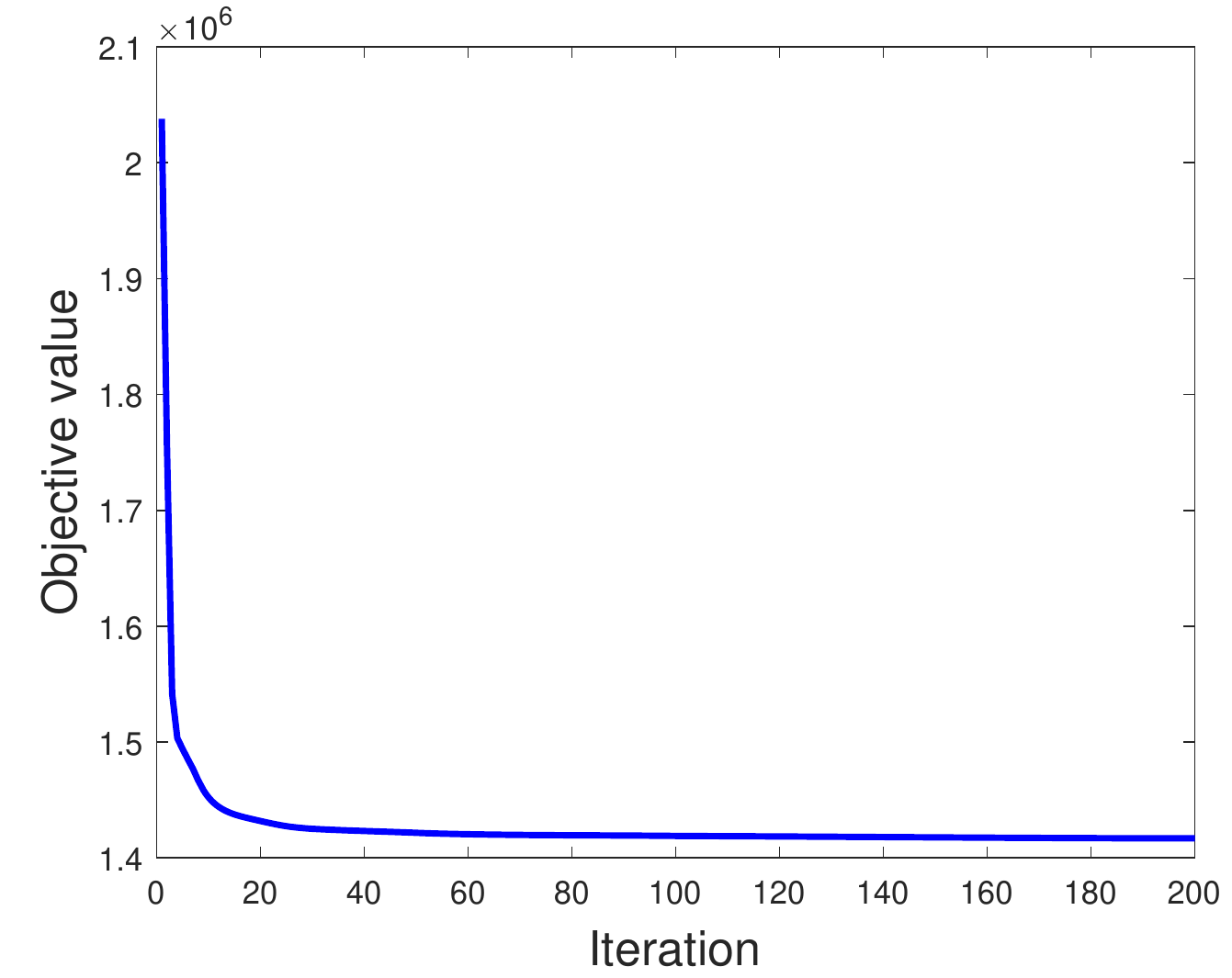}}
\hspace{1pt}
\subfigure[JAFFE]{
\includegraphics[width=0.23\textwidth]{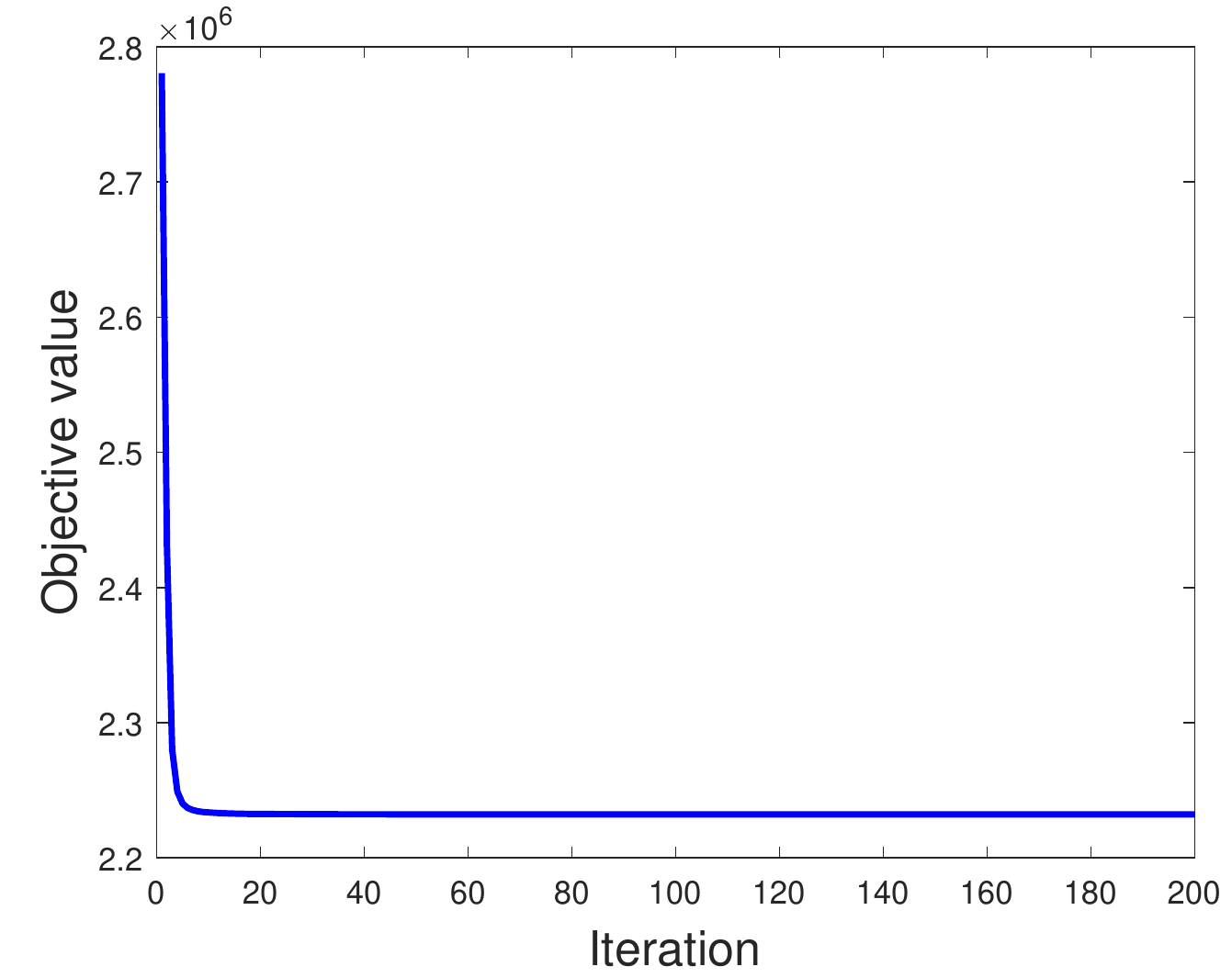}}
\end{center}

\begin{center}
\subfigure[UMIST]{
\includegraphics[width=0.23\textwidth]{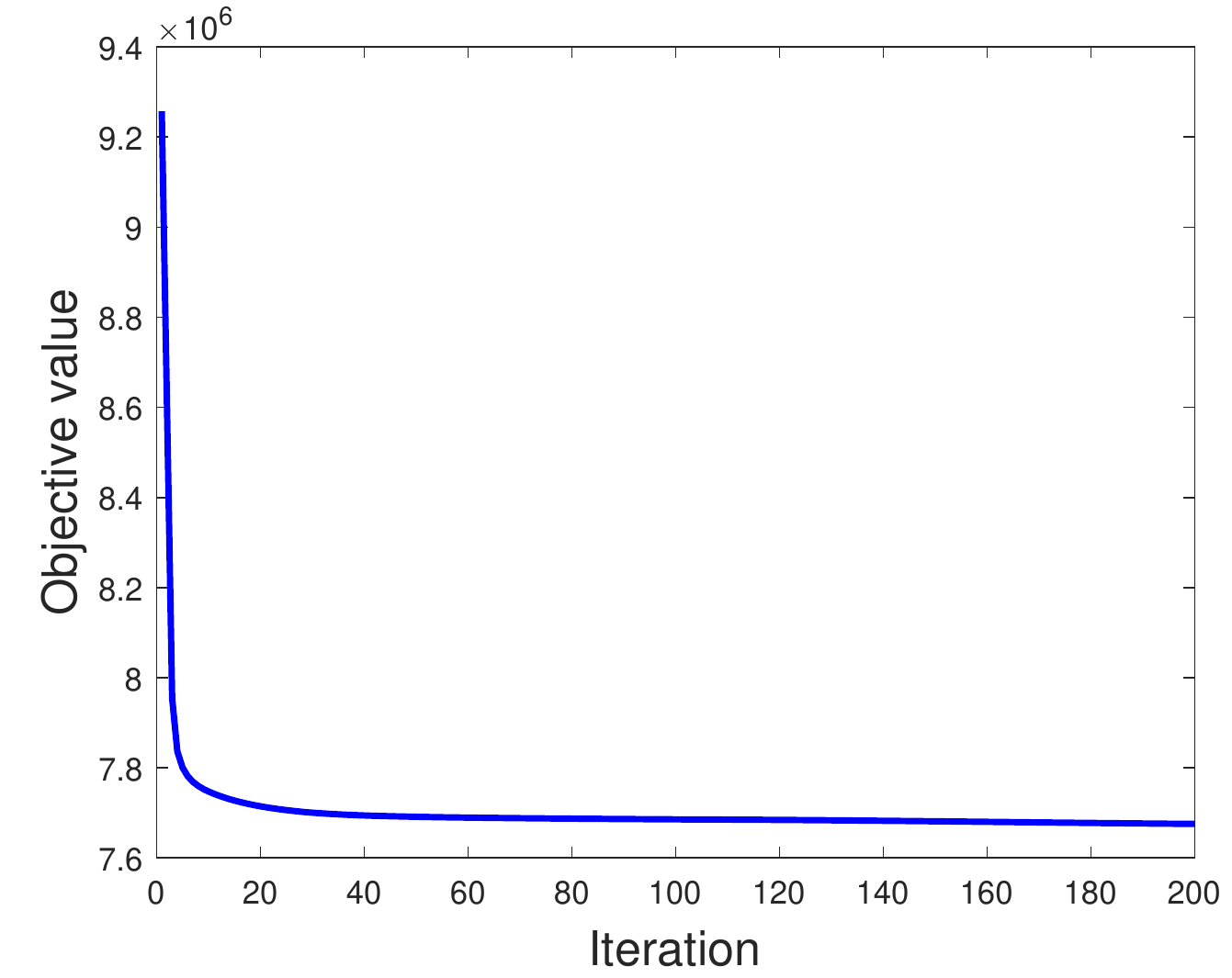}}
\hspace{1pt}
\subfigure[Mfeat]{
\includegraphics[width=0.23\textwidth]{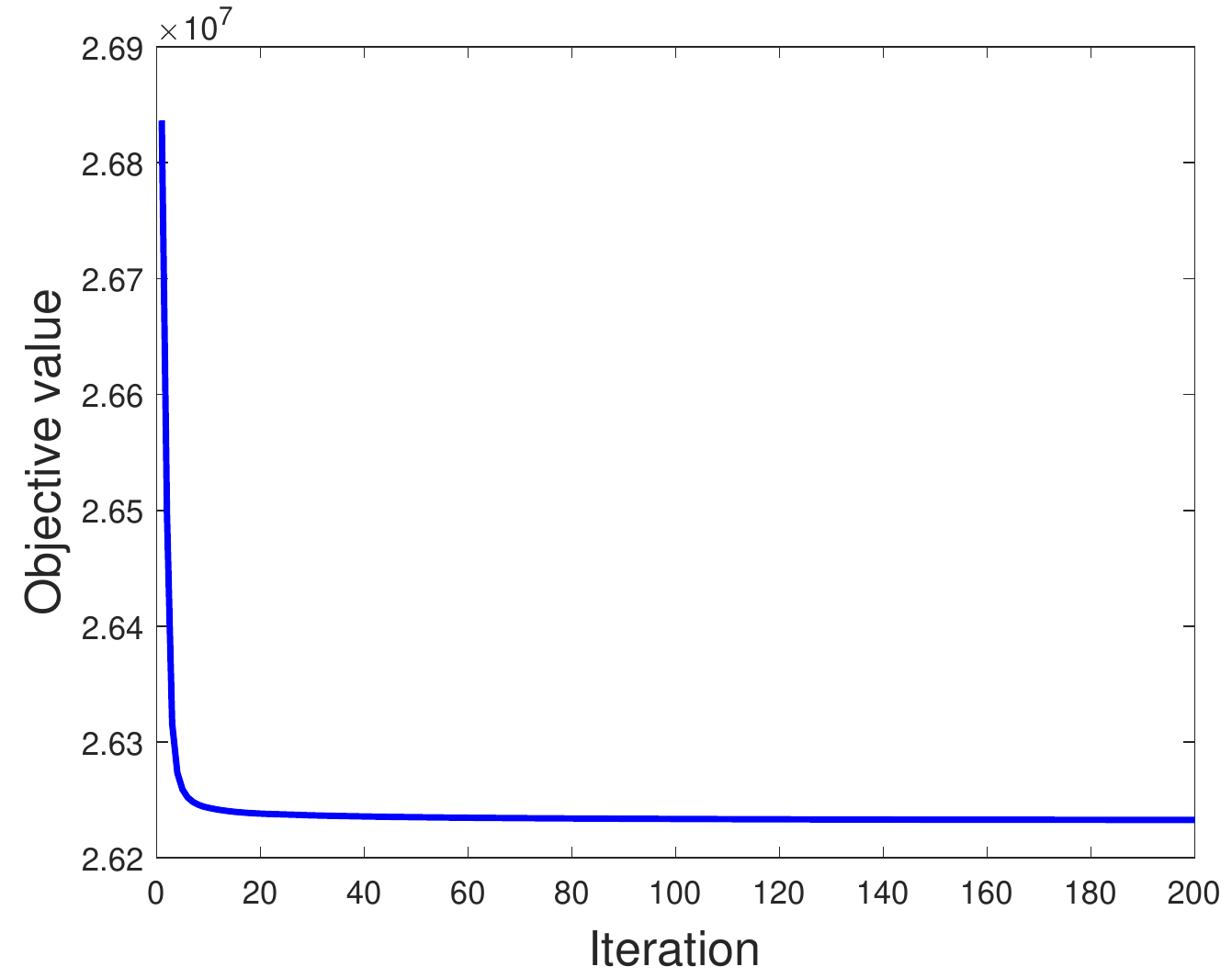}}
\hspace{1pt}
\subfigure[BA]{
\includegraphics[width=0.23\textwidth]{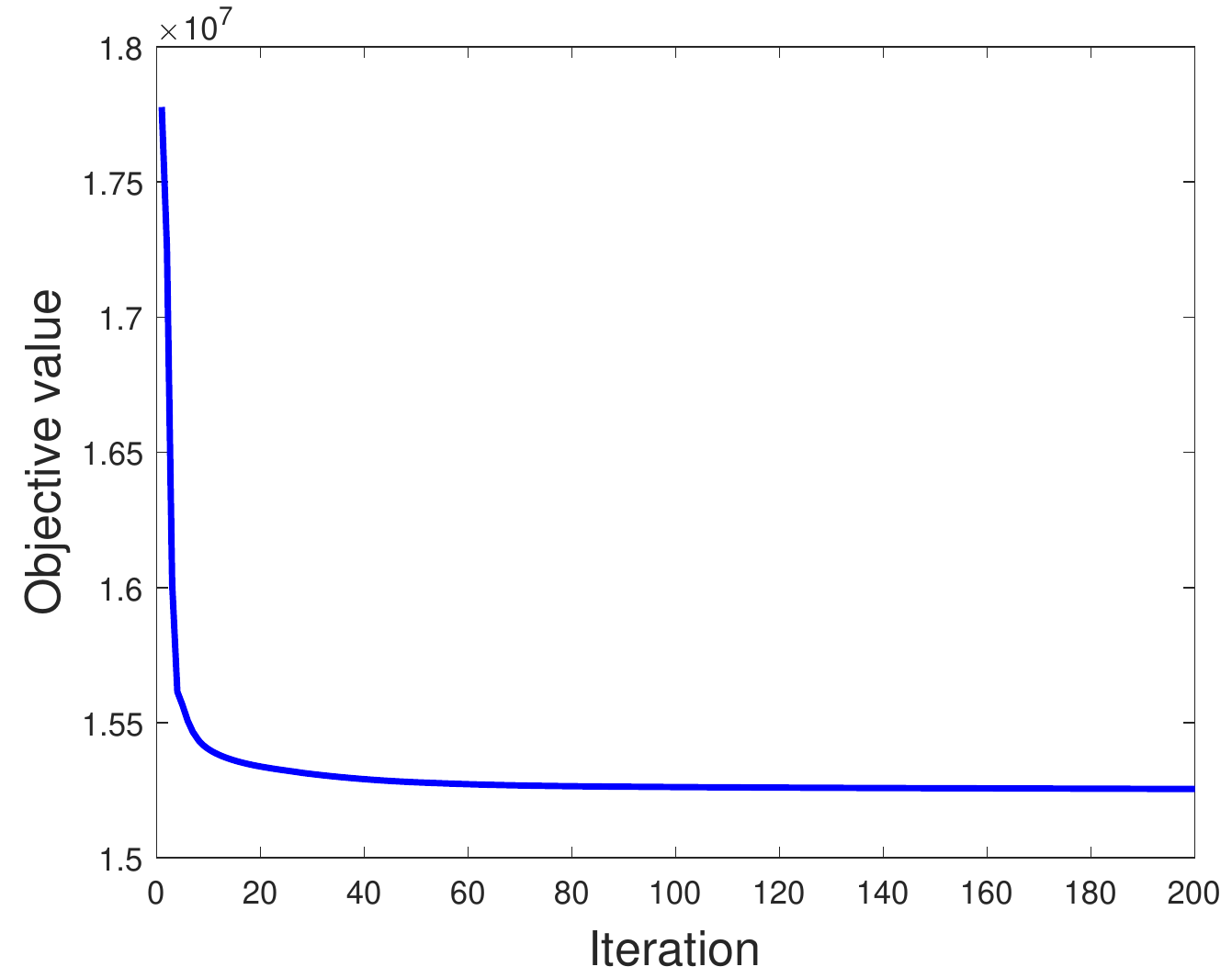}}
\hspace{1pt}
\subfigure[Movement]{
\includegraphics[width=0.23\textwidth]{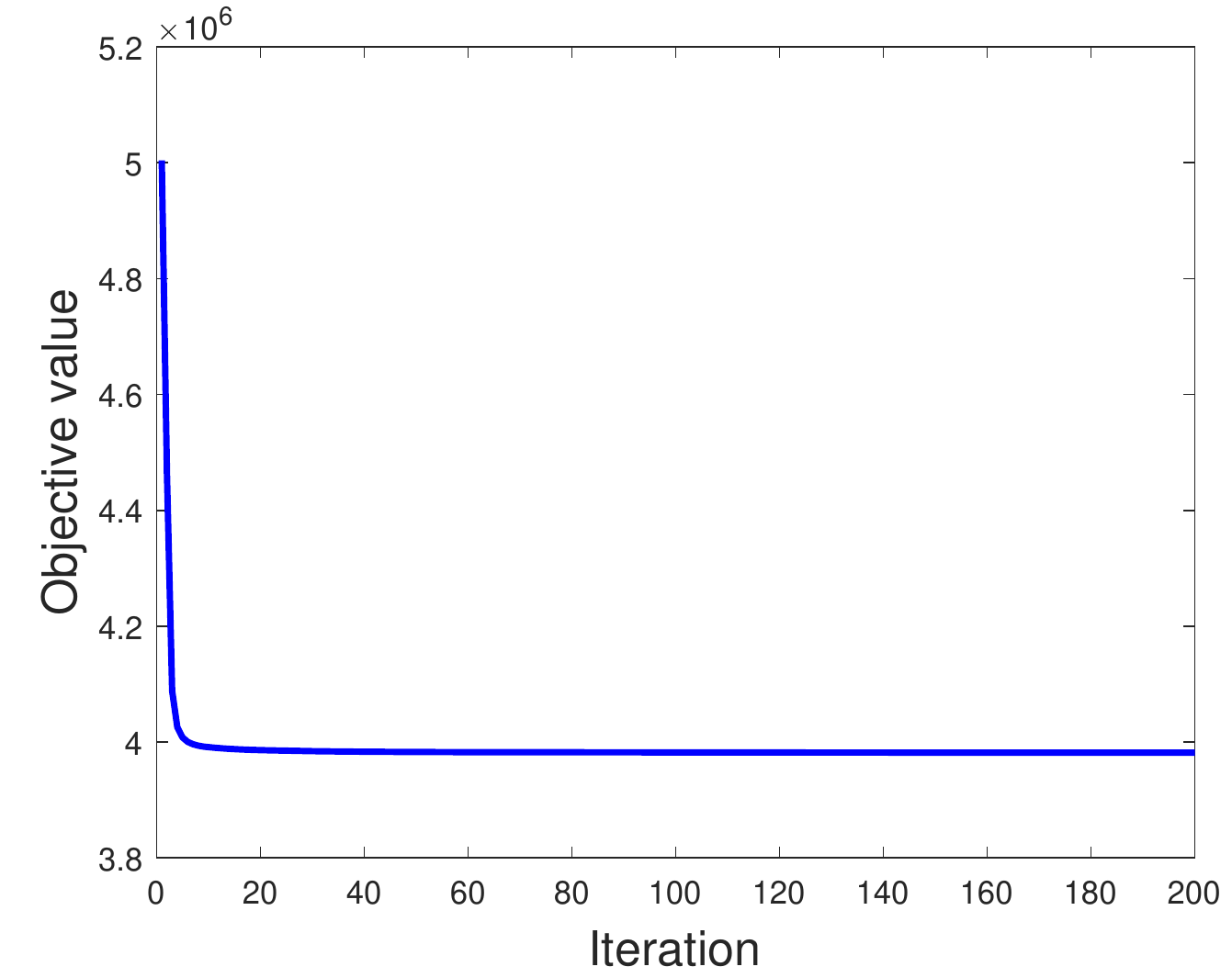}}
\end{center}
\caption{Convergence curves of G-EMMF.}
\label{convcurveg}
\end{figure*}

\textbf{Performance}: the clustering results on different datasets are exhibited in Table~\ref{emmftable}. The proposed EMMF shows best performance in terms of ACC and NMI. On these datasets, most samples obey the normal distribution. The performance of PNMF and GSNMF are unsatisfying due to the lack of clustering interpretation. Specifically, their coefficient matrices do not contain the clear cluster structure. $\ell _{2,1}$-NMF and Hx-NMF show better results than NMF, which demonstrates the advantage of improving robustness. EMMF outperforms all the competitors because it moves the centroids towards the samples with small approximation errors. In this way, most of the samples are well represented by the centroids, and the clustering performance is improved. The convergence curves of EMMF are also given in Fig.~\ref{convcurve}. The optimization algorithm converges within 40 iterations on all the datasets, which ensures the efficiency.

To further prove the robustness of EMMF, we introduce large outliers into the real-world datasets. For each dataset, randomly generated vectors are added into the data matrix. The elements of the vectors are within the range of $[0, 10\times \sigma]$, where $\sigma$ is the largest value in the original data matrix. We perform all the methods methods on the outlier datasets with different number of outliers ($[0,\frac{n}{3}]$), and show the clustering results of the original samples in Fig.~\ref{outliercurve}. As the outlier number increases, the performance of NMF and NMF-DIV drops dramatically, especially on YALE and JAFFE. $\ell _{2,1}$-NMF and Hx-NMF show similar robustness by depressing the outliers. The ACC curves of EMMF are more stable than the competitors on all the datasets, because the outliers have less effect on the updating of centroids. Therefore, EMMF is able to handle the data with large outliers.

\begin{figure*}
\begin{center}

\subfigure[COIL20]{
\includegraphics[width=0.23\textwidth]{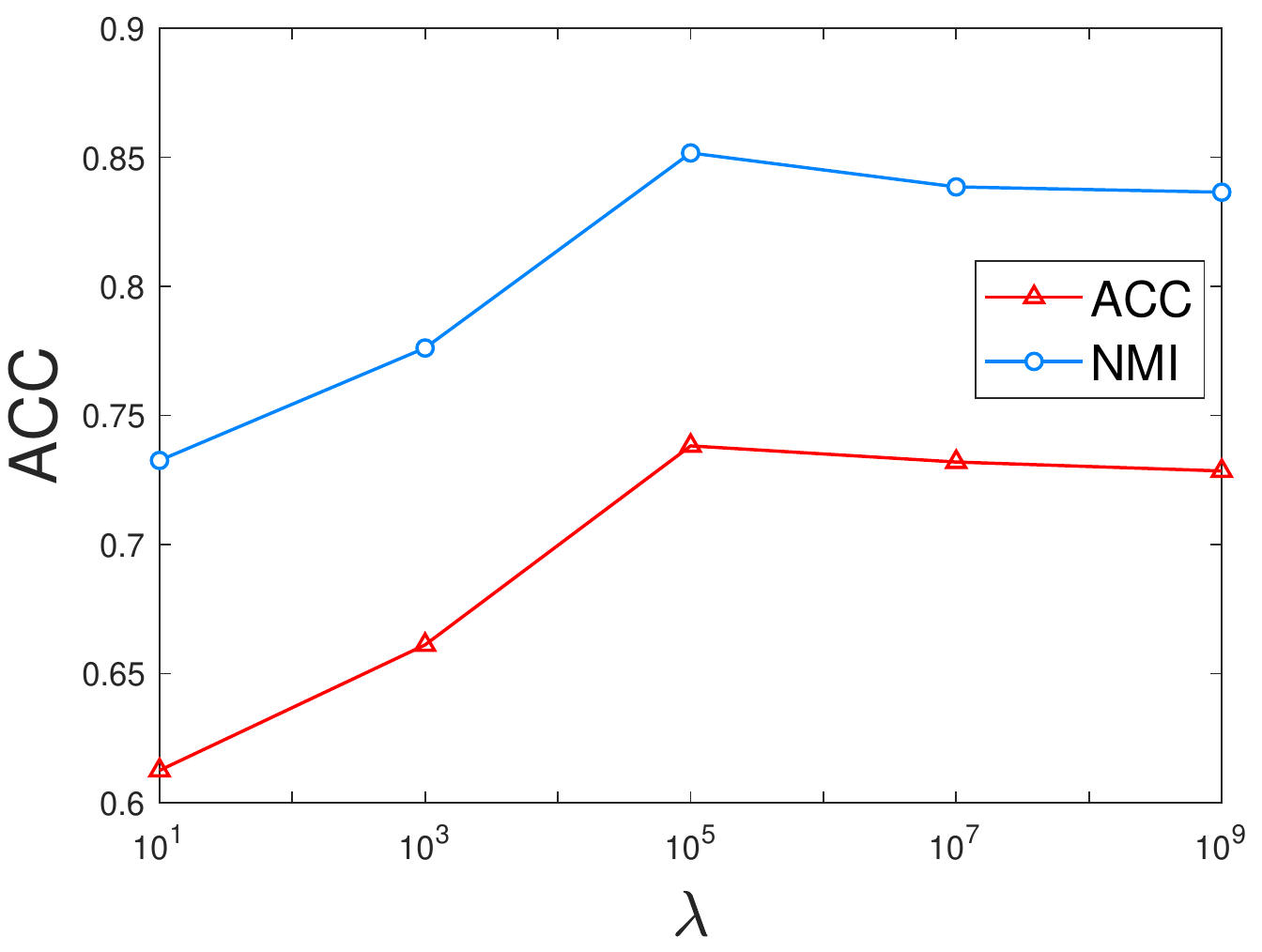}}
\hspace{1pt}
\subfigure[COIL100]{
\includegraphics[width=0.23\textwidth]{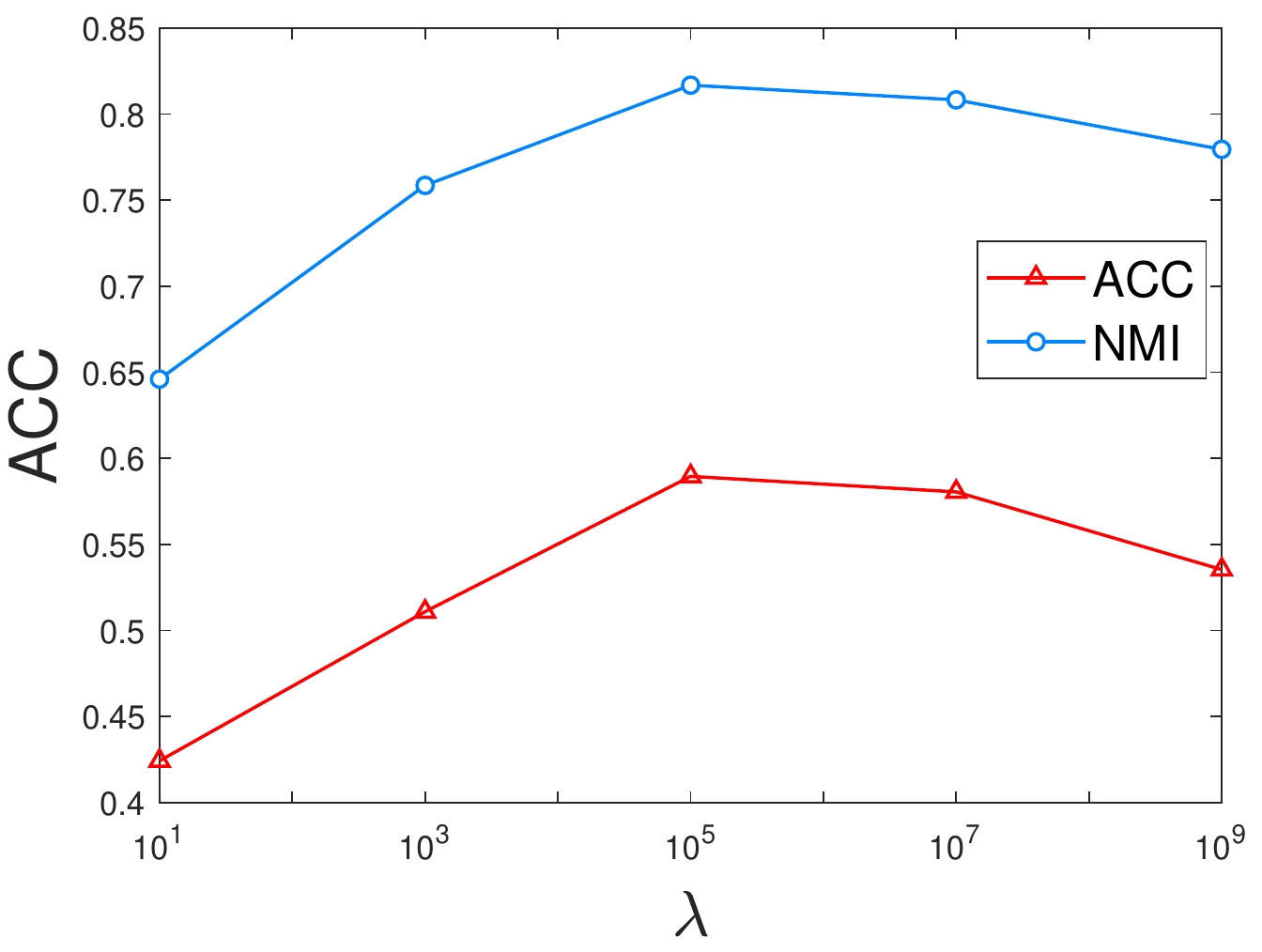}}
\hspace{1pt}
\subfigure[YALE]{
\includegraphics[width=0.23\textwidth]{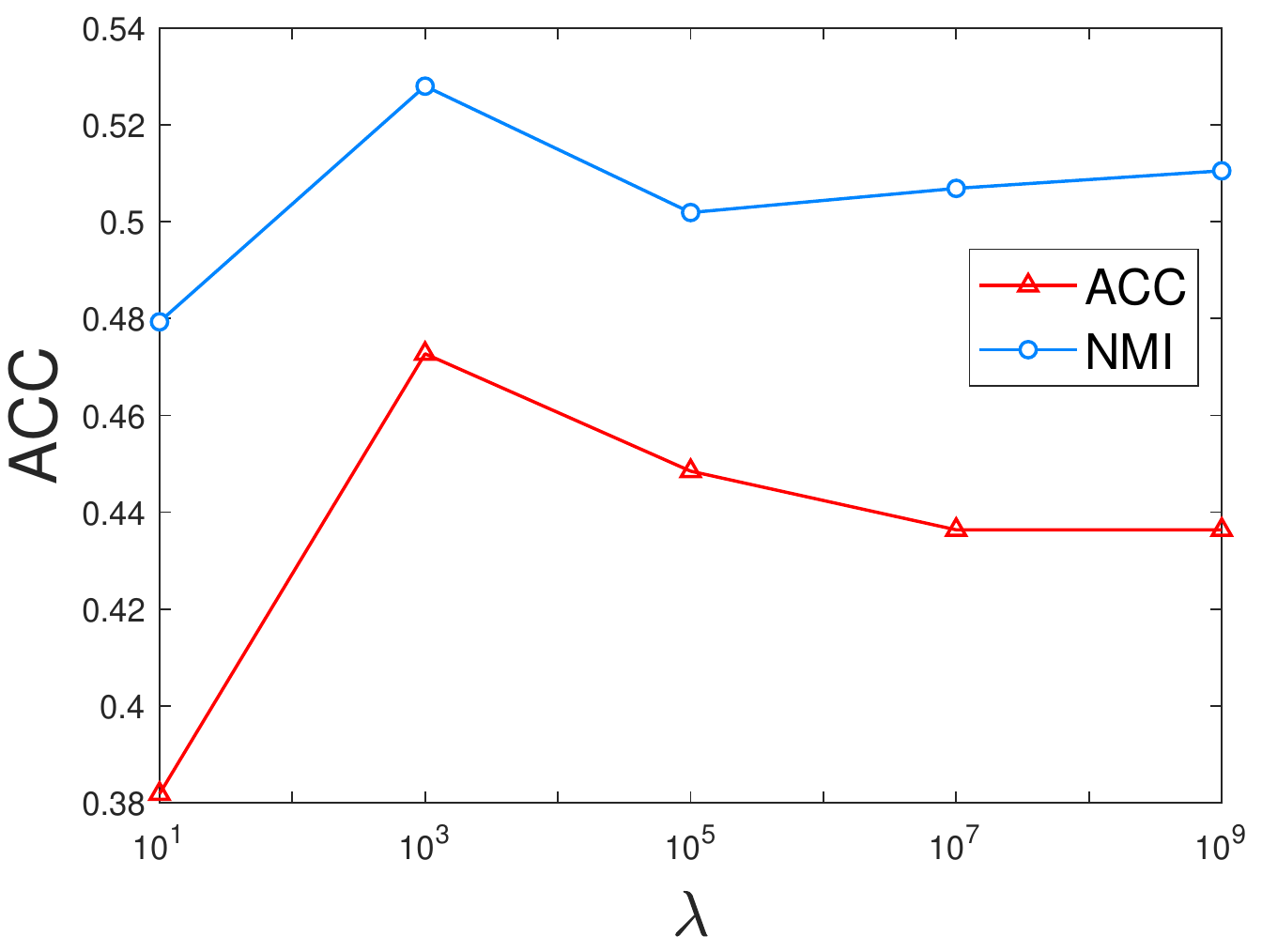}}
\hspace{1pt}
\subfigure[JAFFE]{
\includegraphics[width=0.23\textwidth]{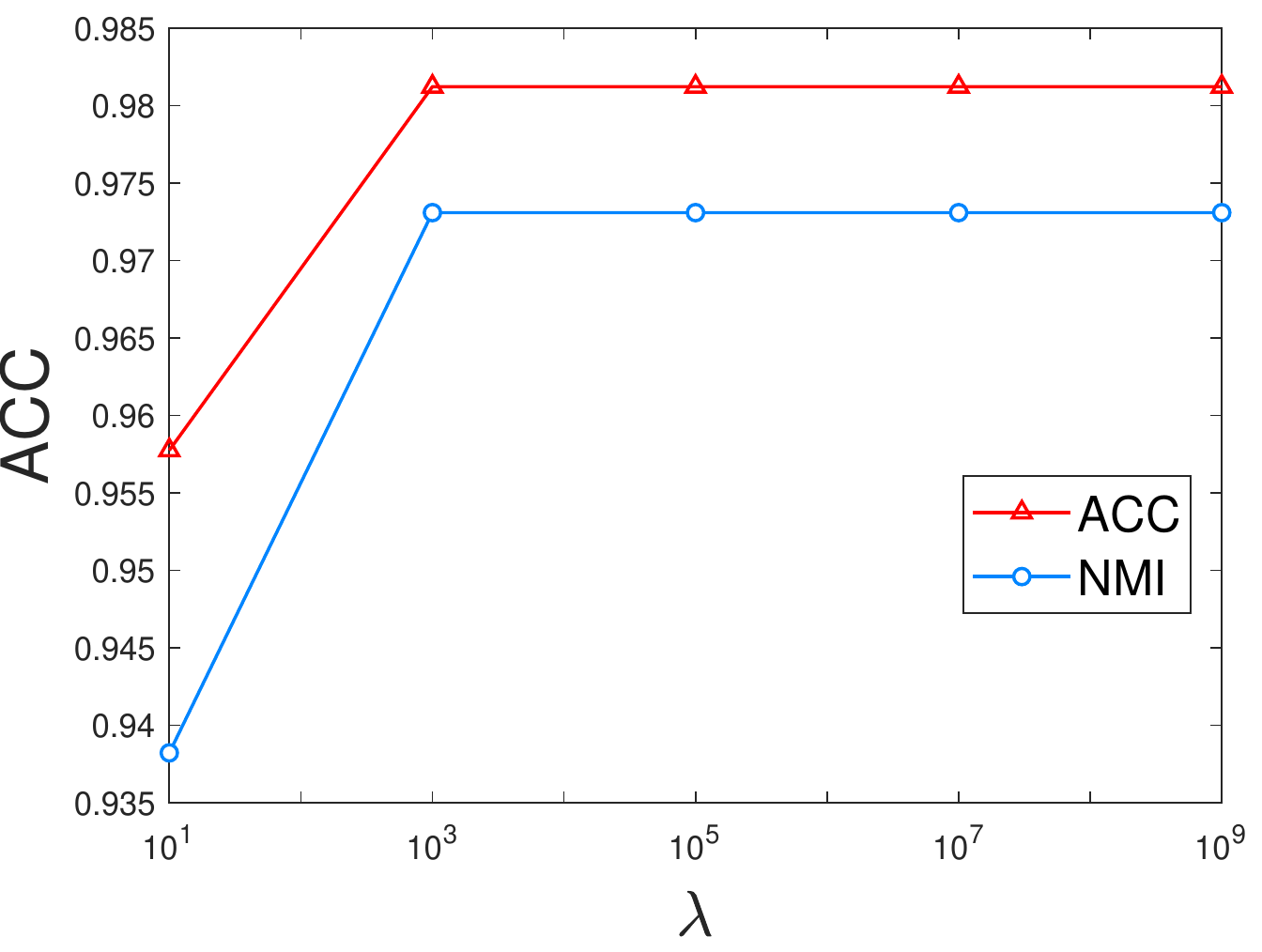}}
\end{center}

\begin{center}
\subfigure[UMIST]{
\includegraphics[width=0.23\textwidth]{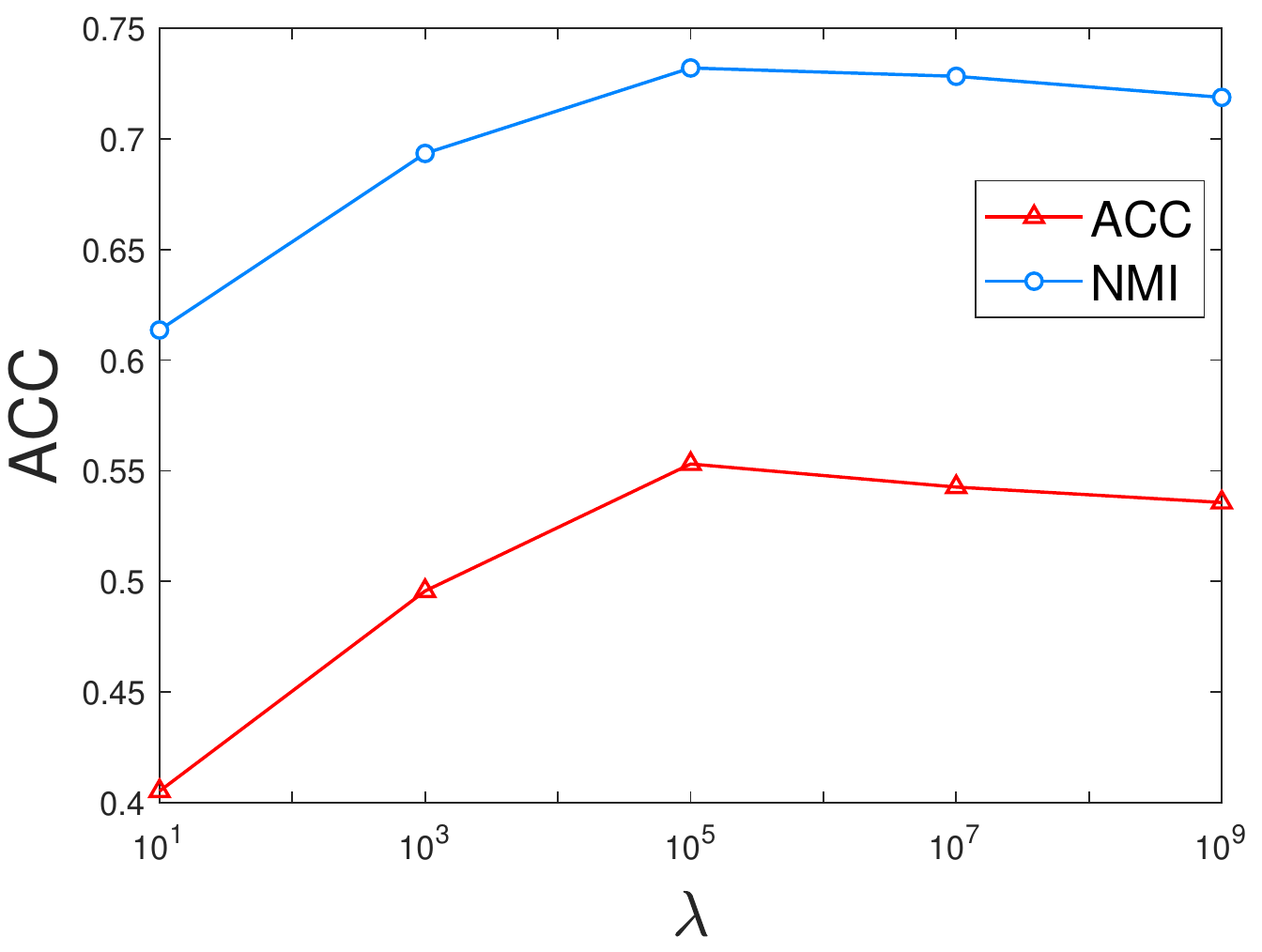}}
\hspace{1pt}
\subfigure[Mfeat]{
\includegraphics[width=0.23\textwidth]{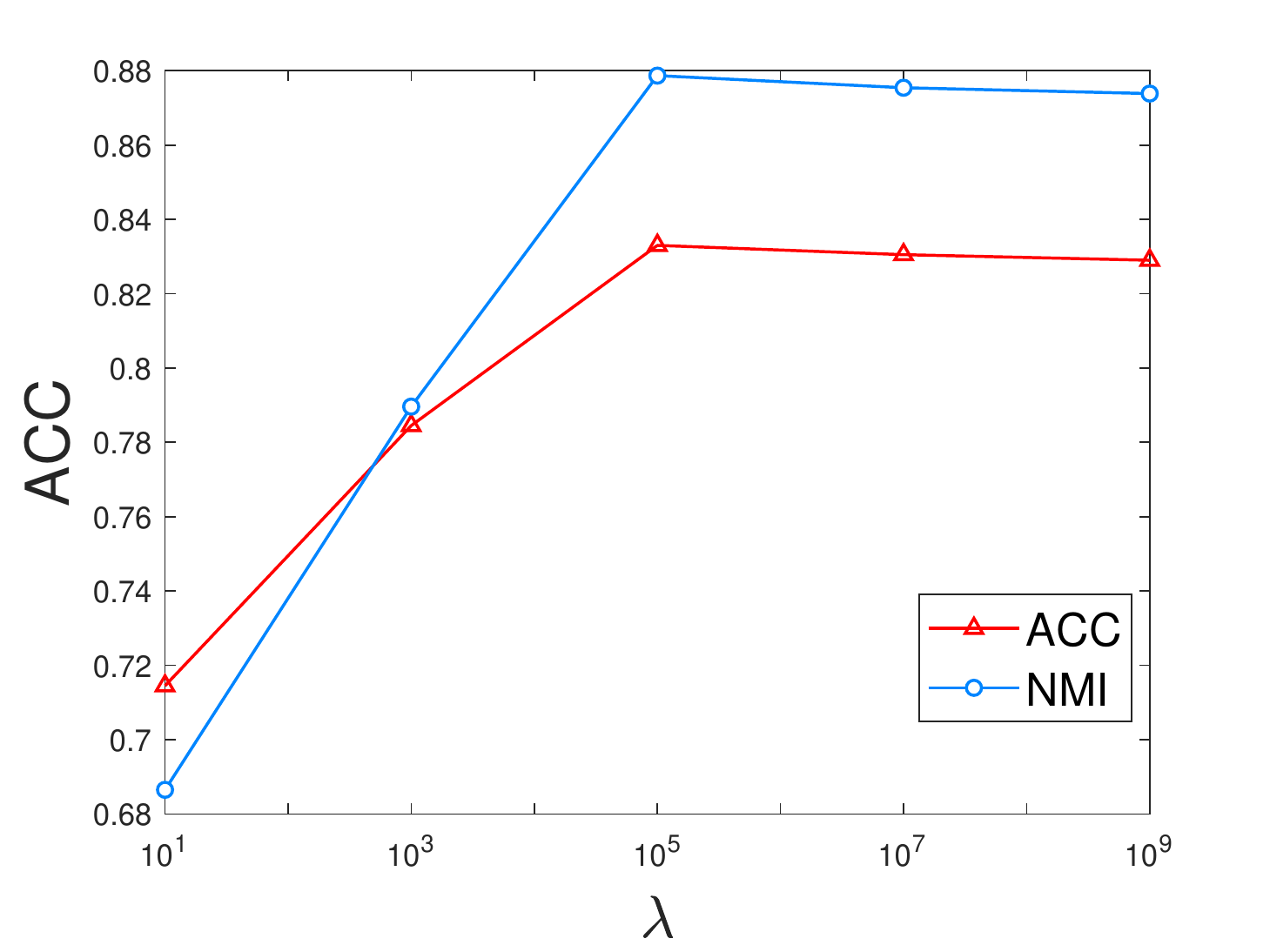}}
\hspace{1pt}
\subfigure[BA]{
\includegraphics[width=0.23\textwidth]{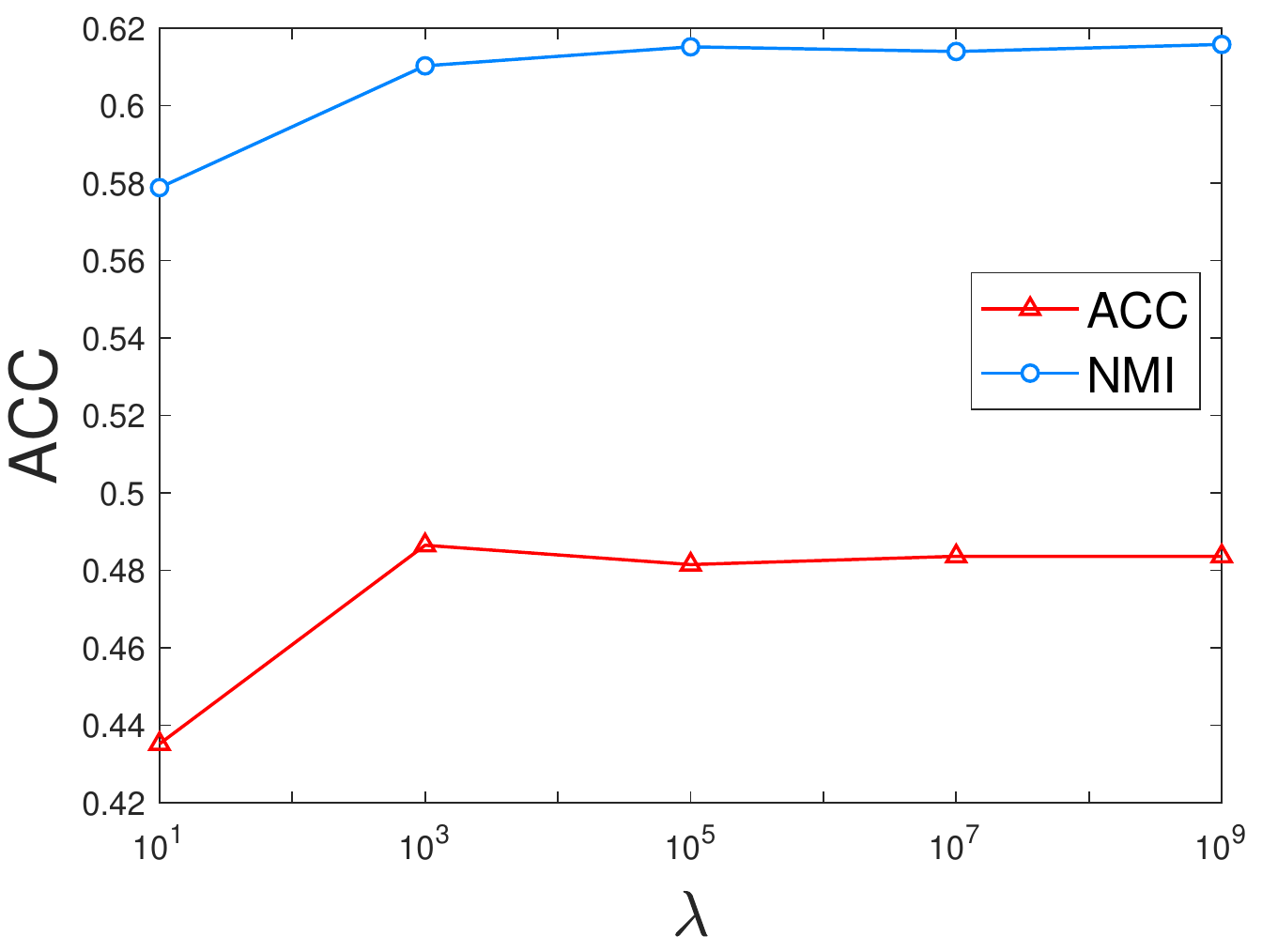}}
\hspace{1pt}
\subfigure[Movement]{
\includegraphics[width=0.23\textwidth]{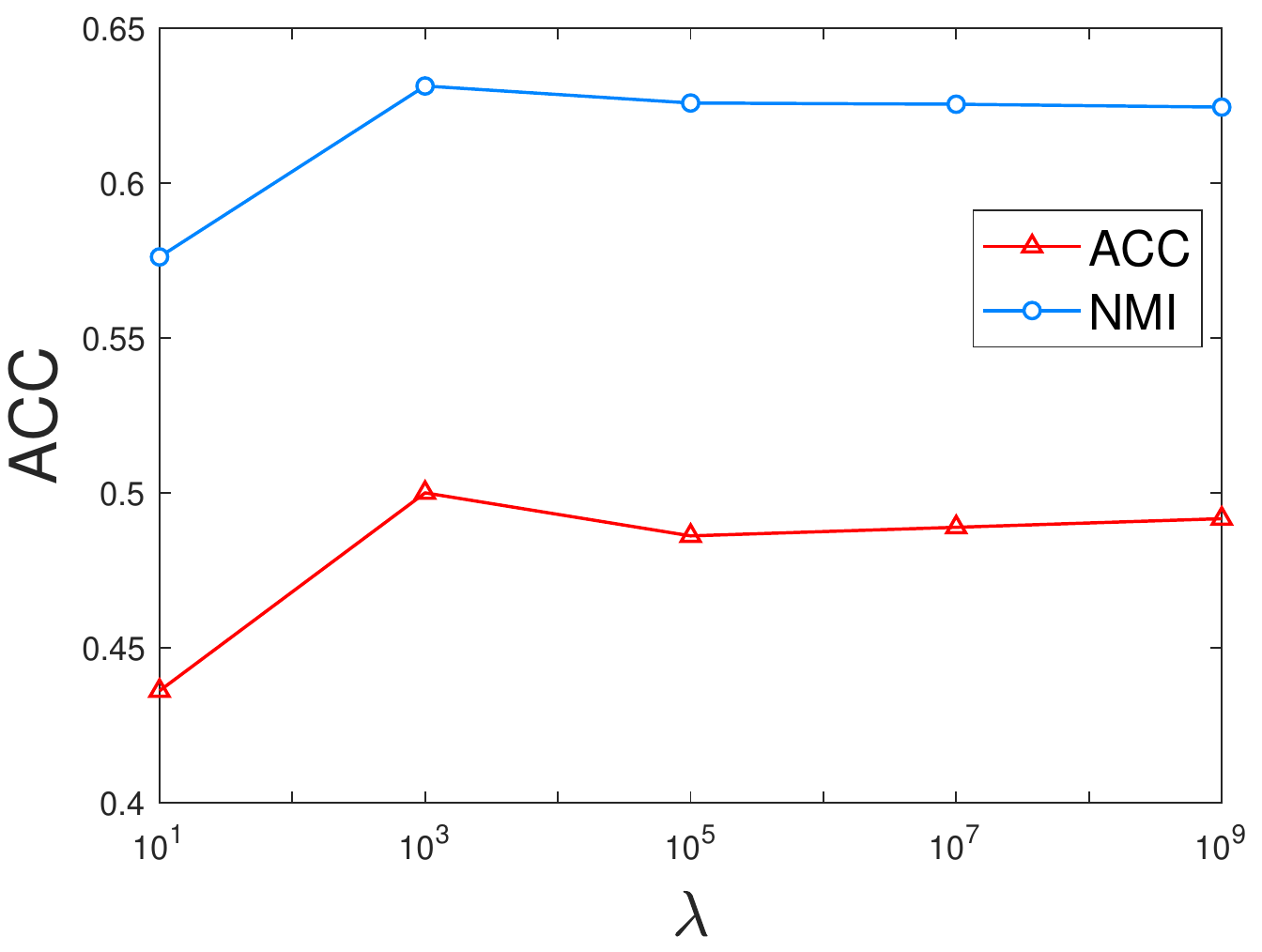}}
\end{center}
\caption{Performance of G-EMMF with different value of $\lambda$.}
\label{lambdacurve}
\end{figure*}

\section{Evaluation of G-EMMF}
\label{sec:expergemmf}

The effectiveness of G-EMMF is validated through experiments. The datasets used in this part are the same as in Section~\ref{sec:experemmfreal}.

\begin{figure}
\begin{center}
\subfigure[outliers with 4$\times$4 block noises]{
\includegraphics[width=0.45\textwidth]{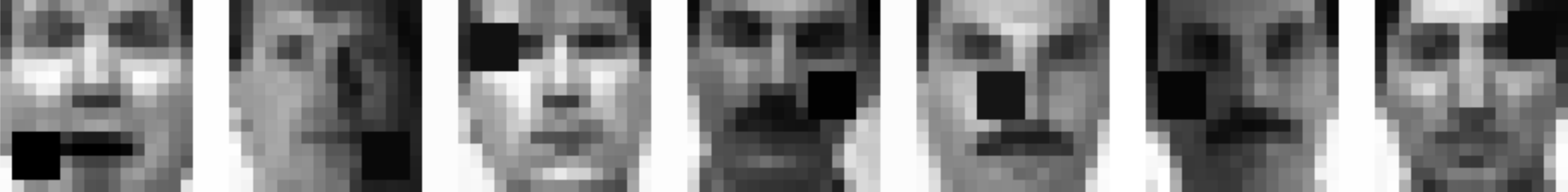}}

\subfigure[outliers with 6$\times$6 block noises]{
\includegraphics[width=0.45\textwidth]{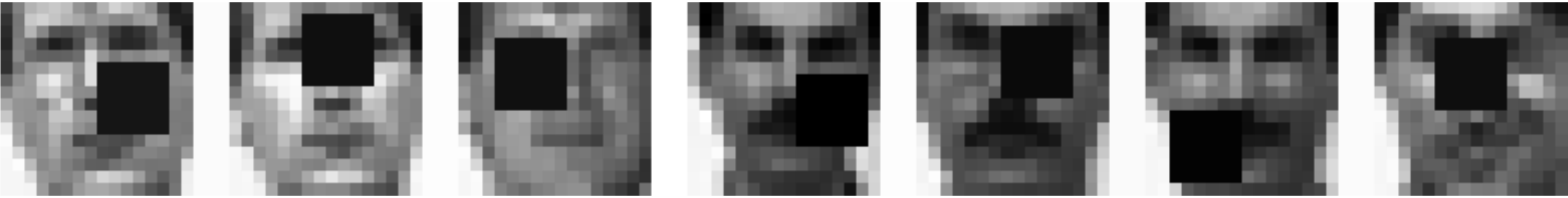}}

\subfigure[outliers with 8$\times$8 block noises]{
\includegraphics[width=0.45\textwidth]{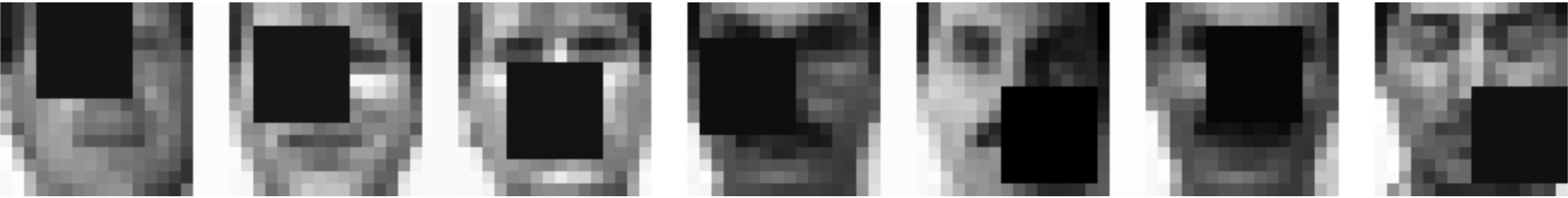}}

\subfigure[outliers with 10$\times$10 block noises]{
\includegraphics[width=0.45\textwidth]{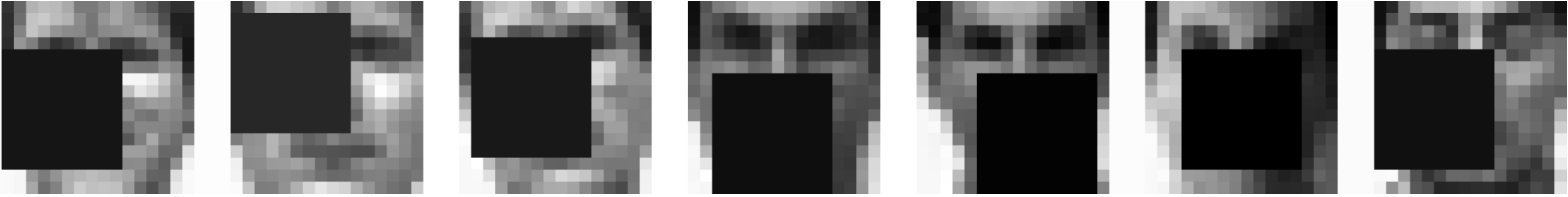}}
\end{center}
\caption{Illustration of outliers for graph regularized NMF methods..}
\label{occsample}
\end{figure}

\begin{table}
\caption{Performance of graph regularized NMF methods on YALE datasets with outliers. Best results are in bold face.}
\label{gemmftable-outlier}
\centering
\renewcommand\arraystretch{1.2}
\small
\begin{tabular}{|p{0.6cm}<{\centering}|p{1.4cm}<{\centering}|p{1.0cm}<{\centering}|p{1.0cm}<{\centering}|p{1.0cm}<{\centering}|p{1.0cm}<{\centering}|}

\hline
\multirow{7}*{ACC}& & 4$\times$4 & 6$\times$6 & 8$\times$8&10$\times$10\\
\cline{2-6}
&RMNMF  &0.4242 &0.3818 &0.3697 &0.3212\\
\cline{2-6}
&NLCF &0.3697 &0.3939 &0.3212 &0.3273\\
\cline{2-6}
&LCF &0.3576 &0.3697 &0.3333 &0.3333\\
\cline{2-6}
&LSNMF &0.2727 &0.1879 &0.1636 &0.1818 \\
\cline{2-6}
&SRMCF  &0.4242 &0.3939 &0.3455 &0.3455\\
\cline{2-6}
&NMFAN  &0.3636 &0.3818 &0.3333 &0.3515\\
\cline{2-6}
& G-EMMF  &\textbf{0.4364} &\textbf{0.4121} &\textbf{0.4061} &\textbf{0.3939}\\
\hline
\hline
\multirow{7}*{NMI}& & 4$\times$4 & 6$\times$6 & 8$\times$8&10$\times$10\\
\cline{2-6}
&RMNMF  &0.4818 &0.4298 &0.4464 &0.4041\\
\cline{2-6}
&NLCF &0.4439 &0.4273 &0.3623 &0.3775\\
\cline{2-6}
&LCF &0.4506 &0.4398 &0.3749 &0.3869\\
\cline{2-6}
&LSNMF   &0.2613 &0.1696 &0.1178 &0.1789\\
\cline{2-6}
&SRMCF  &0.4703 &0.4207 &0.4005 &0.3843\\
\cline{2-6}
&NMFAN  &0.4333 &0.4079 &0.3720 &0.3850\\
\cline{2-6}
& G-EMMF   &\textbf{0.5063} &\textbf{0.4708} &\textbf{0.4609} &\textbf{0.4519}\\
\hline
\end{tabular}

\end{table}

\textbf{Competitors:} six graph-regularized NMF methods are used for comparison, including
\begin{itemize}
\item RMNMF~\cite{rmnmf}: robust manifold NMF, which integrates the spectral clustering term into the objective of $\ell _{2,1}$-NMF,
\item NLCF~\cite{nlcf}: nonnegative local coordinate factorization, which uses local coordinate learning to encode the data structure.
\item LCF~\cite{lcf}: local coordinate concept factorization, which combines local coordinate learning and convex matrix factorization.
\item LSNMF~\cite{lsnmf}: local-centroids structured NMF, which employs multiple local centroids to represent a cluster.
\item SRMCF~\cite{srmcf}: self-representative manifold concept factorization, which optimizes the data graph during performing convex matrix factorization.
\item NMFAN~\cite{nmfan}: NMF with adaptive neighbors, which learns the local data relationship adaptively.
\end{itemize}
The best parameters of the competitors are found by searching the grid $\{10^{-5},10^{-4},\cdots,10^{5}\}$. The graph regularization parameter $\lambda$ in G-EMMF is found within the grid $\{10^{1},10^{3},\cdots,10^{9}\}$.  The methods are initialized with $k$-means, and the $0-1$ weighting 5 nearest neighbor graph is constructed as the similarity graph. The stop criteria for all the methods is set as 500 maximum iterations. After repeating the methods twenty times, the average ACC and NMI are reported.

\textbf{Performance}: the results of the graph regularized NMF methods are given in Table~\ref{gemmftable}. RMNMF fixes the data graph during optimization. The outliers may affect the graph quality, and further influence the clustering results. Compared with RMNMF, the other competitors show better performance because they learn the data relationship iteratively, such that the graph quality is improved. Similar with RMNMF, the proposed G-EMMF also relies on the input graph. It still achieves the best performance because the noises in the graph has less effect on the learned $\mathbf{U}$. In addition, compared with the results of EMMF in Table~\ref{emmftable}, G-EMMF improves the clustering performance on all the datasets, which demonstrates the necessity of exploiting data structure.

The convergence curves are shown in Fig.~\ref{convcurveg}, which verifies the effectiveness of the optimization algorithm. Besides, to investigate the effect of parameter $\lambda$ on the clustering results, we plot the curves of ACC and NMI with varying $\lambda$, as shown in Fig.~\ref{lambdacurve}. The clustering performance is not very sensitive to $\lambda$ within a wide range. When $\lambda$ becomes very large, the performance tends to decreases because the matrix approximation error increases. 

The outlier datasets used for the graph-regularized methods are different from those used in Section~\ref{sec:experemmfreal}, because LSNMF is inapplicable when the outliers are directly added into the data matrix. Given the faces images from YALE, we randomly select three images from each class and add block noises on them, so these images become outliers, as shown in Fig.~\ref{occsample}. The clustering results are given in Table~\ref{gemmftable-outlier}. We can see RMNMF outperforms several competitors due to the robust formulation. LSNMF shows unsatisfying results since it aims to learn a graph with exact $c$ connected components, which is unrealistic on the data with outliers. G-EMMF outperforms the competitors on different scale of block noises.

\section{Conclusion and future work}
\label{sec:conclusion}
This paper proposes an Entropy Minimizing Matrix Factorization (EMMF) framework. A novel matrix factorization formulation with entropy loss is designed. Instead of approximating all the samples, the proposed model pursues an imbalance residue distribution, and the outliers with relative large errors are not taken into consideration. In this way, the outliers have less effect on the learned centroids. In addition, the graph regularized EMMF is introduced to handle the data with complex structures. The models can be solved by the suggested optimization algorithms efficiently. Experiments on various datasets demonstrate the robustness of the proposed methods, and show their applicability on data without outliers. Comparison with the state-of-the-arts validates the superiorities of our methods.

In the future work, we are desired to develop the deep model of EMMF, and apply it in large-scale clustering tasks. We also want to incorporate the graph information into EMMF without introducing any additional parameter.

%\section*{Acknowledgments}
%This work was supported by The National Key Research and Development Program of China under Grant 2018YFB1107400, and The National Natural Science Foundation of China under Grant 61871470.

%\bibliographystyle{IEEEtran}
%\bibliography{main}

\begin{thebibliography}{10}
\providecommand{\url}[1]{#1}
\csname url@samestyle\endcsname
\providecommand{\newblock}{\relax}
\providecommand{\bibinfo}[2]{#2}
\providecommand{\BIBentrySTDinterwordspacing}{\spaceskip=0pt\relax}
\providecommand{\BIBentryALTinterwordstretchfactor}{4}
\providecommand{\BIBentryALTinterwordspacing}{\spaceskip=\fontdimen2\font plus
\BIBentryALTinterwordstretchfactor\fontdimen3\font minus
  \fontdimen4\font\relax}
\providecommand{\BIBforeignlanguage}[2]{{%
\expandafter\ifx\csname l@#1\endcsname\relax
\typeout{** WARNING: IEEEtran.bst: No hyphenation pattern has been}%
\typeout{** loaded for the language `#1'. Using the pattern for}%
\typeout{** the default language instead.}%
\else
\language=\csname l@#1\endcsname
\fi
#2}}
\providecommand{\BIBdecl}{\relax}
\BIBdecl

\bibitem{paatero1994positive}
P.~Paatero and U.~Tapper, ``Positive matrix factorization: A non-negative
  factor model with optimal utilization of error estimates of data values,''
  \emph{Environmetrics}, vol.~5, no.~2, pp. 111--126, 1994.

\bibitem{nmf}
D.~Lee and H.~Seung, ``Algorithms for nonnegative matrix factorization,''
  \emph{Advances in Neural Information Processing Systems}, vol.~13, no.~6, pp.
  556--562, 2000.

\bibitem{xiaofeigraph}
Z.~Li, J.~Tang, and X.~He, ``Robust structured nonnegative matrix factorization
  for image representation,'' \emph{{IEEE} Transactions on Neural Networks and
  Learning Systems}, vol.~29, no.~5, pp. 1947--1960, 2018.

\bibitem{docu}
Y.~E. Salehani, E.~Arabnejad, A.~Rahiche, A.~Bakhta, and M.~Cheriet,
  ``Msdb-nmf: Multispectral document image binarization framework via
  non-negative matrix factorization approach,'' \emph{{IEEE} Transactions on
  Image Processing}, vol.~29, pp. 9099--9112, 2020.

\bibitem{zhangzihan}
Z.~Zhang, Q.~Wang, and Y.~Yuan, ``Hyperspectral unmixing {VIA} {L1/4}
  sparsity-constrained multilayer {NMF},'' in \emph{{IEEE} International
  Geoscience and Remote Sensing Symposium}, 2019, pp. 2143--2146.

\bibitem{recom}
J.~Chen, C.~Wang, S.~Zhou, Q.~Shi, J.~Chen, Y.~Feng, and C.~Chen, ``Fast
  adaptively weighted matrix factorization for recommendation with implicit
  feedback,'' in \emph{{AAAI} Conference on Artificial Intelligence}, 2020, pp.
  3470--3477.

\bibitem{dingequivalence}
C.~Ding and X.~He, ``On the equivalence of nonnegative matrix factorization and
  spectral clustering,'' in \emph{{SIAM} International Conference on Data
  Mining}, 2005, pp. 606--610.

\bibitem{rmnmf}
J.~Huang, F.~Nie, H.~Huang, and C.~Ding, ``Robust manifold nonnegative matrix
  factorization,'' \emph{ACM Transactions on Knowledge Discovery from Data},
  vol.~8, no.~3, p.~11, 2014.

\bibitem{ldanmf}
X.~Li, M.~Chen, and Q.~Wang, ``Discrimination-aware projected matrix
  factorization,'' \emph{{IEEE} Transactions on Knowledge and Data
  Engineering}, vol.~32, no.~4, pp. 809--814, 2020.

\bibitem{lcf}
H.~Liu, Z.~Yang, J.~Yang, Z.~Wu, and X.~Li, ``Local coordinate concept
  factorization for image representation,'' \emph{{IEEE} Transactions on Neural
  Networks and Learning Systems}, vol.~25, no.~6, pp. 1071--1082, 2014.

\bibitem{srmcf}
S.~Ma, L.~Zhang, W.~Hu, Y.~Zhang, J.~Wu, and X.~Li, ``Self-representative
  manifold concept factorization with adaptive neighbors for clustering,'' in
  \emph{International Joint Conference on Artificial Intelligence,}, 2018, pp.
  2539--2545.

\bibitem{lccf}
D.~Cai, X.~He, and J.~Han, ``Locally consistent concept factorization for
  document clustering,'' \emph{{IEEE} Transactions on Knowledge and Data
  Engineering}, vol.~23, no.~6, pp. 902--913, 2011.

\bibitem{cf}
W.~Xu and Y.~Gong, ``Document clustering by concept factorization,'' in
  \emph{{ACM} {SIGIR} Conference on Research and Development in Information
  Retrieval}, 2004, pp. 202--209.

\bibitem{tnnconcept}
M.~Chen and X.~Li, ``Concept factorization with local centroids,'' \emph{IEEE
  Transactions on Neural Networks and Learning Systems}, DOI:
  10.1109/TNNLS.2020.3027068,2020.

\bibitem{gnmf}
D.~Cai, X.~He, J.~Han, and T.~Huang, ``Graph regularized nonnegative matrix
  factorization for data representation,'' \emph{{IEEE} Transactions on Pattern
  Analysis and Machine Intelligence}, vol.~33, no.~8, pp. 1548--1560, 2011.

\bibitem{lefeigraph}
L.~Zhang, Q.~Zhang, B.~Du, J.~You, and D.~Tao, ``Adaptive manifold regularized
  matrix factorization for data clustering,'' in \emph{International Joint
  Conference on Artificial Intelligence}, 2017, pp. 3399--3405.

\bibitem{chenneru}
M.~Chen, Q.~Wang, and X.~Li, ``Adaptive projected matrix factorization method
  for data clustering,'' \emph{Neurocomputing}, vol. 306, pp. 182--188, 2018.

\bibitem{tnnrse}
M.~Chen and X.~Li, ``Robust matrix factorization with spectral embedding,''
  \emph{IEEE Transactions on Neural Networks and Learning Systems}, DOI:
  10.1109/TNNLS.2020.3027351,2020.

\bibitem{missing1}
X.~He, J.~Tang, X.~Du, R.~Hong, T.~Ren, and T.~Chua, ``Fast matrix
  factorization with nonuniform weights on missing data,'' \emph{{IEEE}
  Transactions on Neural Networks and Learning Systems}, vol.~31, no.~8, pp.
  2791--2804, 2020.

\bibitem{missing2}
N.~Kallus, X.~Mao, and M.~Udell, ``Causal inference with noisy and missing
  covariates via matrix factorization,'' in \emph{Advances in Neural
  Information Processing Systems}, 2018, pp. 6921--6932.

\bibitem{missing3}
Z.~Lin, C.~Xu, and H.~Zha, ``Robust matrix factorization by majorization
  minimization,'' \emph{{IEEE} Transactions on Pattern Analysis and Machine
  Intelligence}, vol.~40, no.~1, pp. 208--220, 2018.

\bibitem{missing4}
S.~Ohsawa, Y.~Obara, and T.~Osogami, ``Gated probabilistic matrix
  factorization: Learning users' attention from missing values,'' in
  \emph{International Joint Conference on Artificial Intelligence}, 2016, pp.
  1888--1894.

\bibitem{deep1}
S.~Xu, C.~Zhang, and J.~Zhang, ``Bayesian deep matrix factorization network for
  multiple images denoising,'' \emph{Neural Networks}, vol. 123, pp. 420--428,
  2020.

\bibitem{deep2}
S.~Wei, J.~Wang, G.~Yu, C.~Domeniconi, and X.~Zhang, ``Multi-view multiple
  clusterings using deep matrix factorization,'' in \emph{{AAAI} Conference on
  Artificial Intelligence}, 2020, pp. 6348--6355.

\bibitem{l1nmf}
Q.~Ke and T.~Kanade, ``Robust l\({}_{\mbox{1}}\) norm factorization in the
  presence of outliers and missing data by alternative convex programming,'' in
  \emph{{IEEE} Conference on Computer Vision and Pattern Recognition}.\hskip
  1em plus 0.5em minus 0.4em\relax {IEEE} Computer Society, 2005, pp. 739--746.

\bibitem{l21nmf}
D.~Kong, C.~Ding, and H.~Huang, ``Robust nonnegative matrix factorization using
  l21-norm,'' in \emph{{ACM} Conference on Information and Knowledge
  Management}.\hskip 1em plus 0.5em minus 0.4em\relax {ACM}, 2011, pp.
  673--682.

\bibitem{hxnmf}
Q.~Wang, X.~He, X.~Jiang, and X.~Li, ``Robust bi-stochastic graph regularized
  matrix factorization for data clustering,'' \emph{{IEEE} Transactions on
  Pattern Analysis and Machine Intelligence}, DOI: 10.1109/TPAMI.2020.3007673,
  2020.

\bibitem{gaocap}
H.~Gao, F.~Nie, T.~W. Cai, and H.~Huang, ``Robust capped norm nonnegative
  matrix factorization: Capped norm {NMF},'' in \emph{{ACM} International
  Conference on Information and Knowledge Management}.\hskip 1em plus 0.5em
  minus 0.4em\relax {ACM}, 2015, pp. 871--880.

\bibitem{lefeigraph2}
N.~Zhao, L.~Zhang, B.~Du, Q.~Zhang, J.~You, and D.~Tao, ``Robust dual
  clustering with adaptive manifold regularization,'' \emph{{IEEE} Transactions
  on Knowledge and Data Engineering}, vol.~29, no.~11, pp. 2498--2509, 2017.

\bibitem{robust1}
C.~F{\'{e}}votte and N.~Dobigeon, ``Nonlinear hyperspectral unmixing with
  robust nonnegative matrix factorization,'' \emph{{IEEE} Transactions Image
  Processing}, vol.~24, no.~12, pp. 4810--4819, 2015.

\bibitem{dingorth}
C.~Ding, T.~Li, W.~Peng, and H.~Park, ``Orthogonal nonnegative matrix
  t-factorizations for clustering,'' in \emph{Proceedings of the 12th ACM
  SIGKDD international conference on Knowledge discovery and data mining},
  2006, pp. 126--135.

\bibitem{correntropy}
L.~Du, X.~Li, and Y.~Shen, ``Robust nonnegative matrix factorization via
  half-quadratic minimization,'' in \emph{{IEEE} International Conference on
  Data Mining}, 2012, pp. 201--210.

\bibitem{l12}
L.~Yang, L.~Jing, and M.~K. Ng, ``Robust and non-negative collective matrix
  factorization for text-to-image transfer learning,'' \emph{{IEEE}
  Transactions Image Processing}, vol.~24, no.~12, pp. 4701--4714, 2015.

\bibitem{l12nmf2}
R.~Huang, X.~Li, and L.~Zhao, ``Spectral-spatial robust nonnegative matrix
  factorization for hyperspectral unmixing,'' \emph{{IEEE} Transactions on
  Geoscience and Remote Sensing}, vol.~57, no.~10, pp. 8235--8254, 2019.

\bibitem{hypersurface}
A.~Hamza and D.~Brady, ``Reconstruction of reflectance spectra using robust
  nonnegative matrix factorization,'' \emph{{IEEE} Transactons on Signal
  Processing}, vol.~54, no.~9, pp. 3637--3642, 2006.

\bibitem{shannon}
C.~Shannon, ``A mathematical theory of communication,'' \emph{The Bell system
  technical journal}, vol.~27, no.~3, pp. 379--423, 1948.

\bibitem{logsum}
T.~Cover and J.~Thomas, \emph{Elements of information theory {(2.} ed.)}.\hskip
  1em plus 0.5em minus 0.4em\relax Wiley, 2006.

\bibitem{dingconvex}
C.~Ding, T.~Li, and M.~Jordan, ``Convex and semi-nonnegative matrix
  factorizations,'' \emph{{IEEE} Transactions on Pattern Analysis and Machine
  Intelligence}, vol.~32, no.~1, pp. 45--55, 2010.

\bibitem{sym}
D.~Kuang, H.~Park, and C.~Ding, ``Symmetric nonnegative matrix factorization
  for graph clustering,'' in \emph{{SIAM} International Conference on Data
  Mining}, 2012, pp. 106--117.

\bibitem{yale}
X.~He, S.~Yan, Y.~Hu, P.~Niyogi, and H.~Zhang, ``Face recognition using
  laplacianfaces,'' \emph{{IEEE} Transactions on Pattern Analysis and Machine
  Intelligence}, vol.~27, no.~3, pp. 328--340, 2005.

\bibitem{jaffe}
M.~Lyons, J.~Budynek, and S.~Akamatsu, ``Automatic classification of single
  facial images,'' \emph{{IEEE} Transactions on Pattern Analysis and Machine
  Intelligence}, vol.~21, no.~12, pp. 1357--1362, 1999.

\bibitem{umist}
D.~Graham and N.~Allinson, ``Characterising virtual eigensignatures for general
  purpose face recognition,'' \emph{Face Recognition Form Theory to
  Applications}, vol. 163, no.~2, pp. 446--456, 1998.

\bibitem{uci}
\BIBentryALTinterwordspacing
D.~Dua and C.~Graff, ``{UCI} machine learning repository,'' 2017. [Online].
  Available: \url{http://archive.ics.uci.edu/ml}
\BIBentrySTDinterwordspacing

\bibitem{ba}
P.~Belhumeur, J.~Hespanha, and D.~Kriegman, ``Eigenfaces vs. fisherfaces:
  Recognition using class specific linear projection,'' \emph{{IEEE}
  Transactions on Pattern Analysis and Machine Intelligence}, vol.~19, no.~7,
  pp. 711--720, 1997.

\bibitem{pnmf}
T.~Brouwer, J.~Frellsen, and P.~Li{\`{o}}, ``Comparative study of inference
  methods for bayesian nonnegative matrix factorisation,'' in \emph{European
  Conference on Machine Learning and Knowledge Discovery in Databases}, vol.
  10534, 2017, pp. 513--529.

\bibitem{gsnmf}
J.~Pan and N.~Gillis, ``Generalized separable nonnegative matrix
  factorization,'' \emph{{IEEE} Transactions on Pattern Analysis and Machine
  Intelligence}, DOI: 10.1109/TPAMI.2019.2956046, 2019.

\bibitem{kmeans}
J.~Macqueen, ``Some methods for classification and analysis of multivariate
  observations,'' in \emph{Berkeley Symposium on Mathematical Statistics and
  Probability}, 1967, pp. 281--297.

\bibitem{nlcf}
Y.~Chen, J.~Zhang, D.~Cai, W.~Liu, and X.~He, ``Nonnegative local coordinate
  factorization for image representation,'' \emph{{IEEE} Transactions on Image
  Processing}, vol.~22, no.~3, pp. 969--979, 2013.

\bibitem{lsnmf}
H.~Gao, F.~Nie, and H.~Huang, ``Local centroids structured non-negative matrix
  factorization,'' in \emph{{AAAI} Conference on Artificial Intelligence},
  2017, pp. 1905--1911.

\bibitem{nmfan}
S.~Huang, Z.~Xu, and F.~Wang, ``Nonnegative matrix factorization with adaptive
  neighbors,'' in \emph{International Joint Conference on Neural Networks},
  2017, pp. 486--493.

\end{thebibliography}

%\begin{IEEEbiography}[{\includegraphics[width=1in,height=1.25in,clip,keepaspectratio]{Chen.pdf}}]{Mulin Chen} received the B.E. degree in software engineering and the Ph.D. degree in computer application technology from Northwestern Polytechnical University, Xi'an, China, in 2014 and 2019 respectively. He is currently a researcher with the Center for OPTical IMagery Analysis and Learning (OPTIMAL), Northwestern Polytechnical University, Xi'an, China.
% His current research intersts include computer vision and machine learning.
%\end{IEEEbiography}
%
%
%\begin{IEEEbiographynophoto} {Xuelong Li} (M'02-SM'07-F'12) is currently a Full Professor with the  Center for OPTical IMagery Analysis and Learning (OPTIMAL), Northwestern Polytechnical University, Xi'an 710072, China.

\ifCLASSOPTIONcaptionsoff
  \newpage
\fi

\end{document}